\documentclass[fleqn,usenatbib]{mnras}

\usepackage{subcaption}
\usepackage{amsmath}
\usepackage{amssymb}

\usepackage{newtxtext,newtxmath}
\usepackage[T1]{fontenc}
\usepackage{multirow}

\DeclareRobustCommand{\VAN}[3]{#2}
\let\VANthebibliography\thebibliography
\def\thebibliography{\DeclareRobustCommand{\VAN}[3]{##3}\VANthebibliography}

\usepackage{graphicx}

\title[bending delays in PSR-BH (Kerr) binaries]{On the effect of the light bending phenomenon for a pulsar in a binary with a Kerr black hole }

\author[Debnath et al.]{Jyotijwal Debnath,$^{1, 2}$\thanks{E-mail: jdebnath@imsc.res.in}
and Manjari Bagchi,$^{1,2}$ \\
$^{1}$The Institute of Mathematical Sciences, C. I. T. campus, Taramani, Chennai, 600113, India\\
$^{2}$Homi Bhabha National Institute, Training School Complex, Anushakti Nagar, Mumbai 400094, India}

\date{Accepted XXX. Received YYY; in original form ZZZ}

\pubyear{2024}

\begin{document}
\label{firstpage}
\pagerange{\pageref{firstpage}--\pageref{lastpage}}
\maketitle

\defcitealias{dbb23}{Paper-I}

\begin{abstract}

We study the effect of light-bending on the signal of a pulsar in binaries with rotating black hole companions, focusing on stellar mass black holes. We show that the impacts of various parameters on the bending delays visually match with those observed for a non-rotating black holes, because the magnitude of the spin as well as the orientation of the spin axis of the black hole introduce changes in the nanosecond order and other parameters do so in the microsecond order. Consequently, the distortion of the beam and the resulting changes in the pulse shape are minimally influenced by spin-related parameters of the black hole. We also investigate the impact of various parameters on the difference of the delays with and without the spin of the black hole and notice nanosecond scale discontinuities at orbital phases where the path of the light ray changes its direction with respect to the direction of the spin of the black hole. Moreover, as in the Schwarzschild case, the bending delays become irregular (on the microsecond scale) near the superior conjunction. We also explore the effect of bending on the pulse profiles and bending delays if the companion of the pulsar is a rotating super-massive black hole. We find significant enhancement and change in the shape of the profiles at and near the superior conjunction in comparison to stellar mass black holes. Moreover, bending delays are about three orders of magnitude higher than those in case of the stellar mass black holes.

\end{abstract}

\begin{keywords}
pulsars: general -- stars: neutron -- stars: black holes -- gravitation -- binaries: general -- stars: rotation
\end{keywords}

\section{Introduction}

Rotation powered radio pulsars, in particular, the old and recycled millisecond pulsars are so extremely stable rotators that they are considered as the most precise clocks in the Universe. This property make them useful for various scientific purposes, e.g., the detection of low-frequency gravitational waves \citep[and references therein]{agazie24}, construction of a relativistic deep space positioning system \citep{bcm11, trc11}, testing various theories of gravity \citep{bt14}, etc. However, one first needs to eliminate various delays experienced by the pulsar signal including the clock delays, the dispersion delay due to the interstellar medium, the delay due to the motion of the pulsar as well as the observer (the earth), the curvature of the spacetime around the massive planets in the solar system, etc \citep{lorimer04}. Signals from pulsars in binary systems experience some additional delays due to the orbital motion of the pulsar as well as the curvature of the spacetime near the companion \citep{lorimer04}. The curvature of the spacetime near the companion affects the pulsar signal in multiple ways. The first is the propagation delay or the Shapiro delay and the second one is the light bending delay that arises due to the starting point of the photons in the pulsar beam appearing to be different than their actual starting points \citep[and references therein]{dbb23}. The bending delay occurs in two ways \citep[and references therein]{dbb23}, the first one is the `longitudinal bending delay' due to the apparent change in the longitudinal coordinates of the light rays and the second one is the `latitudinal bending delay' due to the apparent change in the (co)-latitudinal coordinates of the light rays.

Similar to the Shapiro delay, the total bending delays are also stronger for edge-on systems, especially at the orbital phases near the superior conjunction. So, it might be difficult to decouple these delays \citep{hkc22}. However, the value of the Shapiro delay depends only on the orbital parameters and the masses of the pulsar and the companions. On the other hand, in addition to these parameters, the orientation of the spin and the magnetic axes of the pulsar and the orientation of the spin axis (if any) of the companion might affect the amount of bending delays. Conversely, we will show in this paper that the magnitude of the spin of the companion does not impart in perceptible effect on the phenomenon of bending. As the bending delays are expected to be larger for pulsar-black hole binaries, it is wiser to study these delays for such systems.

All of the pioneering works investigating bending delays \citep{sch90, dk95, RL06} were based on the theory of gravitational lensing in Schwarzschild spacetime. In this framework, it is assumed that the path of the light ray is confined in the plane of the lens and the bending angle is directly proportional to the mass of the gravitating body and inversely proportional to the impact parameter of the light ray. In our previous work \citep[][hereafter Paper-I]{dbb23}, we demonstrated the features of the bending delay in the signal of a pulsar with a Schwarzschild black hole (stellar mass) companion by solving the equations for null geodesics in Schwarzschild spacetime as given by \citet{chandrasekhar84}. More specifically, we noticed that although the above formalism was sufficient for double neutron star systems, it was inadequate for pulsar-black hole binaries. We also showed that in addition to the delay, the phenomenon of bending leads to a distortion of the beam resulting in a change in the pulse profile, especially when both of the orbital phase and the inclination angle are close to $90^{\circ}$. The use of Schwarzschild black hole was not unreasonable, as via modelling stellar evolution in binaries, \citet{css21} found that about $80 \%$ of pulsar-black hole systems have the black hole as the first born object, and in that case, the spin of the  black hole can be ignored based on the study by \citet{fm19}. However, these studies do not consider pulsar-black hole binaries born via dynamical formation channel, which might be the more dominant in the dense regions like globular clusters and near the Galactic centre. Moreover, considering the uncertainties involved in the theories and computations of stellar evolution, one can not rule out the existence of pulsar-black hole binaries where the rotation of the black hole is significant. As the possibility of a pulsar with a rotating black hole as the companion remains and in the present paper, we investigate the effect of the spin of the black hole companion on the effects of the light bending phenomenon. As explained in \citetalias{dbb23}, the motion of the gravitating body (the companion of the pulsar) is ignored. 

There are some earlier studies on the effects of the spin of the companion on the propagation delay of the signal of the pulsar. The first one was done by \citet{Laguna1997}. Afterward, \citet{wk1999} studied this delay, which they named the frame-dragging propagation delay. The effect of the companion's spin on the propagation delay was also studied by \citet{Tartaglia:2005} and \citet{RMT2005}. \citet{km02} performed analytical calculations related to the propagation of electromagnetic waves in time-varying gravitational fields of arbitrary-moving and spinning bodies, which can be applied to pulsar-black hole binaries.

\citet{RL06} were the first to incorporate the bending effect on the frame-dragging propagation delay. However, all these studies so far have not included full general relativistic treatment. \citet{BE22} performed a study of the exact propagation delay with full general relativistic treatment, but they did not explore the bending delays. In the present paper, we study the effects of light bending phenomenon on the signal of a pulsar in a binary system with a stellar mass black hole companion using the formalism devised by \citet{KDL20} and \citet{GSL20} to solve the equations for the null geodesics in Kerr spacetime. It is obvious that the spin of the pulsar would not affect the bending phenomenon as it happens near the companion. However, it plays a significant role in the geometry, i.e., in determining the initial direction of the pulsar beam. 

This paper is organised as follows, in section \ref{sec:analytic}, we describe the analytical formalism of our work that includes a description of various coordinate frames needed, definitions of some important vectors and angles in those frames, the definitions of various bending delays, and the equations for the null geodesic in the Kerr spacetime. In section \ref{sec:numerical}, we present our numerical results to quantitatively demonstrate the effects of the spin of the black hole on the light bending phenomenon for a range of values of other parameters, when the black hole is chosen as a stellar mass one. In section \ref{sec:testcaseSupermassive}, we present the numerical results for some test cases for pulsar with a super-massive black hole as its binary companion. As expected, we see some interesting features when the pulsar is at or near the superior conjunction configuration. These happen due to the strong bending. In section \ref{sec:appendix}, these features have been explained with a simplified geometry as per the lensing theory as the full general relativistic treatment and the lensing theory becomes equivalent in the near superior conjunction configurations. The paper is ended with a short conclusive discussion in section \ref{sec:conclu}.

\section{Analytical set up}
\label{sec:analytic}
 
As presented in \citetalias{dbb23}, to understand bending delays in the signals of pulsars in binary systems, we need to model the beam of the pulsar, the motion of the pulsar in its orbit, and the propagation of the light rays in the curved spacetime. For the first two topics, we use almost the same approach as in \citetalias{dbb23}, and summarise below for the sake of completeness. The third point is modelled differently and is explained in Section \ref{subsec:FD_bending_delay}.

\subsection{Modelling the beam geometry of the pulsar}
\label{subsec:beamgeometry}

First, we present the model for the beam geometry. Following \citetalias{dbb23}, we use two Cartesian coordinate systems, the first is the ${\rm X_I Y_I Z_I}$ frame or the `I-frame' and the second is the ${\rm X_m Y_m Z_m}$ frame, or the `m-frame'. The origin of the I-frame is at the centre of the pulsar with the spin axis along the ${\rm Z_I}$ direction. We choose the ${\rm Z_I X_I}$ plane along the line of sight (LoS) that makes an angle $\zeta_L$ (co-latitude) with the ${\rm Z_I}$ axis. We define the positive direction of the LoS as pointing from the pulsar to the observer and the unit vector along this direction is denoted by $\widehat{N}_{\rm I}$.

\begin{figure}
	\includegraphics[width=100mm]{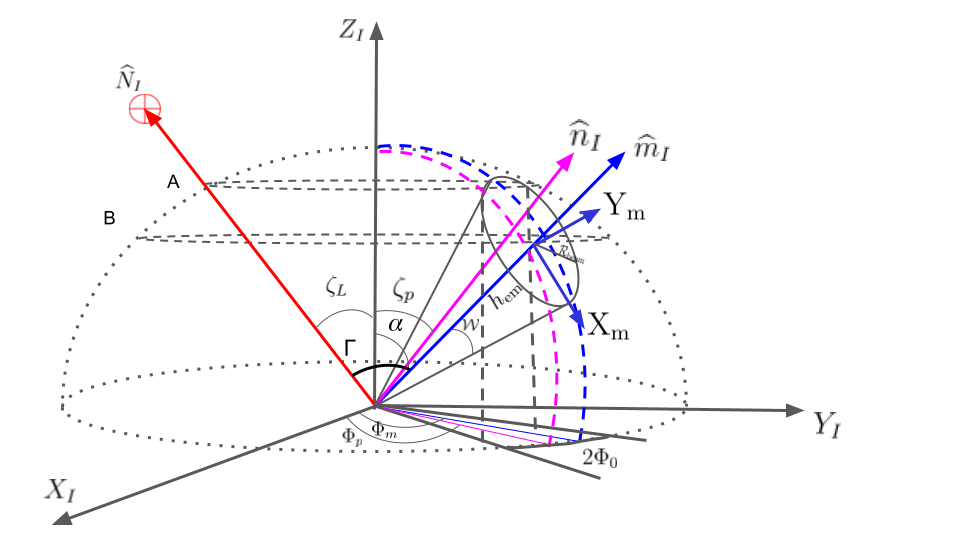}
    \caption{Reference frames, angles, and unit vectors used to model the geometry of the beam of the pulsar, as explained in the text.}
    \label{fig:spin_axis_frame}
\end{figure}

The origin of the m-frame is at the emission height with the ${\rm Z_m}$ axis being the magnetic axis. The ${\rm X_m Y_m}$ plane represents the cross-section of the beam, and the ${\rm Z_m X_m}$ plane makes an angle $\Phi_m$ with the ${\rm Z_I X_I}$ plane. These two frames and various unit vectors and angles at an arbitrary time $t$ are demonstrated in Fig. \ref{fig:spin_axis_frame}. 

As shown in Fig. \ref{fig:spin_axis_frame}, $\widehat{m}_{\rm I}$ is a unit vector along the magnetic axis, which is also the axis of the beam with a half-opening angle $\mathcal{W}$. Note that, in this figure, the beam is represented by a single cone for the sake of simplicity, although we use a core-double cone model for the beam \citepalias{dbb23}. As the pulsar spins, $\widehat{m}_{\rm I}$ rotates around the ${\rm Z_I}$-axis with a period $P_{\rm s}$ along a trajectory shown by a dashed line labelled B in Fig. \ref{fig:spin_axis_frame}. We assume that at the time zero ($t=0$), $\widehat{m}_{\rm I}$ lies in the ${\rm X_I Z_I}$ plane, giving $\Phi_m(t) = \frac{2\pi}{P_{\rm s}} t$.

The angle between the magnetic axis (the ${\rm Z_m}$-axis along which $\widehat{m}_{\rm I}$ has been defined) and the spin axis (the ${\rm Z_I}$-axis) is denoted by $\alpha$, and the angle between $\widehat{N}_{\rm I}$ and $\widehat{m}_{\rm I}$ is denoted by $\Gamma$. For a given pulsar, angles $\alpha$ and $\mathcal{W}$ remain constant over time, while $\Gamma$ and $\Phi_m$ vary. The angle $\zeta_L$ also vary slowly due to the free precession of the pulsar caused by its ellipticity \citep{jones12}. This effect has been ignored in the present work.

Although all of the light rays in the beam rotates around the ${\rm Z_I}$ axis with the period $P_{\rm s}$, only the ones with a co-latitude of $\zeta_L$ align with the LoS once in a full rotation and are visible (without bending). The trajectory of these light rays is shown with a dashed line labelled A in Fig. \ref{fig:spin_axis_frame}. At any given time, different light rays on this trajectory have different phases. The phase of the any light ray is defined as the angle its projection on the ${\rm X_I Z_I}$ plane makes with the ${\rm X_I}$-axis. Note, in the rest frame of the pulsar, A is the trajectory of the LoS.

However, all light rays in the beam will experience bending if there is a nearby gravitating object, such as a binary companion. In Fig. \ref{fig:spin_axis_frame}, the unit vector $\widehat{n}_{\rm I}$ represents a generic light ray with phase $\Phi_p$ and co-latitude $\zeta_p$. This light ray with co-latitude $\zeta_p$ would never fall on the LoS without bending and would not be visible unless bending makes it appear to originate from a co-latitude $\zeta_L$. The unit vectors $\widehat{m}_{\rm I}$ and $\widehat{n}_{\rm I}$ can be expressed as:
\begin{equation}
\widehat{m}_{\rm I} = [\cos \Phi_m \sin\alpha, \, \sin \Phi_m \sin\alpha, \, \cos\alpha] 
\end{equation}
and
\begin{equation}
\widehat{n}_{\rm I} = [\cos \Phi_p \sin\zeta_p, \, \sin \Phi_p \sin\zeta_p, \, \cos\zeta_p] .
\end{equation}

From Fig. \ref{fig:spin_axis_frame}, it is clear that the pulsar will be observable only if the angle $\zeta_L$ satisfies the condition $|\alpha - \mathcal{W} | \leq |\zeta_L| \leq |\alpha + \mathcal{W} |$. The duration of observability, or the pulse width, is determined by the time for which the LoS remains within the beam, given by the condition $\Gamma \leq \mathcal{W}$. The values of the angles $\alpha$ and $\zeta_L$ determine how LoS would cut the cross-section of the beam due to the rotation of the beam.

In this paper, we use the same models for the emission height ($h_{\nu, {\rm em}}$, $\nu$ being the frequency of the emitted radiation) and the intensity distribution on the cross section of the beam as described in \citetalias{dbb23}. The pulse shape comes from the intensity distribution over the line where the LoS cuts the cross-section of the beam (the ${\rm X_m Y_m}$ plane). More specifically, we choose a core, inner cone, and outer cone model of the beam so that we get a five component profile if the LoS travels across the cross-section of the beam through its centre. The two outer most components of the profile come from the outer cone of the beam, the next two pulse components come from the inner cone of the beam and the middle component of the pulse profile come from the core of the beam.

If $\Phi_0$ is the half-width of the observed pulse, then $2\Phi_0$ is the difference in phases between the points where the LoS enters and exits the beam in the rest frame of the pulsar, as shown in Figure \ref{fig:spin_axis_frame}. 

As explained in \citetalias{dbb23}, any unit vector $\widehat{k}_{\rm m}$ in the m-frame can be converted to a unit vector $\widehat{k}_{\rm I}$ in the I-frame using:
\begin{equation}
\label{eq:unitvecINbANDm}
 \widehat{k}_{\rm I} =  R_z(\Phi_m) R_y(\alpha) \widehat{k}_{\rm m},
\end{equation}
where $R_z(\theta)$ and $R_y(\theta)$ are the active rotation matrices for the rotation angle $\theta$ about the Z-axis and Y-axis, respectively, of any right-handed Cartesian coordinate system. Similarly, $R_x(\theta)$ would imply the active rotation matrix for the rotation angle $\theta$ about the X-axis. If the position vector of the starting point of a light ray on the ${\rm X_m Y_m}$ plane is denoted by ${\vec{\mathcal{R}}_{\rm m}}$, then the unit vector along the light ray in the m-frame is given by \citepalias{dbb23}:
\begin{equation}
\label{eq:nm}
\widehat{n}_{\rm m} = (h_{\nu, {\rm em}} \, \widehat{m}_{\rm I} + {\vec{\mathcal{R}}_{\rm m}}) (|h_{\nu, {\rm em}} \, \widehat{m}_{\rm I} + {\vec{\mathcal{R}}_{\rm m}}|)^{-1}.
\end{equation}
The vector ${\vec{\mathcal{R}}_{\rm m}}$ can be determined if the coordinates of the light rays on the ${\rm X_m Y_m}$ plane are known, and vice versa. As explained in \citetalias{dbb23}, we generate these coordinates for a number of light rays in order to produce a core-double cone intensity distribution on the ${\rm X_m Y_m}$ plane. For each ${\vec{\mathcal{R}}_{\rm m}}$, we find $\widehat{n}_{\rm m}$ using Eq. (\ref{eq:nm}) and then obtain $\widehat{n}_{\rm I}$ from $\widehat{n}_{\rm m}$ using Eq. (\ref{eq:unitvecINbANDm}). Note that in Eq. (\ref{eq:nm}), $\widehat{n}_{\rm m}$ is defined at the centre of the pulsar (where $\widehat{n}_{\rm I}$ is also defined), not at the height $h_{\nu, \rm em}$ where the light rays are generated.

\subsection{Modelling the orbital geometry and the orbital motion of the pulsar}
\label{subsec:orbitalgeometrymotion}

To model the orbital geometry of a binary pulsar, we first introduce the ${\rm X_I^\prime Y_I^\prime Z_I^\prime}$ frame, which is obtained by shifting the ${\rm X_I Y_I Z_I}$ frame (as defined in the previous section) to the barycentre of the binary along the line joining the pulsar and the companion in such a way that the ${\rm X_I^\prime}$ axis is parallel to the ${\rm X_I}$ axis, ${\rm Y_I^\prime}$ axis is parallel to the ${\rm Y_I}$ axis, and ${\rm Z_I^\prime}$ axis is parallel to the ${\rm Z_I}$ axis. Hence, the unit vector along the spin axis of the pulsar, $\widehat{S}_p$, which is physically along the ${\rm Z_I}$-axis, can be shifted and taken along the ${\rm Z_I^{\prime}}$-axis.

The second frame is the ${\rm X_s Y_s Z_s}$ frame or the `s-frame', which is centred at the barycentre of the binary and the ${\rm Z_s}$-axis is along the negative direction of the LoS. The ${\rm X_s}$-axis is the line of the ascending node (AN) and the ${\rm X_s}{\rm Y_s}$ plane is the plane of the sky. 

The third frame is the ${\rm X_b} {\rm Y_b} {\rm Z_b}$ frame or the `b-frame', which is also centred at the barycentre. In this frame, the pulsar orbits its companion in the ${\rm X_b Y_b }$ plane and the ${\rm X_b}$-axis is along the periastron of the orbit. Hence, the ${\rm Z_b}$-axis is aligned along the orbital angular momentum. The angle between the ${\rm Z_s}$-axis and the ${\rm Z_b}$-axis is denoted by $i$, which is the inclination of the orbit with respect to the sky plane. The angle between the ${\rm X_s}$-axis and the ${\rm X_b}$-axis is $\omega$, which is known as the longitude of the periastron. The orbital phase of the pulsar is defined as $A_T + \omega$ where $A_T$ is the true anomaly. 

\begin{figure*}
\centering
\includegraphics[width=0.75\linewidth]{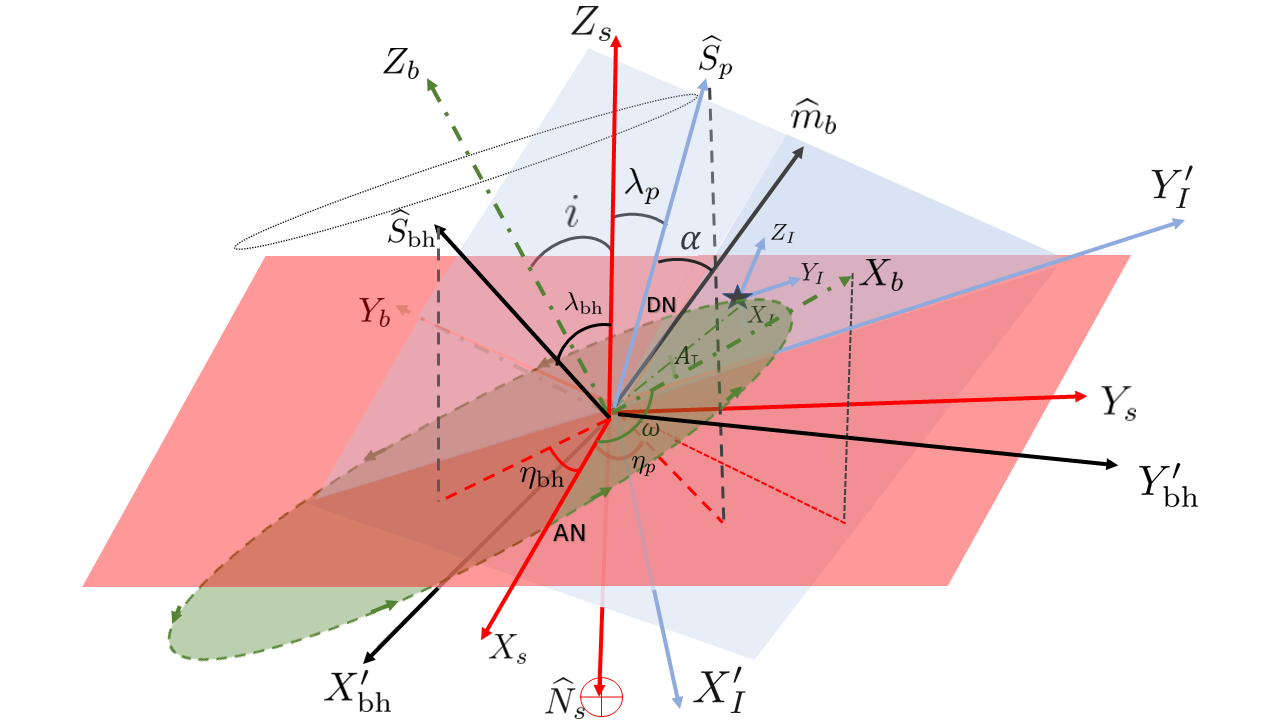}
\caption{The orbital geometry of a pulsar (denoted by a $\star$) in a binary system. The sky plane (the ${\rm X_s}{\rm Y_s}$ plane) is illustrated in light red, while the orbital plane (the ${\rm X_b Y_b}$ plane) is shown in light green. The blue plane represents the ${\rm X_I^{\prime} Y_I^{\prime}}$ plane. The unit vector along the spin axis of the pulsar is denoted by $\widehat{S}_p$ and the unit vector along the spin axis of the companion is denoted by $\widehat{S}_{\rm bh}$, both have been shown after shifting parallely to the barycentre. The ascending node is labelled as AN, and the descending node as DN. $\omega$ is the longitude of periastron, $A_T$ is the true anomaly, and $i$ is the inclination angle. The angles $\lambda_p$, $\eta_p$, $\lambda_{\rm bh}$, and $\eta_{bh}$ have been defined in the text. }
\label{frames}
\end{figure*}

We depict these frames, various angles, and unit vectors in Fig. \ref{frames} where the sky plane or the ${\rm X_s}{\rm Y_s}$ plane is illustrated in light red, the orbital plane (the ${\rm X_b Y_b}$ plane) is illustrated in light green and the ${\rm X_I^{\prime} Y_I^{\prime}}$ plane is shown in light blue. Additionally, the ascending node (AN) and the descending node (DN) have also been marked in this figure. The pulsar is marked with a black star in the figure.

There are two additional frames, the first one is the ${\rm X_p Y_p Z_p}$ frame of the `p-frame' and the second one is the ${\rm X_c Y_c Z_c}$ frame or the `c-frame'. The origin of the first frame lies at the centre of the pulsar while the origin of the second frame lies at the centre of the companion. None of these two frames are shown in the figure. These two frames are connected to the b-frame through parallel shifts along the line joining the pulsar and the companion. As the b-frame, the p-frame, and the c-frame are parallel to each other, any direction vector calculated in one frame would be the same in the other two. Additionally, the direction of the LoS is the same in all these three frames.

Now, we need another frame to specify the direction of the companion's spin axis, similar to the I-frame of the pulsar. This frame is centred at the companion and is called the `bh-frame' or the $\rm X_{bh} Y_{bh} Z_{bh}$ frame (not shown in Fig. \ref{frames}). The $\rm Z_{bh}$-axis is taken in along the spin axis of the companion and the $\rm Y_{bh}$-axis is taken in the sky-plane. The orientation of the $\rm X_{bh}$ and the $\rm Y_{bh}$ axes depend on the orientation of the spin axis of the companion. The bh-frame can be translated to the barycenter of the orbit along the line joining the pulsar and the companion, giving another frame, the $\rm X_{bh}^\prime Y_{bh}^\prime Z_{bh}^\prime$ frame. Note that the c-frame and the bh-frame are both centred at the companion. However, the $\rm Z_{bh}$-axis aligns with the spin of the companion, while the $\rm Z_{c}$-axis aligns with the orbital angular momentum of the companion. The unit spin vector of the companion is denoted by $\widehat{S}_{\rm bh}$, which can be taken along the $\rm Z_{bh}^\prime$ axis too. It is obvious that the ${\rm \rm X_{bh} Y_{bh} Z_{bh}}$ frame and the  ${\rm X_{bh}^\prime Y_{bh}^\prime Z_{bh}^\prime}$ frames are almost the same frame as the barycentre is inside the black hole in case of a pulsar$-$black hole binary. For the same reason, the b-frame and the c-frame are also almost the same. 

The angle between ${\rm Z_I}$-axis or the unit pulsar-spin vector $\widehat{S}_p$ and the ${\rm Z_s}$-axis is denoted by $\lambda_p$ giving $|\lambda_p|= 180^\circ - |\zeta_L|$. The projection of $\widehat{S}_p$ in the sky plane makes an angle $\eta_p$ with the ${\rm X_s}$-axis. Similarly, the angle between  $\widehat{S}_{\rm bh}$ and the ${\rm Z_s}$-axis is denoted by $\lambda_{\rm bh}$ and the projection of $\widehat{S}_{\rm bh}$ in the sky plane makes an angle $\eta_{\rm bh}$ with the ${\rm X_s}$-axis. $\eta_{\rm bh}$ is also the angle between the $\rm Y_{s}$-axis and the $\rm Y_{bh}$-axis, i.e., it specifies the direction of the $\rm X_{bh}$ and the $\rm Y_{bh}$ axes. 

As we study the pulsar for at most one full orbit, we ignore the post-Keplerian parameters and model its orbital motion in the b-frame by solving Kepler’s equations to obtain the true anomaly. This gives us the position vector of the pulsar at any given time in the b-frame ($\vec{r}_b$). From that, the position vectors at that time in the c-frame ($\vec{r}_c$) and in the bh-frame ($\vec{r}_{\rm bh}$) can be calculated. As the light rays are emitted by the pulsar, $\vec{r}_b$ is also the position vector of the light ray at the time of the emission. Here, $|\vec{r}_b |$ is the distance between the pulsar and the barycentre and $|\vec{r}_c | = |\vec{r}_{\rm bh} | =  (1 + M_P/M_c) |\vec{r}_b |$ is the distance between the pulsar and the companion. The unit vectors along these position vectors are $\widehat{{r}}_c$, $\widehat{{r}}_b$, and $\widehat{{r}}_{\rm bh}$. Since the c-frame and the b-frame are parallel to each other, we have $\widehat{{r}}_c = \widehat{{r}}_b$. On the other hand,
\begin{equation}
\label{eq:r_in_bh_frame}
    \vec{r}_{\rm bh}= R_y (-\lambda_{\rm bh}) \, R_z (-\eta_{\rm bh}) \, R_x (i)\, R_z (\omega) \vec{r}_c .
\end{equation}
Moreover,
\begin{equation}
\label{eq:r_in_s_frame}
    \vec{r}_{\rm s}= R_x (i) \, R_z (\omega) \, \vec{r}_b .
\end{equation} where $\vec{r}_{\rm s}$ is the position vector of the pulsar in the s-frame. From Fig. \ref{frames}, we can write
\begin{equation}
\label{eq:Sbh_with_angles}
    \widehat{S}_{\rm bh}= [\cos \eta_{\rm bh} \sin \lambda_{\rm bh}, \sin \eta_{\rm bh} \sin \lambda_{\rm bh}, \cos  \lambda_{\rm bh}] , 
\end{equation} 
\begin{equation}
\label{eq:rb_with angles}
    \widehat{r}_{b} = [\cos A_T, \sin A_T, 0] ,
\end{equation} and
\begin{equation}
\label{eq:Ns_vector}
    \widehat{N}_{s} = [0, 0, -1] .
\end{equation}

Using $\vec{{r}}_b$, we construct one more frame, the $\rm X_TY_TZ_T$ frame or the T-frame. We choose the $\rm Y_T$-axis along $\vec{r}_{\rm b}$ and the $\rm Z_T$ axis is along $\widehat{r}_{\rm b}\times \widehat{n}_{\rm b}$, where the initial direction of the light ray $\widehat{n}_I$ (see Fig. \ref{fig:spin_axis_frame}) has been transformed into b-frame as \citepalias{dbb23}:
\begin{equation}
\label{eq:n_in_b_frame}
    \widehat{n}_{b} = R_z(-\omega)R_x(-i)R_z(\eta_p)R_y(\lambda_p) \, \widehat{n}_{\rm I}   .
\end{equation} The above construction means that the pulsar lies on the ${\rm Y_T}$-axis, leading to the the polar and azimuthal angles of the light rays in the T-frame at the time of emissions being $90^{\circ}$.

To model the orbital evolution as discussed in section \ref{subsec:orbitandspinevolution}, we need the angle between the spin-axis of the pulsar and the orbital angular momentum, which we define as $\xi_p$. With the help of geometry, we find a relation between $\xi_p$ with other angles mentioned earlier as:

\begin{equation}
\label{eq:xip}
\xi_p=\cos ^{-1} \left(    \cos \lambda_p \, \cos i - \sin \lambda_p \, \sin \eta_p \sin i   \right) . 
\end{equation}

\subsection{Modelling the evolution of the orbit and the spins}
\label{subsec:orbitandspinevolution}

A relativistic binary emits gravitational waves leading to a decrease in both of the orbital size and the eccentricity. We model the rate of the change of the orbital period ($\dot{P}_b$) and the rate of change of the eccentricity ($\dot{e}$) up to the first post-Newtonian order following the standard expressions  \citep{peters1964,lorimer04} that are independent of the spin of the pulsar as well as the spin of the companion.

The precession of the orbit and the spin axes of the pulsar and the black hole are affected by their masses, spins, the orbital separation and the eccentricity of the orbit. In the present work, we have assumed the spin vector of the black hole is aligned with the orbital angular momentum vector and hence the spin-precession of the black hole can be ignored. The precession of the spin vector of the pulsar around the orbital angular momentum is $\dot{\overrightarrow{\mathcal{S}}}_{p} = \overrightarrow{\Omega}_{\rm \mathcal{S}_p} \times \overrightarrow{S}_{p}$ where ${\overrightarrow{\mathcal{S}}}_{p}$ is the spin vector of the pulsar and $\overrightarrow{\Omega}_{\rm \mathcal{S}_p}$ is the velocity of precession of the spin vector which is the summation of the PN terms and the Lense-Thirring term, i.e., $\overrightarrow{\Omega}_{\rm \mathcal{S}_p} = \overrightarrow{\Omega}_{\rm 1PN_p} + \overrightarrow{\Omega}_{\rm LT_p} $ (up to the 1PN order). From \citet{BarkerOconnell1975}, the 1PN term can be written as:
\begin{equation}
\label{eq:spinprecfreq1PN}
\overrightarrow{\Omega}_{\rm 1PN_p} = A_{\rm 1PN_p} \, \widehat{L} ,
\end{equation} with 
\begin{equation}
\label{eq:spinprec1PN}
A_{\rm 1PN_p} = \left( \frac{G}{c^3} \right)^{2/3} \, \frac{\sigma_b^{2/5}}{(1-e^2)} \, \frac{M_{c} (4 M_p + 3 M_{c})}{2 (M_p + M_{c})^{4/3}} ,
\end{equation} where $M_p$ is the mass of the pulsar, $M_c$ is the mass of the companion, $G$ is the Gravitational constant, $c$ is the speed of light in vacuum, and $\sigma_b = 2 \pi {P_b}^{-1}$. The Lense-Thirring term is given by: 
\begin{equation}
\label{eq:spinprecfreqLT}
\overrightarrow{\Omega}_{\rm LT_p} = A_{\rm LT_p} \, \left[ \widehat{S}_{\rm bh} - 3 (\widehat{L} . \widehat{S}_{\rm bh}) \widehat{L} \right]   =  - 2 \, A_{\rm LT_p} \, \widehat{L}  ,
\end{equation} where $ \widehat{L}$ is the unit vector along the orbital angular momentum and $\widehat{\mathcal{S}}_{\rm bh}$ is the unit vector along the spin-axis of the black hole. In the second step of Eq. (\ref{eq:spinprecfreqLT}), we have used the fact that $\widehat{\mathcal{S}}_{\rm bh} \parallel \widehat{L}$. We also know that, 

\begin{equation}
\label{eq:spinprecLT}
 A_{\rm LT_p} = \frac{G}{c^3} \, \tilde{a} \, \frac{\sigma_b^2}{(1-e^2)^{3/2}} \, \frac{M_{c}^2}{2 (M_p + M_{c} ) } . 
\end{equation} where $\tilde{a}$ is the spin parameter of the black hole related to its spin angular momentum $\mathcal{S}_{\rm bh}$ by the relation $\mathcal{S}_{\rm bh} = \tilde{a} \, GM_c^2 / c$. $\tilde{a}$ takes a positive value for the prograde rotation of the black hole (with respect to its orbital rotation) and a negative value for the retrograde rotation of the black hole in such a way that $|\tilde{a}|$ remains in the range $0-1$.

To investigate the precession of the orbit, we calculate the rate of the change of the longitude of the periastron, i.e, $\dot{\omega}$, using the PN term (up to the 2nd order) as well as the Lense-Thirring terms due to the spin of the pulsar as well as the spin of the black hole using the expressions given in \citet{bagchi2017} as $\dot{\omega}=\dot{\omega}_{\rm PN} + \dot{\omega}_{\rm LT,p} + \dot{\omega}_{\rm LT, bh}$. The PN term is given by:
\begin{equation}
\dot{\omega}_{{\rm PN}} =  ~\frac{3 \beta_0^2~ \sigma_b}{1-e^2} [1 + \beta_0^2 f_{0}  ] ,
\label{eq:per_advPN}
\end{equation} where
\begin{equation}
\beta_0~=~\frac{(G M_T \sigma_b )^{1/3}}{c}  ,
\label{eq:beta0}
\end{equation} with $M_T = M_{p} + M_{c}$, and 
\begin{multline}
f_{0}~=~\frac{1}{1-e^2}\left( \frac{39}{4} \mathscr{X}_p^2+\frac{27}{4} \mathscr{X}_{c}^2+15 \mathscr{X}_p \mathscr{X}_{c} \right)  \\ - \left( \frac{13}{4} \mathscr{X}_p^2+\frac{1}{4} \mathscr{X}_{c}^2+\frac{13}{3} \mathscr{X}_p \mathscr{X}_{c} \right) ~,
\label{eq:f0p}
\end{multline} where $\mathscr{X}_p~=~M_p/M_T$. The Lense-Thirring terms are given by
\begin{subequations}
\begin{equation}
\dot{\omega}_{\rm LT, p}  = -   \frac{3 \beta_0^3 ~ \sigma_b }{1-e^2} [ \beta_{sp} \, g_{sp}   ] ,
\label{eq:per_advLTp}
\end{equation} and
\begin{equation}
\dot{\omega}_{\rm LT, c}  = -   \frac{3 \beta_0^3 ~ \sigma_b }{1-e^2} [ \tilde{a} \, g_{sc} ] .
\label{eq:per_advLTc}
\end{equation}
\label{eq:per_advLT}
\end{subequations} In Eq. (\ref{eq:per_advLTp}), we have,
\begin{equation}
\beta_{sp}  = \frac{c \mathcal{S}_p}{G M_p^2} = \frac{2 \pi c I_p} {G M_p^2} ,
\label{eq:per_advbetasp}
\end{equation} where $\mathcal{S}_p$ is the magnitude of the spin vector of the pulsar, $\vec{\mathcal{S}}_p = 2 \pi P_s^{-1} \, I_p \, \widehat{\mathcal{S}}_{\rm p} $ where $I_p$ is the moment of inertia of the pulsar. $I_p$ of the pulsar depends on its mass, spin and the Equation of State of the matter \citep[and references therein]{bagchi2010} and can be taken as $1.8 \times 10^{45} \, {\rm g ~ cm^2}$ for a neutron star of mass $1.8~{\rm M_{\odot}}$. We also know \citep{bagchi2017} that,
\begin{equation}
g_{sp}  = \frac{\mathscr{X}_p \left(4 \mathscr{X}_p+ 3 \mathscr{X}_c \right)}{6(1-e^2)^{1/2} {\rm sin^2} i} \times \left[ (3 ~{\rm sin^2} i-1)~ {\rm cos} \, \xi_p +{\rm cos} \, i~ {\rm cos} \, \lambda_p \right] .
\label{eq:per_advgsp}
\end{equation} The angles $\xi_p$ and $\lambda_p$ are defined in Section \ref{subsec:orbitalgeometrymotion}. With our choice of the spin of the black hole being aligned with the orbital angular momentum, we can write,
\begin{equation}
g_{sc}  = \frac{\mathscr{X}_c \left(4 \mathscr{X}_c+ 3 \mathscr{X}_{p}\right)}{3(1-e^2)^{1/2}} .
\label{eq:per_adgsc}
\end{equation}

The Lense-Thirring effect causes a change in the orbital inclination angle too. As we have taken the spin of the black hole perpendicular to the orbital plane, it would not contribute to the change in the orbital inclination. However, the spin of the pulsar would contribute. Following \citet{damourschafer1988}, we can write,
\begin{equation}
\frac{di}{dt} = \frac{\sigma_b^2} {2 (M_p + M_c) c^2}  \, \sum_{j={\rm p, bh}}  \vec{\mathcal{S}}_{i} . \widehat{\Upsilon } = \frac{\pi I_p \, \sigma_b^2} {(M_p + M_c) c^2}  \, \widehat{\mathcal{S}}_{\rm p} . \widehat{ \Upsilon } \, ,
\end{equation} where $\widehat{ \Upsilon }$ is a unit vector along the line of node. The change in $i$ will result in the change in $\widehat{L}$ about which $ \widehat{S}_{\rm p}$ precess. 

There are more recent works on the precession of spin and the orbit, which are more appropriate when a millisecond pulsar orbits around a super-massive black hole in a very tight binary. We will discuss those works when we present numerical results of a test case of pulsar - super-massive black hole binary in Sec \ref{sec:testcaseSupermassive}.

The precession of the spin of the pulsar induces changes in the values of the angles $\lambda_p$ and $\eta_p$. The change in the values of of these angles change the value of $\xi_p$ through Eq. (\ref{eq:xip}), and the changes in $\lambda_p$ and $\xi_p$ affect the value of $\dot{\omega}_{\rm LT, p}$ through Eqs. (\ref{eq:per_advLTp}) and (\ref{eq:per_advgsp}).

Additionally, the change in the value of $\lambda_p$ affects the observability condition that the LoS must pass through the cross-section of the beam once in one rotation for a neutron star to be visible as a pulsar. We use the half-opening angle where the intensity of the outer cone becomes 10 per cent of the maximum intensity of this component (at 1.4 GHz), i.e., $\mathcal{W}_{1.4, 10}^{\rm out}$, in the observability condition \citepalias{dbb23}: 
\begin{equation}
|\alpha - \mathcal{W}_{1.4, 10}^{\rm out}| \leq |\zeta_L| \leq |\alpha + \mathcal{W}_{1.4, 10}^{\rm out} | \, ,
\end{equation} which can also be written as 
\begin{equation}
 180^{\circ} - |\mathcal{W}_{1.4, 10}^{out} + \alpha | \leq | \lambda_p | \leq  180^{\circ} - | \mathcal{W}_{1.4, 10}^{out} - \alpha | \, ,
\end{equation} as $|\lambda_p| = 180^{\circ} - |\zeta_L|$.

If we choose the spin of the black hole along the orbital angular momentum, there would be no precession of the spin of the black hole and the values of the angles $\lambda_{\rm bh}$ and $\eta_{\rm bh}$ would remain constant. This is the configuration we use when we study the effect of the light bending phenomenon for a pulsar with a super-massive black hole companion. This is also the configuration we use in most of our investigations on the effect of the light bending phenomenon for a pulsar with a stellar mass black hole companion, except a few cases where we investigate the impact of the change in the values of either $\lambda_{\rm bh}$ or $\eta_{\rm bh}$. However, in such cases, the precession of the spin of the black hole is negligibly small.

We ignored the non relativistic terms in $\dot{P}_b$, $\dot{e}$, $\dot{\omega}$, $\dot{\overrightarrow{\mathcal{S}}}_{p}$, and $di/dt$. A comprehensive discussions on these can be found in \citet{lorimer04}.

\subsection{Definitions and quantifications of the light bending effect}
\label{subsec:deflightbending}

The light rays from the pulsar propagate through the curvature of the spacetime produced by the companion. This curvature affects the propagation time and alters the direction of the light rays. This effect is strongest near the superior conjunction, i.e., when the inclination of the orbit $i=90^\circ$ and the orbital phase $A_T + \omega = 90^\circ$. The change in the direction of the light rays causes the latitudinal and longitudinal bending delays \citep {dk95, RL06}. 

Following \citetalias{dbb23}, the longitudinal bending delay is defined as:
\begin{equation}
\label{longdelay}
\tau_{\text{long}} = \Delta \Phi \times \frac{P_s}{2\pi} ,
\end{equation} where $\Delta \Phi = \Phi_{\text{without}} - \Phi$ is the difference in the longitude (or phase) of the light rays due to the bending of a light ray that has a phase $\Phi$ when affected by bending but would have the phase $\Phi_{\text{without}}$ if there was no bending.

Similarly, the latitudinal bending delay is defined as:
\begin{equation}
\label{latdelay}
\tau_{\text{lat}} = \Delta \Phi_0 \times \frac{P_s}{2\pi} , 
\end{equation} where  $\Delta \Phi_0$ is the change in the half-width of the beam $\Phi_0$ (see Fig. \ref{fig:spin_axis_frame}) caused by the change in the value of the co-latitude due to the bending as $\Delta \zeta = \zeta_{\text{without}} - \zeta_L$. Here, the co-latitude of the light ray after bending is $ \zeta_L$ in such a way that the same light ray would have the co-latitude $\zeta_{\text{without}} $ if the bending was absent. The expression for $\Delta \Phi_0$ involves $\Delta \zeta$, $ \zeta_L$, $\alpha$, and $\Phi_0$ while the expression for $\Phi_0 $ involves $\mathcal{W}$, $\alpha$, and $\zeta_L$. These expressions are available in \citetalias{dbb23}.

\subsection{Effect of the spin of the companion in bending delay}
\label{subsec:FD_bending_delay}

The spin of the companion affects the propagation and the direction of the light rays \citep{BE22}. The light rays that move in directions co-rotating with the spin of the companion bend more than the the light rays that move in the direction counter-rotating with the spin of the companion. This phenomenon affects both the longitudinal and latitudinal bending delays resulting in frame dragging bending delays. The longitudinal frame dragging bending delay is defined as:
\begin{equation}
    {\rm FD}_{\rm long} = \left( \tau_{\rm long, \tilde{a}} \right) -  \left(  \tau_{\rm long, 0} \right) ,
\end{equation} where $\tau_{\rm long, \tilde{a}}$ and $\tau_{\rm long, 0}$ are the longitudinal bending delays considering the spin and ignoring the spin of the companion, respectively. We use Eq. (\ref{longdelay}) to estimate the values of $\tau_{\rm long, \tilde{a}}$ and $\tau_{\rm long, 0}$

Similarly, the latitudinal frame dragging bending delay is defined as:
\begin{equation}
     {\rm FD}_{\rm lat} = \left( \tau_{\rm lat, \tilde{a}} \right) -  \left(  \tau_{\rm lat, 0} \right) ,
\end{equation} where $\tau_{\rm lat, \tilde{a}}$ and $\tau_{\rm lat, 0}$ are the latitudinal bending delays considering the spin and ignoring the spin of the companion, respectively. Eq. (\ref{latdelay}) is used to estimate the values of $\tau_{\rm lat, \tilde{a}}$ and $\tau_{\rm lat, 0}$.

\subsection{Full Genera relativistic treatment to study the effect of the spin of the companion on the bending delay}
\label{subsec:bending_spin_full_GR}

In the Kerr spacetime, the trajectories of the light rays are governed by a set of null geodesic equations that are commonly studied in Boyer-Lindquist coordinates where the coordinates $x^{\mu}$ are $x^0=ct$, $x^1=r$, $x^2=\theta$, $x^3=\phi$. For each of these coordinates as well as other parameters, we use appropriate subscripts to represent the relevant reference frame, e.g., the subscript `T' is used for the T-frame, the subscript `bh' is used for the bh-frame, etc. Additionally, the subscript `s' is used to mean the values at the source (the pulsar) and the subscript `o' is used to mean the values at the observation point. The integral forms of the null geodesic in Boyer-Lindquist coordinates are given by \citet{GSL20}:
\begin{equation}
\label{geodesic11}
I_r=G_\theta\ , 
\end{equation} and
\begin{equation}
\label{geodesic22}
\phi_o-\phi_s=I_\phi+D_\lambda G_\phi\ ,
\end{equation}
where we have defined a set of parameters as follows:

\begin{subequations}
\label{parameters_geodesic}

\begin{align}
\label{R_rparticle}
\mathcal{R}(r) = ~ & r^4+ \left[ \left(\tilde{a}\frac{GM_c}{c^2}\right)^2-D_\lambda^2-D_q \right] r^2 \nonumber \\
 & + 2  \left( \frac{G M_c}{c^2} \right) \, \left[ D_q  + \left\{ D_\lambda - \left(\tilde{a}\frac{GM_c}{c^2} \right) \right \} ^2 \right] r  \nonumber \\
& -\left(\tilde{a}\frac{GM_c}{c^2}\right)^2D_q- \left ( \frac{M_c^2 c^4}{E^2} \right) r^2\Delta ,
\end{align}

\begin{align}
\label{theta_thetaparticle}
\Theta(\theta) = ~ & D_q+\left(\tilde{a}\frac{GM_c}{c^2}\right)^2\cos ^2 \theta -D_\lambda^2 \cot ^2 \theta \nonumber \\ & - \left( \frac{M_c^2c^4}{E^2} \right) \left(\tilde{a}\frac{GM_c}{c^2}\right)^2\cos ^2 \theta ,
\end{align}

\begin{equation}
\label{I_r}
I_r = \fint_r\frac{dr}{ {\rm sign}(p_r) \sqrt{\mathcal{R}(r)}} ,
\end{equation}

\begin{equation}
\label{I_phi}
I_\phi = \fint_r\frac{\left(\tilde{a}\frac{GM_c}{c^2}\right)\left((2GM_c/c^2)r-D_\lambda \left(\tilde{a}\frac{GM_c}{c^2}\right)\right)}{{\rm sign}(p_r) \sqrt{\mathcal{R}(r)}\Delta}dr ,
\end{equation}

\begin{equation}
\label{G_theta}
G_\theta = \fint_\theta\frac{d\theta}{{\rm sign}(p_\theta ) \sqrt{\Theta(\theta)}} ,
\end{equation}

\begin{equation}
\label{G_phi}
G_\phi = \fint_\theta\frac{\csc ^2 \theta}{{\rm sign}(p_\theta) \sqrt{\Theta(\theta})}d\theta ,
\end{equation}

\begin{equation}
\label{D_lambda}
D_{\lambda} = c L_z/E ,
\end{equation}

\begin{equation}
\label{D_q}
D_q = \mathcal{L}/E^2 .
\end{equation}

\end{subequations}
In the above equations, the symbol $\fint$ implies the integrals to be the path integrals along the trajectory connecting the coordinates of the source to the coordinates of the observer. $E$ (the energy), $L_z$ (the component of the angular momentum of the light ray along the relevant Z-axis), and $ \mathcal{L}$ are the constants separating the polar and the azimuthal part of the geodesic equation \citep{pro16} making $D_q$ and $D_\lambda$ two constants of the motion. $M_c$ is the mass of the gravitating object. As defined in subsection  \ref{subsec:orbitandspinevolution}, $\tilde{a}$ is the dimensionless spin parameter $\mathcal{S}_{\rm bh} = \tilde{a} \, GM_c^2 / c$ of the gravitating body and $\mathcal{S}_{\rm bh}$ is its spin angular momentum. The components of the momentum of the light rays in the temporal, radial, polar, and the azimuthal directions are denoted by $p_t$, $p_r$, $p_\theta$, $p_\phi$, respectively. 

We also use the conventional parameters used in the Kerr metric:
\begin{equation}
\label{def_deltakerr}
\Delta=r^2-2\frac{GM_c}{c^2}r+\left(\tilde{a}\frac{GM_c}{c^2}\right)^2 ,
\end{equation} 

\begin{equation}
\label{def_rhokerr}
\rho^2=r^2+\left(\tilde{a}\frac{GM_c}{c^2}\right)^2\cos ^2 \theta , 
\end{equation} 

\begin{equation}
\label{def_sigmakerr}
\Sigma^2=\left[r^2+\left(\tilde{a}\frac{GM_c}{c^2}\right)^2\right]^2-\left(\tilde{a}\frac{GM_c}{c^2}\right)^2 \, \Delta \, \sin ^2 \theta , 
\end{equation}

\begin{equation}
\label{def_Omegakerr}
 \Omega =  \frac{2\left(\frac{GM_c}{c^2}\right)r\left(\tilde{a}\frac{GM_c}{c^2}\right)}{\Sigma^2} .
\end{equation}

To solve the geodesic equations, we need to express $D_q$ and $D_\lambda$ in terms of quantifiable parameters. The procedure is discussed below. As discussed in section \ref{subsec:orbitalgeometrymotion}, the initial direction of the light ray is on the $\rm X_T Y_T$ plane. Hence, the initial momentum of the light ray in polar direction is $p_{\theta T,\ s}=0$. If $\chi$ is the angle between $\vec{r}_{\rm b}$ and $\widehat{n}_{\rm b}$, then the ratio of the initial azimuthal and the initial temporal momenta of the light ray in the T-frame is $p_{\phi T,\ s}/p_{tT,\ s}=-\sin \chi$. The angular momentum vector of the light ray scaled by $|\vec{r}_{\rm b}| \ p_{t T, \, s}$ is denoted by $(\vec{\tilde{L}}_{T,\ s})$. The components of this vector along the $\rm X_T$, $\rm Y_T$, and $\rm Z_T$ axes are $\tilde{L}_{X_{T, \, s}}$, $\tilde{L}_{Y_{T, \, s}}$, and $\tilde{L}_{Z_{T, \, s}}$, respectively, and are given by:
\begin{subequations}
\label{eq:cartesianangularmomentumtildeTframe}
    \begin{equation}
        \tilde{L}_{X_{T, \, s}} = -   \left[ \frac{p_{\phi_{T, \, s}}} {p_{t_{T, \, s}}} \cos \theta_{T, \, s} \cos \phi_{T, \, s} + \frac{p_{\theta_{T, \, s} }} {p_{t_{T, \, s}}} \sin \phi_{T, \, s}    \right]   ,
    \end{equation}
    \begin{equation}
        \tilde{L}_{Y_{T, \, s}} = -  \left[ \frac{p_{\phi_{T, \, s}}} {p_{t_{T, \, s}}}  \cos \theta_{T, \, s} \sin \phi_{T, \, s} - \frac{p_{\theta_{T, \, s}}} {p_{t_{T, \, s}}} \cos \phi_{T, \, s}  \right]  ,
    \end{equation}
    \begin{equation}
        \tilde{L}_{Z_{T, \, s}} =  \frac{p_{\phi_{T, \, s}}} {p_{t_{T, \, s}}} \sin \theta_{T, \, s}     .
    \end{equation}
\end{subequations} As mentioned at the end of Section \ref{subsec:orbitalgeometrymotion}, the T-frame is chosen in such a way that $\theta_{T, \, s} = 90^{\circ}$ and $\phi_{T, \, s} = 90^{\circ}$. The scaled angular momentum $\vec{\tilde{L}}_{T,\ s}$ is transformed to the bh$^\prime$-frame as: 
\begin{equation}
    \vec{\tilde{L}}_{\rm bh^{\prime}, \, s}= R_{y}(-\lambda_{\rm bh})\, R_{z}(-\eta_{\rm bh}) \, R_{x}(i)\, R_{z}(\omega) \, R_{z}(\varphi_{\rm Tb}) \,  R_{y}(\vartheta_{\rm Tb}) \vec{\tilde{L}}_{T, \, s}  ,
\end{equation}
where, $\vartheta_{\rm Tb}$ is the angle between the $\rm Z_T$-axis and the $Z_b$-axis and $\varphi_{\rm Tb}$ is the angle between the $\rm Y_T$-axis and the $\rm Y_b$-axis. The values of these two angles can be calculated with the help of known unit vectors $\widehat{y}_T$, $\widehat{y}_b$, $\widehat{z}_T$, and $\widehat{z}_b$ \citepalias{dbb23}. As the bh-frame and the bh$^\prime$-frame are connected by a parallel shift, we can write $\vec{L}_{\rm bh, \, s} = \vec{\tilde{L}}_{\rm bh^{\prime}, \, s}$. Then, from $\vec{\tilde{L}}_{\rm bh, \, s}$, we get the ratio of the initial polar and temporal components of momentum ($p_{\theta_{\rm bh, \, s}} / p_{t_{\rm bh, \, s}}$) as well as the ratio of the initial azimuthal and temporal components of the momentum ($p_{\phi_{\rm bh, \, s}} / p_{t_{\rm bh, \, s}}$), in the bh-frame as:
\begin{subequations}
\label{eq:cartesianangularmomentumtilde_bhframeLinearMomentum}
    \begin{equation}
        \frac{p_{\theta_{\rm bh, \, s}}}{p_{t_{\rm bh, \, s}}} =  \tilde{L}_{Y_{\rm bh, \, s}} \, \cos \phi_{\rm bh, \, s}  - \tilde{L}_{X_{\rm bh, \, s}} \,  \sin \phi_{\rm bh, \, s}   , 
    \end{equation} and
    \begin{equation}
        \frac{p_{\phi_{\rm bh, \, s}}} {p_{t_{\rm bh, \, s}}} = \frac{ \tilde{L}_{Z_{\rm bh, \, s}}} {\sin \theta_{\rm bh, \, s} }  ,
    \end{equation}
\end{subequations}
where, $\tilde{L}_{X_{\rm bh, \, s}}$, $\tilde{L}_{Y_{\rm bh, \, s}}$ and $\tilde{L}_{Z_{\rm bh, \, s}}$ are the components of the vector $\vec{\tilde{L}}_{\rm bh, \, s}$ along the ${\rm X_{bh}}$, ${\rm Y_{bh}}$, and ${\rm Z_{bh}}$ directions, respectively. Moreover, $p_{t_{\rm bh, \, s}} = p_{t_{T, s}}$ as the bh-frame and the T-frame are connected by spatial rotations and translation only. 

In terms of the known parameters, the two constants of the motion $D_q$ and $D_\lambda$ are written as \citep{Yang_2013}: 
\begin{subequations}
    \begin{equation}
    \label{Dlambda_calculate}
        D_\lambda=\frac{\sin \theta_{\rm bh}\ (p_{\phi_{\rm bh}}/p_{t_{\rm bh}})}{-\sqrt{\Delta}\rho^2/\Sigma^2 + \Omega \sin \theta_{bh}\ (p_{\phi_{\rm bh}}/p_{t_{\rm bh}})} ,
    \end{equation} and
    \begin{equation}
        \label{Dq_calculate}
        D_q=\left[\frac{D_\lambda^2}{\sin ^2 \theta_{\rm bh}}-\left(\tilde{a}\frac{GM_c}{c^2}\right)^2\right]\cos ^2 \theta_{bh}+\left[\frac{p_{\theta_{\rm bh}}}{p_{t_{\rm bh}}}(1-D_\lambda \Omega)\right]^2\frac{\Sigma ^2}{\Delta}  .    
    \end{equation}
\end{subequations} 
As these are constants of the motion, we can use the values of $p_{\phi_{\rm bh \, s}} / p_{t_{\rm bh \, s}}$ $p_{\theta_{\rm bh \, s}} / p_{t_{\rm bh \, s}}$ obtained using Eqs. (\ref{eq:cartesianangularmomentumtilde_bhframeLinearMomentum}) and replace $\theta_{\rm bh}$ by $\theta_{bh, \, s}$ in Eqs. (\ref{Dlambda_calculate}) and (\ref{Dq_calculate}). The Kerr parameters $\Delta$, $\rho^2$, $\Sigma^2$, $\Omega$ are obtained using $r = r_{\rm bh, s}$ and $\theta = \theta_{\rm bh, \, s}$ for a given value of $\tilde{a}$ in Eqs. (\ref{def_deltakerr}), (\ref{def_rhokerr}), (\ref{def_sigmakerr}), and (\ref{def_Omegakerr}), which are then used to find the values of $D_q$ and $D_\lambda$.

After calculating the constants of the motion, we solve the geodesic Eqs. (\ref{geodesic11}) and (\ref{geodesic22}) with the help of Eqs. (\ref{parameters_geodesic}) as suggested by \citet{KDL20} and \citet{GSL20}. Through this procedure, we obtain the final polar and azimuthal directions ($\theta_{\rm bh,\ f}$ and $\phi_{\rm bh,\ f}$)  of the light rays in the bh-frame. The unit vector along the final direction of the light ray in the bh-frame can be written as:
\begin{equation}
\widehat{n}_{\rm bh,\ f}=[\cos \phi_{\rm bh,\ f}\ \sin \theta_{\rm bh,\ f},\sin \phi_{\rm bh,\ f}\ \sin \theta_{\rm bh,\ f},\cos \theta_{\rm bh,\ f}]\ .
\end{equation}
We transform this to the I-frame and get $\widehat{n}_{\rm I, \, f}$ as:
\begin{equation}
\label{nIinft}
    \widehat{n}_{\rm I,\ f} = R_y(-\lambda_p)R_z(-\eta_p)R_z(\eta_{\rm bh})R_y(\lambda_{\rm bh}) \widehat{n}_{\rm bh, \, f} .
\end{equation}

To estimate the values of the bending delays, first, we note a particular time, at which, in the absence of the bending, a light ray aligns with $\widehat{N}_{\rm I}$, i.e., with the LoS. At this time, the phase of this light ray is zero and the co-latitude is $\zeta_L$. Now, we identify the light ray which would have the phase zero and the co-latitude $\zeta_L$ at that particular time in the presence of bending. We do this with the help of the final directions of the light ray given by Eq. (\ref{nIinft}). After this we find out what was the phase ($\Phi_{\rm p, old}$) and colatitude ($\zeta_{\rm old}$) of that particular light ray. The change in the phase due to bending is $\Delta \Phi_{\rm p} = \Phi_{\rm p, old}$ and the change in the colatitude due to bending as $\Delta \zeta = \zeta_{\rm old} - \zeta_L$. These values are used in Eqs. (\ref{longdelay}) and (\ref{latdelay}).

\section{Numerical study}
\label{sec:numerical}

As already mentioned, we use a core-double cone model for the beam. For each of these components, the half opening angle can be defined separately and each of these depends on the observation frequency and the spin period of the pulsar. Following \citetalias{dbb23}, we model the beam at 1.4 GHz and simulate a number of initial position of the light rays on the ${\rm X_m Y_m}$ plane. We trace the paths of the light rays in the curved spacetime including the effect of the bending following the formalism outlined in Section \ref{subsec:bending_spin_full_GR}. We use the rays that are visible after bending to calculate the values of the bending delays and to construct the distorted beam and the resulting pulse profile.

Due to the absence of any known pulsar-black hole binary, we use hypothetical pulsar-black hole binaries with standard values for relevant parameters that are not too different from the theoretical predictions of one of the models by \citet{css21}, namely ZM$-$001. These parameters include the mass of the pulsar ($M_p$), the mass of the companion ($M_c$), the spin period of the pulsar ($P_{\rm s}$), the rate of change of the spin period of the pulsar ($\dot{P}_{\rm s}$), the orbital period ($P_{\rm b}$), and the orbital eccentricity ($e$). The values of these parameters are given in Table \ref{tab:PSRBH}. Additional parameters needed for our calculations are $i$, $\omega$, $\eta_p$, $\lambda_p$, $\eta_{\rm bh}$, $\lambda_{\rm bh}$ and $\alpha$. The meaning of all these parameters are as explained in Section \ref{sec:analytic}. 

We take $i=87.5^\circ$ in most of our investigations. The values of $\alpha$, $\eta_p$, $\lambda_p$ $\eta_{\rm bh}$, and $\lambda_{\rm bh}$ are not observationally measurable. Unless otherwise mentioned, we choose the value of $\alpha$ as $50^\circ$,  $\lambda_p$ and $\eta_p$ as $130^\circ$ and $45^\circ$, respectively, and we also choose the orbital plane of the pulsar as the equatorial plane of the black hole, i.e., $\eta_{\rm bh}=-90^\circ$ and $\lambda_{\rm bh}=i$. The above choice of $\alpha$ and $\lambda_p$ ensures the fact that the LoS cut the cross-section of the beam through its centre, which requires $\alpha=\zeta_L$.

We set the time $t=0$ when the true anomaly ($A_T$) is zero, and assume that the values of the time-dependent parameters as given in Table \ref{tab:PSRBH} are the values at $t=0$. With the above choice of parameters and with $i=\lambda_{\rm bh}=87.5^{\circ}$ and using the formalism mentioned in Sec. \ref{subsec:orbitandspinevolution}, we find that the rate of change of different parameters are very small and can be ignored over a few orbit. Specifically, $\omega$ changes by $0.00128^{\circ}$, $\eta_p$ changes by $-0.000324^{\circ}$, and $\lambda_p$ changes by $-0.000417^{\circ}$ during one orbital revolution regardless of the spin of the black hole (the spin of the black hole affects the parameters in the lower decimal places). Similarly, $P_b$ changes by $-1.35$ s, $e$ changes by $7.05 \times 10^{-15}$, and $i$ changes by $(8.17 \times 10^{-12})^{\circ}$ over one full orbit regardless of the spin of the black hole. The spin of the pulsar changes by $(6.56 \times 10^{-10})^{\circ}$ over one full orbital revolution.

\begin{table}
\caption{Parameters for a canonical pulsar-black hole binary. These are (from top to bottom), the mass of the pulsar ($M_p$), the mass of the black hole ($M_c$), the spin period of the pulsar ($P_{\rm s}$), the rate of change of the spin period of the pulsar ($\dot{P}_{\rm s}$), the orbital period ($P_{\rm b}$), the orbital eccentricity ($e$), the longitude of the periastron of the orbit ($\omega$), the inclination angle between the orbital plane and the sky plane ($i$), the angle between the spin and the magnetic axes of the pulsar$(\alpha$), the angle between the ${\rm X_s}$-axis and the projection of the spin axis of the pulsar in the sky plane ($\eta_p$), the angle between the ${\rm Z_s}$-axis and the spin axis of the pulsar ($\lambda_p$), the angle between the ${\rm X_s}$-axis and the projection of the spin axis of the black hole in the sky plane ($\eta_{\rm bh}$), and the angle between the ${\rm Z_s}$-axis and the spin axis of the black hole ($\lambda_{\rm bh}$). The values of $e, ~M_c, ~M_p, ~P_{\rm s}, ~ \dot{P}_{\rm s}$, and $P_{\rm b}$ are taken from \citet{css21}. The values of other parameters are chosen arbitrarily, but satisfying the visibility condition $| 180^{\circ} + \mathcal{W}_{1.4, 10}^{out} - \alpha | \leq | \lambda_p | \leq | 180^{\circ} - \mathcal{W}_{1.4, 10}^{out} - \alpha |$. $\mathcal{W}_{1.4, 10}^{out}$ is the half-opening angle of the outer cone where the intensity falls to $10 \%$ of its maximum value at 1.4 GHz observing frequency, and is calculated using the expression given in \citetalias{dbb23}. Moreover, the values of the time-dependent parameters are the values at $t=0$ (when $A_T=0$). In this work, the values ($t=0$) values whenever applicable) of some of the parameters have been varied to understand their effects. For three parameters, two values are quoted, the first one used more often than the second one within parenthesis. For most of the cases, we used $\lambda_{\rm bh}=i$. The captions of various figures state the values of these parameters explicitly.}
\label{tab:PSRBH}
\begin{tabular}{ll}
\hline
 Parameters   & Value    \\
 \hline 
 $M_p$ [${\rm M_{ \odot}}$]  & 1.8\\
 $M_c$ [${\rm M_{\odot}}$]    & 14.5 \\
  $P_{\rm s}$ [s] & 1.67\\
 $\dot{P}_{\rm s}$ [${\rm s \, s}^{-1}$] & $1.38\times 10^{-15}$\\
 $P_{\rm b}$ [days] & 5.5\\
 $e$ &0.35\\ 
 $\omega$ [deg] &  73.804\\
 $i$ [deg] &  87.5 (or 90.0)  \\
   $\alpha$ [deg] & 50.0 \\
   $\eta_p$ [deg] & 45.0 \\
   $\lambda_p$ [deg] & 130.0 \\
   $\eta_{\rm bh}$ [deg] & -90.0 \\
   $\lambda_{\rm bh}$ [deg] & 87.5 (or 90.0) \\
   $\tilde{a}$  & 0.9 (or 0.5) \\
\hline
\end{tabular}
\end{table}

For each of the orbital phase of the pulsar, we need to find the light ray that aligns with the LoS after bending. To find this light ray accurately, we use an iterative method. First, we create an $N \times N$ square grid around the LoS, which has the phase (or the longitude) $\Phi = 0$ and the co-latitude $\zeta_{\rm L}$ ( see Fig. \ref{fig:spin_axis_frame}). Hence, the edge of the grid are at longitudes $\pm \Phi_{\text{div}}$ and co-latitudes $\zeta_{\rm L} \pm \zeta_{\rm L div}$, where $\Phi_{\text{div}}$ and $\zeta_{\rm L div}$ are some pre-defined numbers determining the size of the grid. We create initial positions of the light rays at each of these grid points and trace their paths using the formalism described in Sec. \ref{subsec:bending_spin_full_GR}. We then identify the light ray from this grid that is closest to the LoS after bending. After this, we create another $N \times N$ square grid (smaller in size) around this closest point, generate initial positions of the light rays at each of these grid points, trace their paths and find the light ray (from this new grid) that comes closest to the LoS after bending. We continue this process for 20 iterations, which is more than sufficient, as after 15 iterations, the closest light ray remains the same in two successive iterations. In other words, after 15 iterations, the change in the values of the bending delays found in two successive iterations is less than 0.1 nanoseconds. At the first step, we chose both $\Phi_{\text{div}}$ and $\zeta_{\rm L \text{div}}$ as 0.001 radian and in each step reduce both by a factor of 5. The value of $N$ is kept fixed as 11 throughout. We have also checked that any increase in the values of $\Phi_{\text{div}}$, $\zeta_{\rm L \text{div}}$, and $N$ does not affect the values of the bending delays.

\begin{figure}
  \centering
    \includegraphics[width=0.48\textwidth]{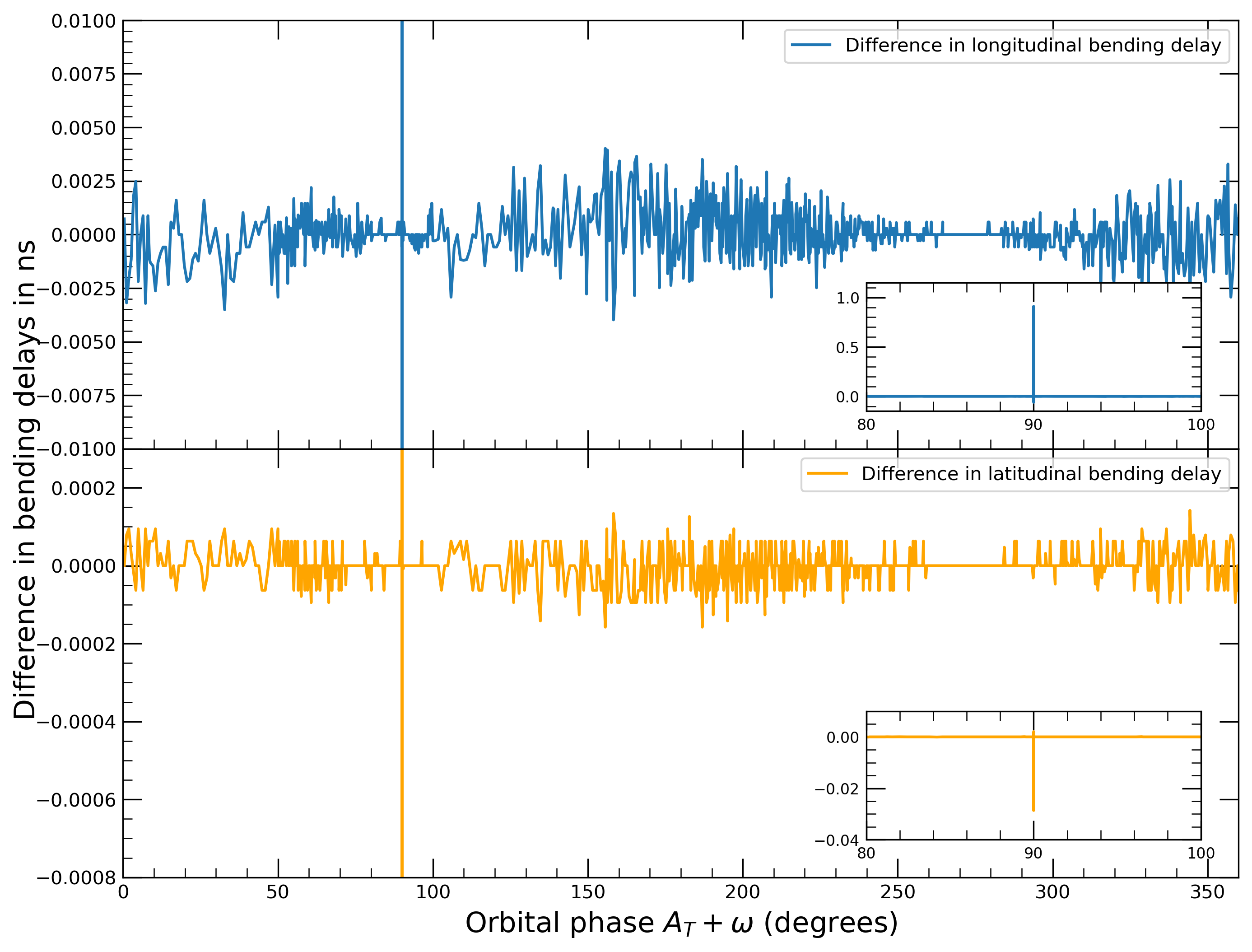}
    \caption{The differences in the bending delays calculated using two different solutions of the null geodesic around a non-rotating black hole, one setting the spin parameter $\tilde{a}=0$ in the Kerr solutions and the other using the Schwarzschild solution. For both of the cases, we use $\lambda_{\rm bh}=i=87.5^{\circ}$, and the values of all other relevant parameters as listed in Table \ref{tab:PSRBH}.}
    \label{fig:Comp_bending_delya_with_ch}
\end{figure}

First, we check whether we get exactly the same results if we set the black hole spin parameter $\tilde{a}=0$ in the formalism followed in this work, i.e., tracing the path of the light rays in the Kerr spacetime around the black hole with those obtained using the formalism adopted in \citetalias{dbb23}, i.e., tracing the path of the light rays in the Schwarzschild spacetime around the black hole. In Fig. \ref{fig:Comp_bending_delya_with_ch}, we plot the orbital phase ($A_T + \omega$) in degrees along the ordinate and the difference in delays (calculated using the two formalism as mentioned above) in nanoseconds along the abscissa, the top panel showing the differences in the longitudinal delays while the bottom panel showing the differences in the latitudinal delays. We see that exactly at $A_T + \omega = 90^{\circ}$, the difference in the longitudinal bending delay is 0.9 nanosecond and latitudinal bending delay is 0.02 nanosecond. At all other orbital phases, the difference in the longitudinal bending delay remains below 0.004 nanosecond and the difference in the latitudinal bending delay remains below 0.0001 nanosecond. Hence, one can conclude that the agreement between the two formalism is satisfactory as it should be from the theoretical point of view. For the rest of the paper, we solve the equations for the null geodesic in the Kerr spacetime even when we set $\tilde{a}$ as zero.

\begin{figure*}
  \centering
  \begin{subfigure}[b]{0.49\textwidth} 
    \centering
    \includegraphics[width=\textwidth]{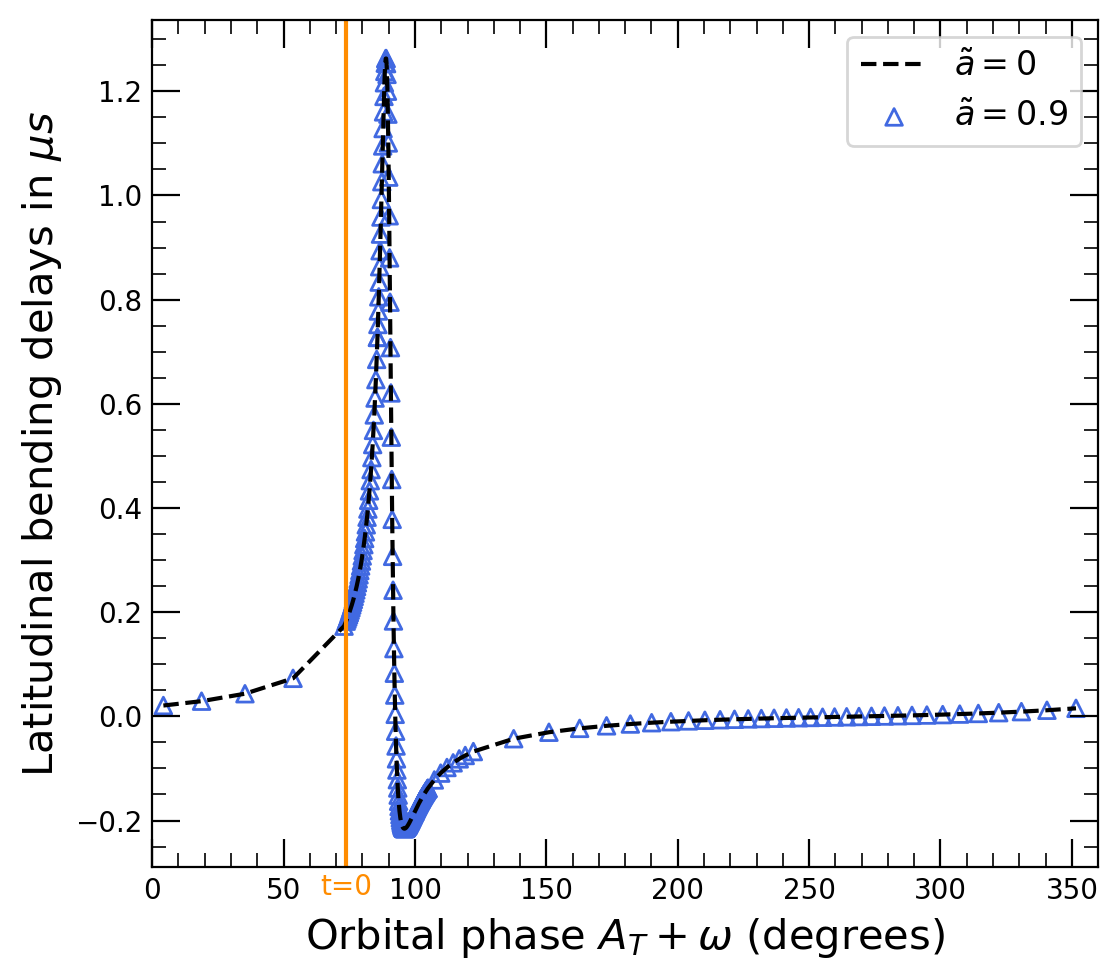}
       \phantomcaption{}
  \label{fig:Latitudinal_bending_delay_a_9} 
  \end{subfigure}
  \hfill
  \begin{subfigure}[b]{0.49\textwidth}  
    \centering
    \includegraphics[width=\textwidth]{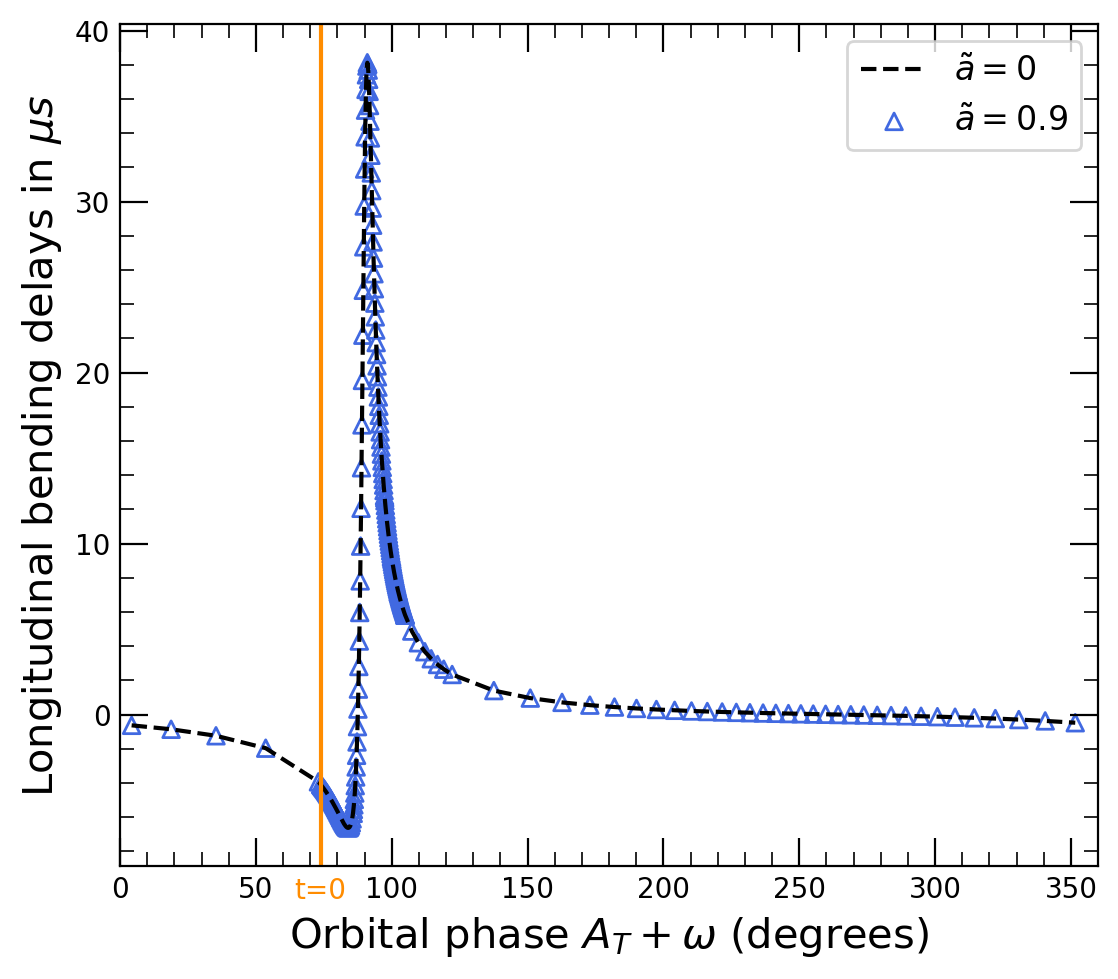}
    \phantomcaption{}
    \label{fig:Longitudinal_bending_delay_a_9}
  \end{subfigure}
  \caption{The comparison of the bending delays for $\tilde{a}=0$ (black dashed line) and $\tilde{a}=0.9$ (blue triangles) over one full orbit for a hypothetical pulsar-black hole binary with $\lambda_{\rm bh}=i=87.5^{\circ}$. The values of all other relevant parameters are the same as those listed in Table \ref{tab:PSRBH}. The left panel shows the latitudinal bending delay, and the right panel shows the longitudinal bending delay. In each of the panels, the vertical orange line corresponds to $A_T=0^{\circ}$ at which we set the time $t=0$. }
  \label{fig:bending_delay_a_9}
\end{figure*}

Next, we aim to quantify the effect of the spin parameter on the values of the bending delays. In Figs. \ref{fig:Latitudinal_bending_delay_a_9} and \ref{fig:Longitudinal_bending_delay_a_9}, we plot the latitudinal and the longitudinal bending delays respectively as functions of the orbital phase for $\tilde{a}=0$ (black dashed line) and $\tilde{a}=0.9$ (blue triangles). Note that, the vertical orange line in each of these figures correspond to $A_T=0^{\circ}$ at which we set the time $t=0$. Hence, the left side of this line actually represent the orbital phase post $360^{\circ}$. For the sake of simplicity $t=0$ line is not shown in any other figure in this section. From Figs. \ref{fig:Latitudinal_bending_delay_a_9} and \ref{fig:Longitudinal_bending_delay_a_9}, we see that the difference is so small that it can not be visualised in these plots. Hence, we decide to rather investigate the differences of the bending delays for zero and non-zero values of the spin parameters. These differences are already defined in Sec. \ref{subsec:FD_bending_delay} as the frame-dragging bending delays (or FD bending delays).

\begin{figure}
\centering
{\hspace*{-0.5cm}\includegraphics[width=1.2\linewidth]{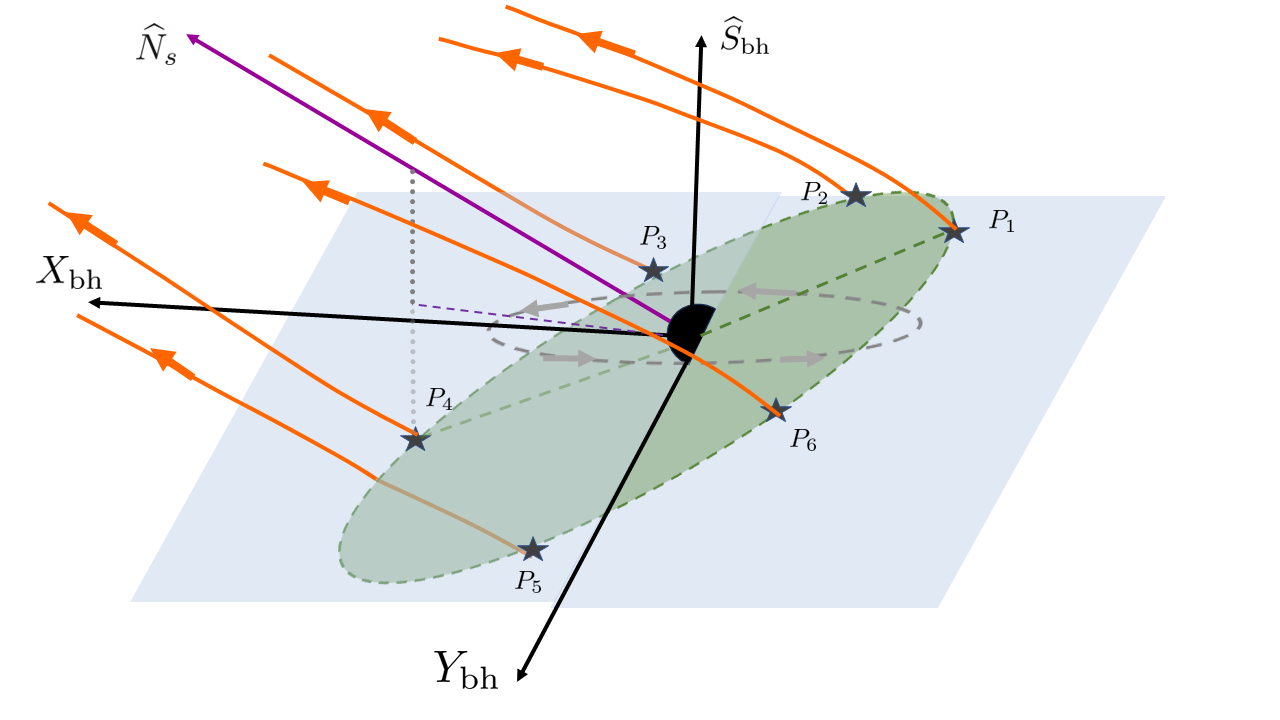}}
\caption{A schematic diagram showing how the directions of the light rays (orange lines) aligning with the LoS (the purple line) vary with respect to the spin of the black hole, represented by a grey dashed circle with arrows. Various positions of the pulsar in its orbit (the light-green plane) are marked with black stars and denoted as $P_1$, $P_2$, $P_3$, $P_4$, $P_5$, and $P_6$. The light-blue plane is the equatorial plane of the black hole while the black hole has been shown (partially) with a black filled half circle (not according to the scale). At positions $P_1$, $P_2$, and $P_3$, the directions of the paths of the light rays is the same as the spin of the black hole. Conversely, at positions $P_4$, $P_5$, and $P_6$, the directions of the paths of the light rays are opposite to the spin of the black hole. The purple dashed line represents the projection of the LoS on the equatorial plane of the black hole and the green dashed line is the projection of the LoS on the orbital plane. Points $P_1$, $P_2$, $P_3$ lie on one side of the green dashed line, while points $P_4$, $P_5$, $P_6$ lie on the opposite side of it. }
\label{fig:schematic_discont}
\end{figure}

\begin{figure*}
  \centering
  \begin{subfigure}[b]{0.49\textwidth}
    \centering
    \includegraphics[width=\textwidth]{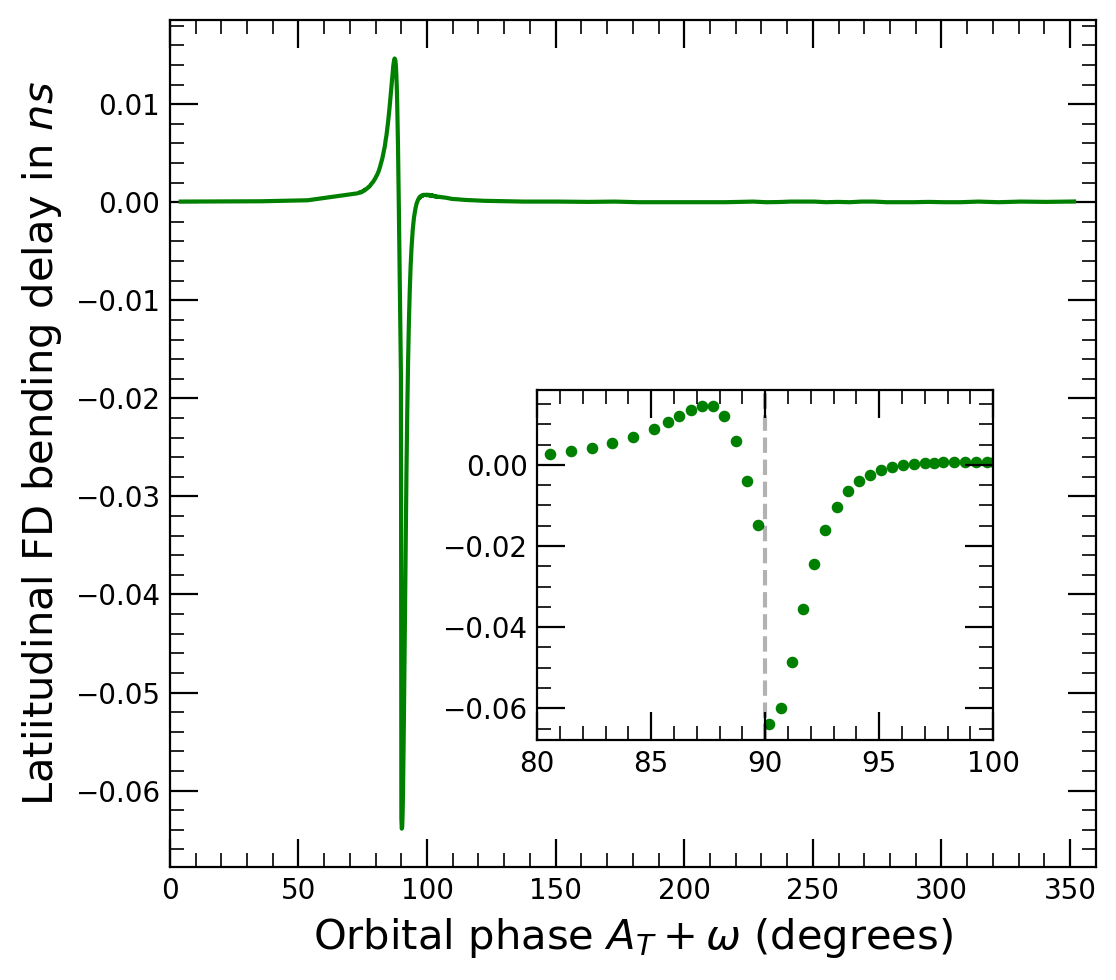}
 \phantomcaption{}
    \label{fig:FD_lati_delay}
  \end{subfigure}
  \hfill
  \begin{subfigure}[b]{0.49\textwidth} 
    \centering
    \includegraphics[width=\textwidth]{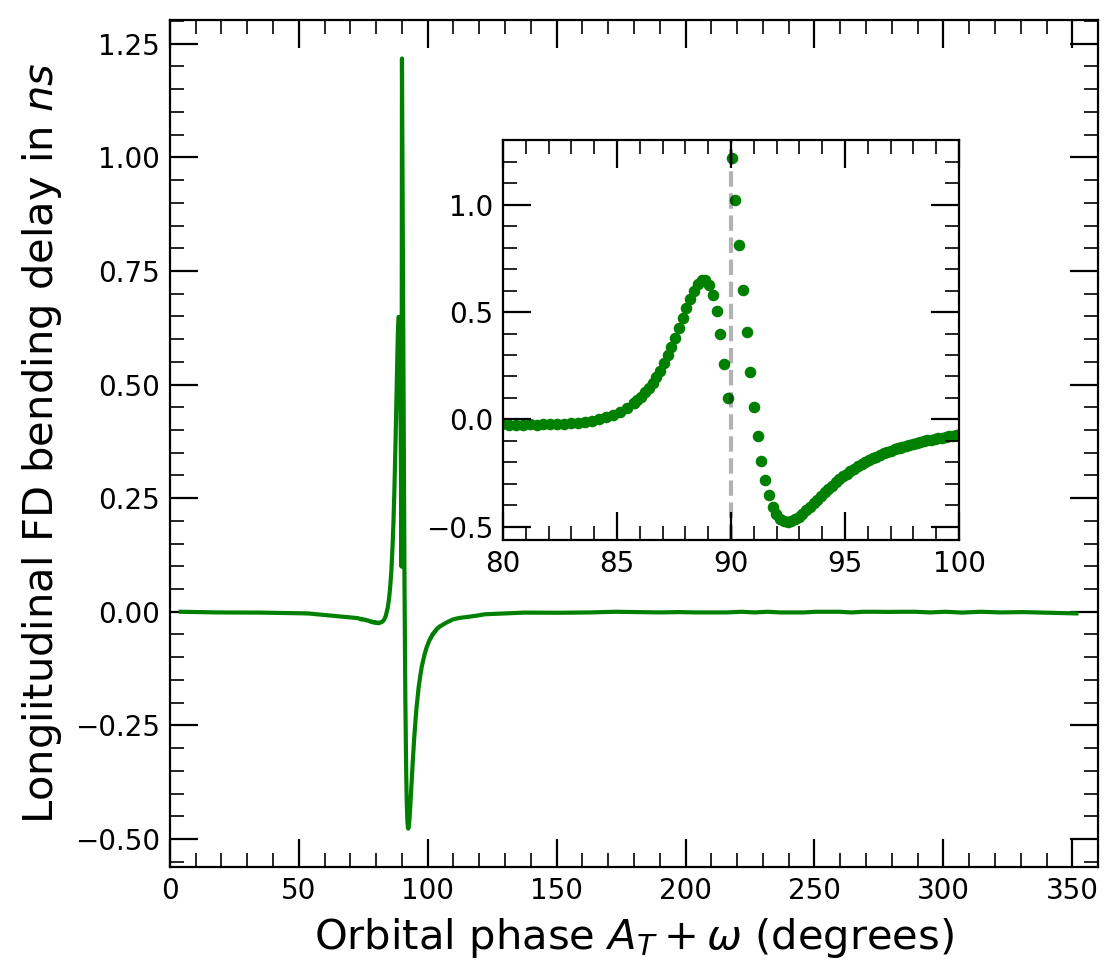}
 \phantomcaption{}
    \label{fig:FD_long_delay}
  \end{subfigure}
  \caption{The FD bending delays for $\tilde{a}=0.9$ over one full orbit for a hypothetical pulsar-black hole binary with $\lambda_{\rm bh}=i=87.5^{\circ}$. The values of all other relevant parameters are the same as those listed in Table \ref{tab:PSRBH}. The left panel shows the latitudinal FD bending delay and the right panel shows the longitudinal FD bending delay. In both of the panels, the grey dashed vertical line in the zoomed subplot represents the location of the discontinuity. }
  \label{fig:FD_delay}
\end{figure*}

Figs. \ref{fig:FD_lati_delay} and \ref{fig:FD_long_delay} show the latitudinal and longitudinal FD bending delays, respectively, both computed for $\tilde{a}=0.9$. The delays are plotted in nanoseconds (ns) along the ordinate, and the orbital phase $A_T + \omega$ in degrees along the abscissa. A discontinuity is observed in both Figs \ref{fig:FD_lati_delay} and \ref{fig:FD_long_delay} at the orbital phase $A_T + \omega = 90^\circ$. The locations of the discontinuities have been marked with vertical grey dashed lines in the zoomed subplots. We find that this discontinuity happens at two orbital phases where $\widehat{r}_s$ comes in the plane containing $\widehat{S}_{\rm bh}$ and $\widehat{N}_s$. As at this two points, the path of the light ray reverses its direction with respect to the spin of the black hole as demonstrated in Fig. \ref{fig:schematic_discont}. The above condition can be expressed analytically as:
\begin{equation}
\label{eq:discontinuity}
    \widehat{N}_{s} \cdot (\widehat{r}_{s} \times \widehat{S}_{\rm bh}) = 0 .
\end{equation}
The vectors are as defined in Sec. \ref{subsec:orbitalgeometrymotion}. Using Eqs. (\ref{eq:r_in_s_frame}), (\ref{eq:Sbh_with_angles}), (\ref{eq:rb_with angles}), and (\ref{eq:Ns_vector}), we find that the condition given in Eq. (\ref{eq:discontinuity}) is satisfied for $A_T+\omega = 90^{\circ}$ and $A_T+\omega = 270^{\circ}$ when $\eta_{\rm bh} = -90^{\circ}$, i.e., the spin axis of the black hole is in the $\rm Y_sZ_s$ plane. The discontinuity can be seen in both the latitudinal and the longitudinal FD bending delays for $A_T+\omega = 90^{\circ}$ in Figs \ref{fig:FD_lati_delay} and \ref{fig:FD_long_delay}. However, the discontinuity at $A_T + \omega = 270^\circ$ is very small because the bending itself is negligible at this orbital phase, thus the effect of the spin on the light bending is not perceptible. 

\begin{figure*}
  \centering
  \begin{subfigure}[b]{0.49\textwidth}
    \centering
    \includegraphics[width=\textwidth]{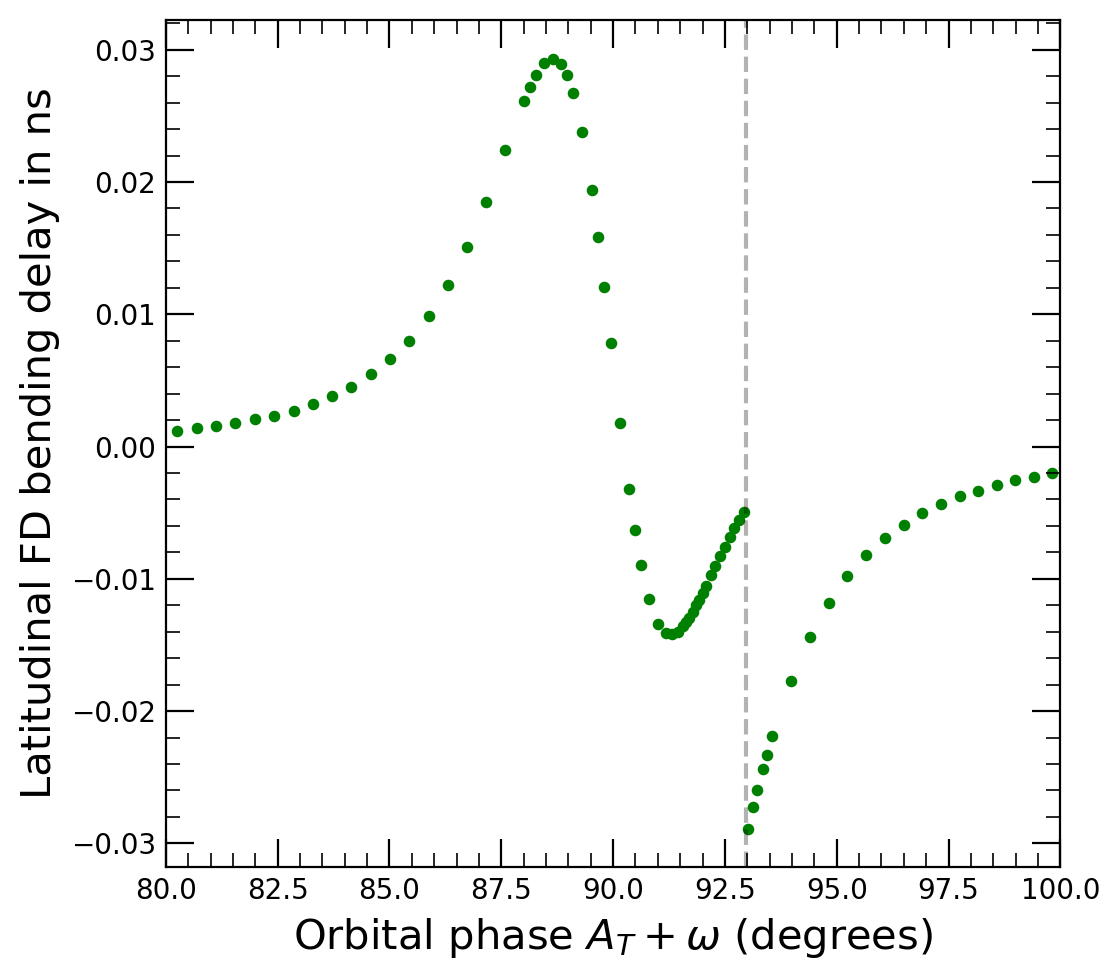}
    \phantomcaption{}
    \label{fig:FD_lat_delay_eta_bh40}
  \end{subfigure}
  \hfill
  \begin{subfigure}[b]{0.49\textwidth} 
    \centering
    \includegraphics[width=\textwidth]{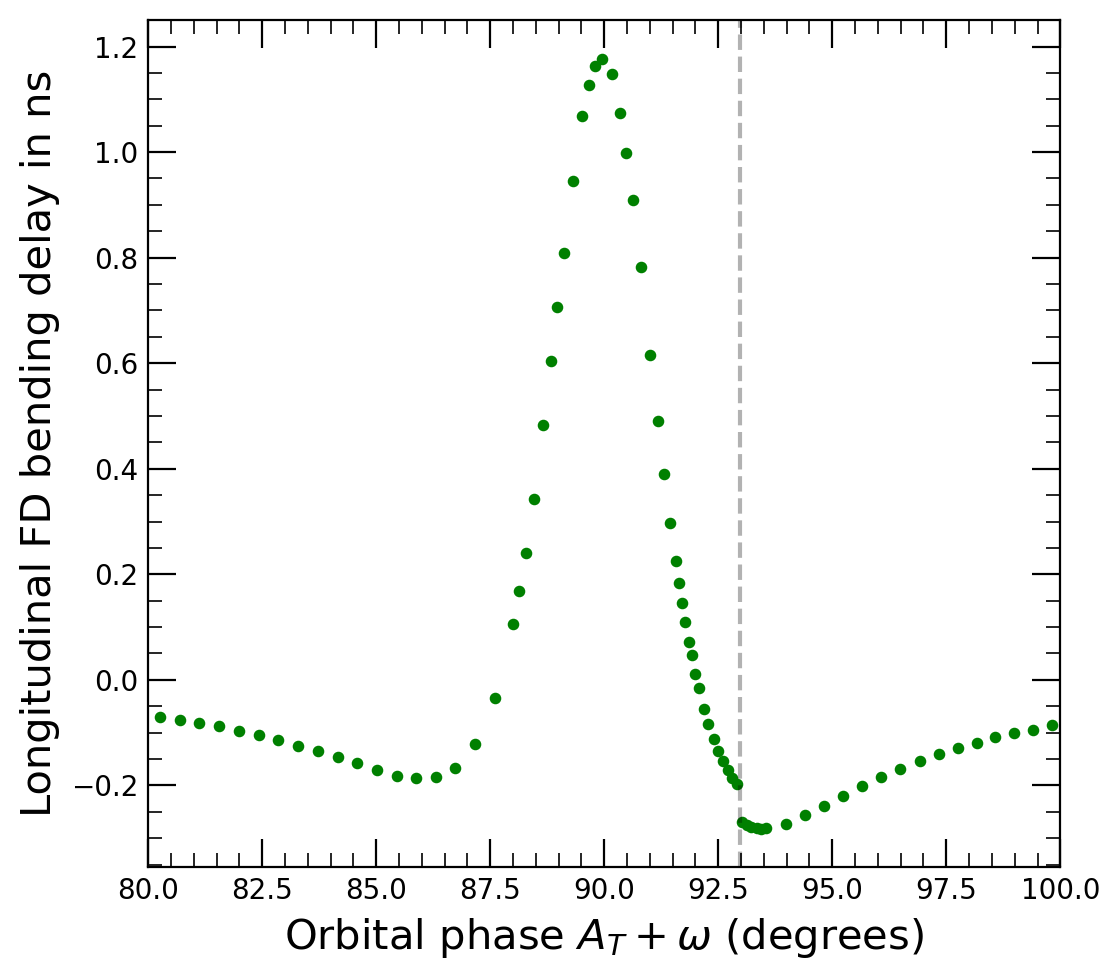}
    \phantomcaption{}
    \label{fig:FD_long_delay_eta_bh40}
  \end{subfigure}
  \caption{The FD bending delays for $\tilde{a}=0.9$ over one full orbit for a hypothetical pulsar-black hole binary. We have chosen $\eta_{\rm bh}=-40^{\circ}$, $\lambda_{\rm bh}=i=87.5^{\circ}$, and the values of all other relevant parameters are the same as those listed in Table \ref{tab:PSRBH}. The left panel shows the latitudinal FD bending delay and the right panel shows the longitudinal FD bending delay. In both of the panels, the grey dashed vertical line represents the location of the discontinuity.  }
  \label{fig:FD_delay_eta_bh40}
\end{figure*}

Similarly, the condition given in Eq. (\ref{eq:discontinuity}) is satisfied for $A_T+\omega = 92.98^{\circ}$ and $A_T+\omega = 272.98^{\circ}$ when $\eta_{\rm bh} = -40^{\circ}$. Figs. \ref{fig:FD_lat_delay_eta_bh40} and \ref{fig:FD_long_delay_eta_bh40} show the latitudinal and the longitudinal FD bending delays for $\eta_{\rm bh} = -40^\circ$ zooming near the larger discontinuity, i.e., in the range of $A_T + \omega$ as $80^{\circ}$-$100^{\circ}$ . As expected, we see the discontinuities at $A_T + \omega = 92.975^\circ$, although the discontinuity is less prominent for the longitudinal FD delay as the delay itself is small at this orbital phase. The discontinuity at $A_T + \omega = 272.98^\circ$ can not be seen as the FD delays themselves are very small and not plotted in this orbital phase. From Fig. \ref{frames}, it is understandable that a change of $\lambda_{\rm bh}$ for a fixed $\eta_{\rm bh} $ does not affect the solution of Eq. (\ref{eq:discontinuity}).

\begin{figure*}
  \centering
  \begin{subfigure}[b]{0.49\textwidth}
    \centering
    \includegraphics[width=\textwidth]{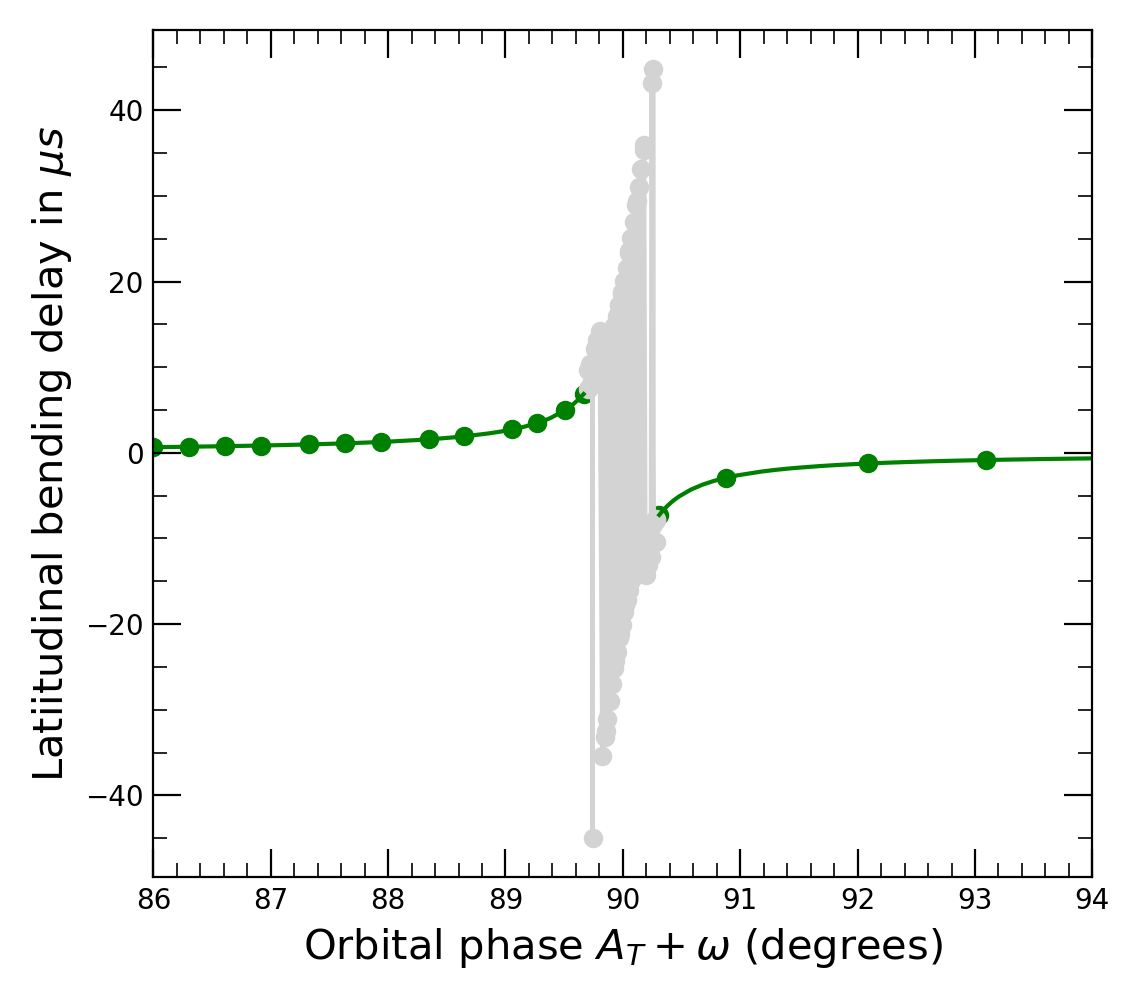}
    \caption{Latitudinal FD bending delays for $\tilde{a} = 0.5$, $\lambda_{\rm bh}=i=90^{\circ}$.}
    \label{fig:lati_delay_i_90}
  \end{subfigure}
  \hfill
  \begin{subfigure}[b]{0.49\textwidth} 
    \centering
    \includegraphics[width=\textwidth]{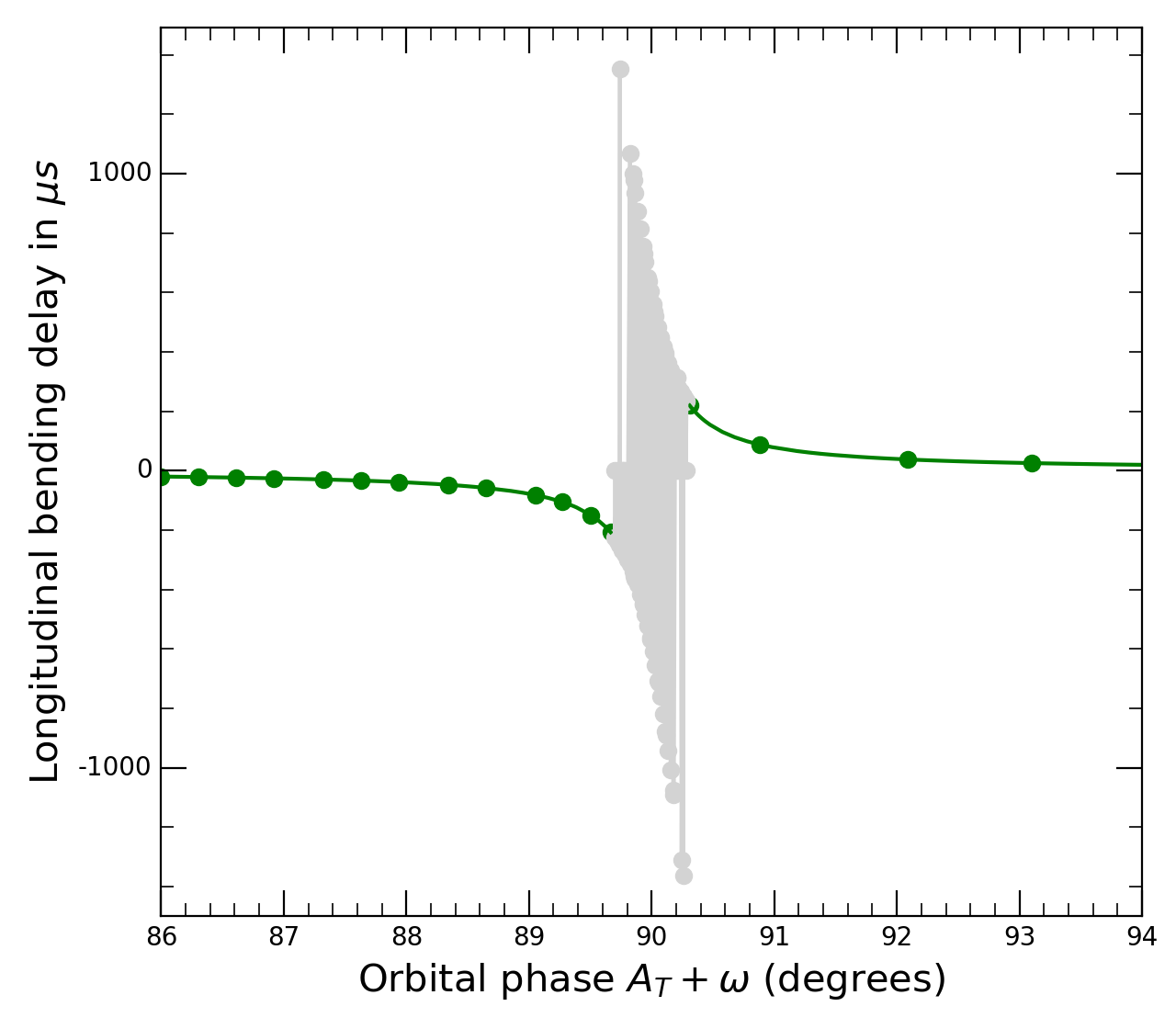}
    \caption{Longitudinal FD bending delays for $\tilde{a} = 0.5$, $\lambda_{\rm bh}=i=90^{\circ}$.}
    \label{fig:long_delay_i_90}
  \end{subfigure}
  \hfill
  \begin{subfigure}[b]{0.49\textwidth} 
    \centering
    \includegraphics[width=\textwidth]{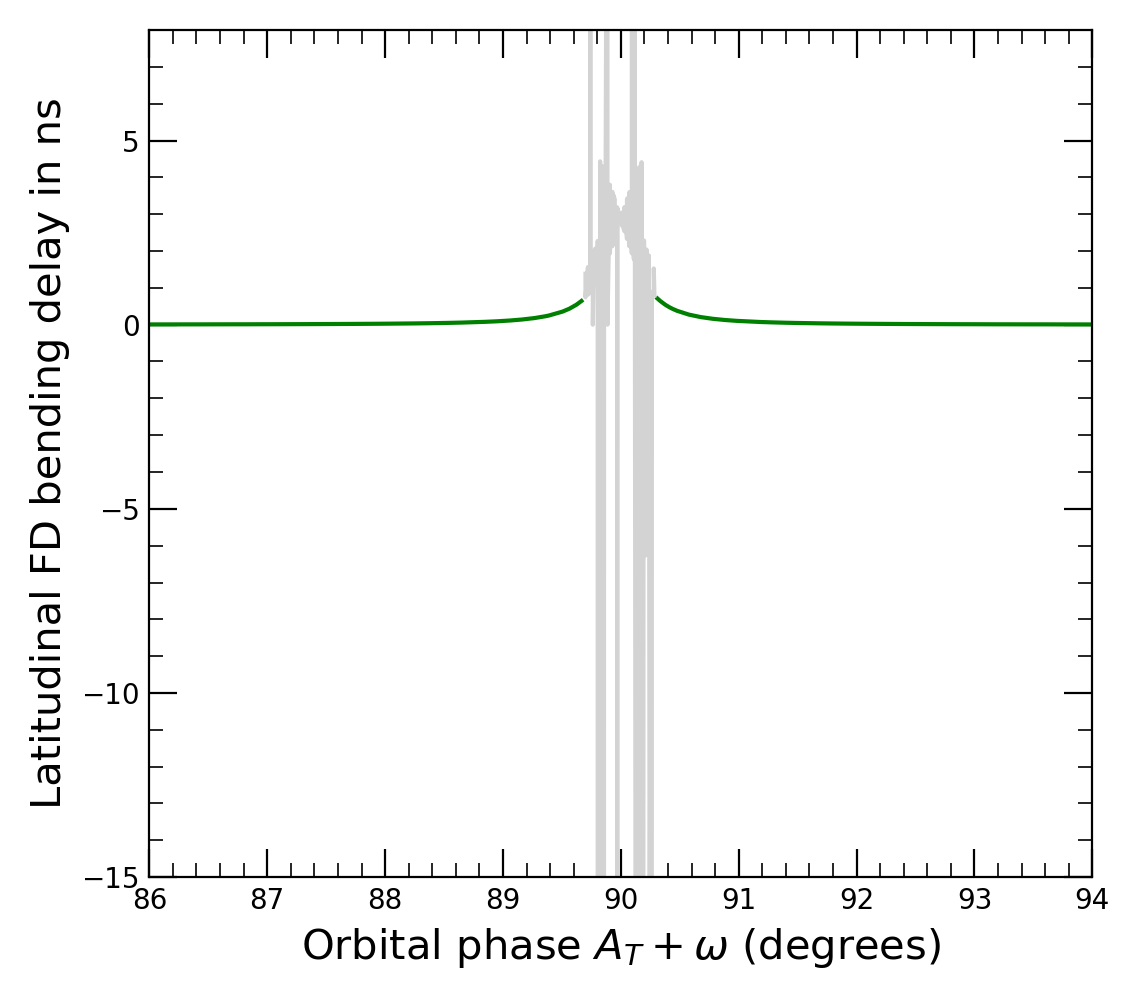}
    \caption{Latitudinal FD bending delays for $\tilde{a} = 0.5$, $\lambda_{\rm bh}=i=90^{\circ}$.}
    \label{fig:FD_lati_delay_i_90}
  \end{subfigure}
  \hfill
  \begin{subfigure}[b]{0.49\textwidth} 
    \centering
    \includegraphics[width=\textwidth]{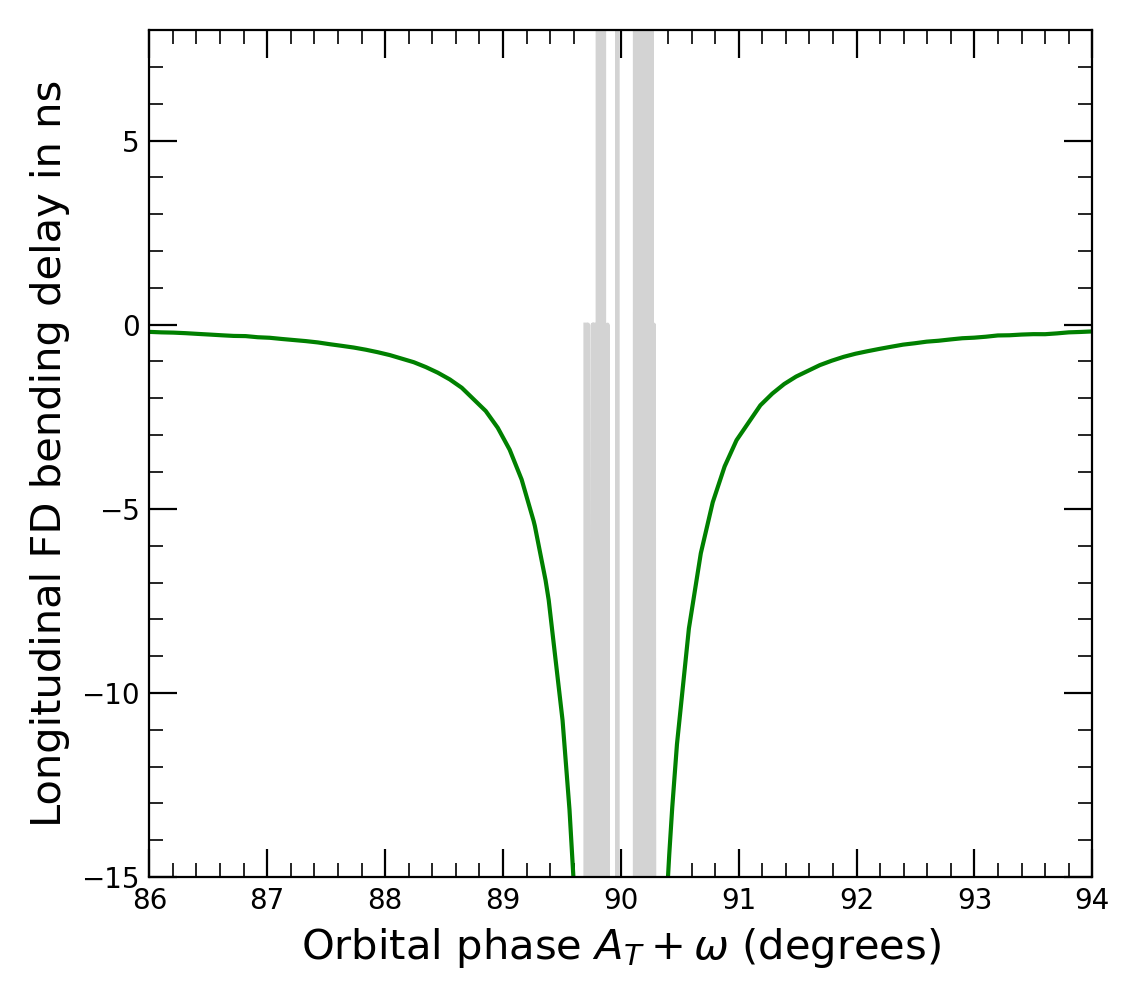}
    \caption{Longitudinal FD bending delays for $\tilde{a} = 0.5$, $\lambda_{\rm bh}=i=90^{\circ}$.}
    \label{fig:FD_long_delay_i_90}
  \end{subfigure}
  \caption{The bending delays and the FD bending delays obtained for a hypothetical pulsar-black hole binary. The upper left panel (panel a) shows the latitudinal bending delay and the upper right panel (panel b) shows the longitudinal bending delay. The lower left panel (panel c) shows the latitudinal FD bending delay and the lower right panel (panel d) shows the longitudinal FD bending delay. We have chosen $\tilde{a} = 0.5$, $\lambda_{\rm bh}=i=90^{\circ}$, and the values of all other relevant parameters are the same as those listed in Table \ref{tab:PSRBH}. The delay curves become irregular between $A_T + \omega =89.7^\circ$ and $A_T + \omega = 90.3^\circ$. The irregular regions are shown in gray. }
  \label{fig:irregularites}
\end{figure*}

One intriguing aspect to investigate is the FD delays for an edge-on orbit ($i=90^\circ$) with the pulsar rotating in the equatorial plane of the black hole, specified by $\eta_{\rm bh}=-90^\circ$ and $\lambda_{\rm bh}=90^\circ$. As discussed in \citetalias{dbb23}, for an edge-on binary, the bending delays exhibit irregularities near the superior conjunction ($A_T + \omega = 90^\circ$). At this point, the cross-section of the pulsar beam reach the observer by crossing the black hole, which acts as a convex lens. This lensing effect converges light rays from a ring-shaped region towards the LoS, causing these irregularities \citepalias{dbb23}. In Figs. \ref{fig:lati_delay_i_90} and \ref{fig:long_delay_i_90}, we plot the latitudinal and longitudinal bending delays, respectively, for $\tilde{a}=0$ (solid line) and $\tilde{a}=0.5$ (green filled circles) for $i=90^\circ$, $\eta_{\rm bh}=-90^\circ$, and $\lambda_{\rm bh}=90^\circ$, zooming near orbital phase where the irregularities are expected. These irregular regions are highlighted in grey. Both delays exhibit irregularities between the orbital phases $A_T + \omega = 89.7^\circ$ and $A_T + \omega = 90.3^\circ$. In Figs. \ref{fig:FD_lati_delay_i_90} and \ref{fig:FD_long_delay_i_90}, we plot the the latitudinal and the longitudinal FD delays, respectively, and see similar irregular regions.

In \citetalias{dbb23}, where we solved the equations for the null geodesic in the Schwarzschild spacetime, we showed that the values of both the longitudinal and the latitudinal bending delays increase with the increase in the values of the companion mass ($M_c$), the orbital inclination angle ($i$), and the orbital eccentricity ($e$), and decrease with the increase of the orbital period ($P_b$). Here we noticed similar trends when we solve the equations for the null geodesic (in each case, for a fixed value of $\tilde{a}$) for the exact same physical reasons. It seems more interesting to explore what happens to the FD bending delays, especially near the orbital phase of $90^{\circ}$ where the magnitudes of the FD bending delays are large (as seen in Fig. \ref{fig:FD_delay}). Hence, we investigate how the FD bending delays change if we change only one parameter at a time keeping the remaining ones fixed as those mentioned in Table. \ref{tab:PSRBH}, with $\tilde{a}=0.5$ and $i=87.5^{\circ}$.

\begin{figure*}
  \centering
  \begin{minipage}{0.33\textwidth}
    \centering
    \includegraphics[width=\textwidth]{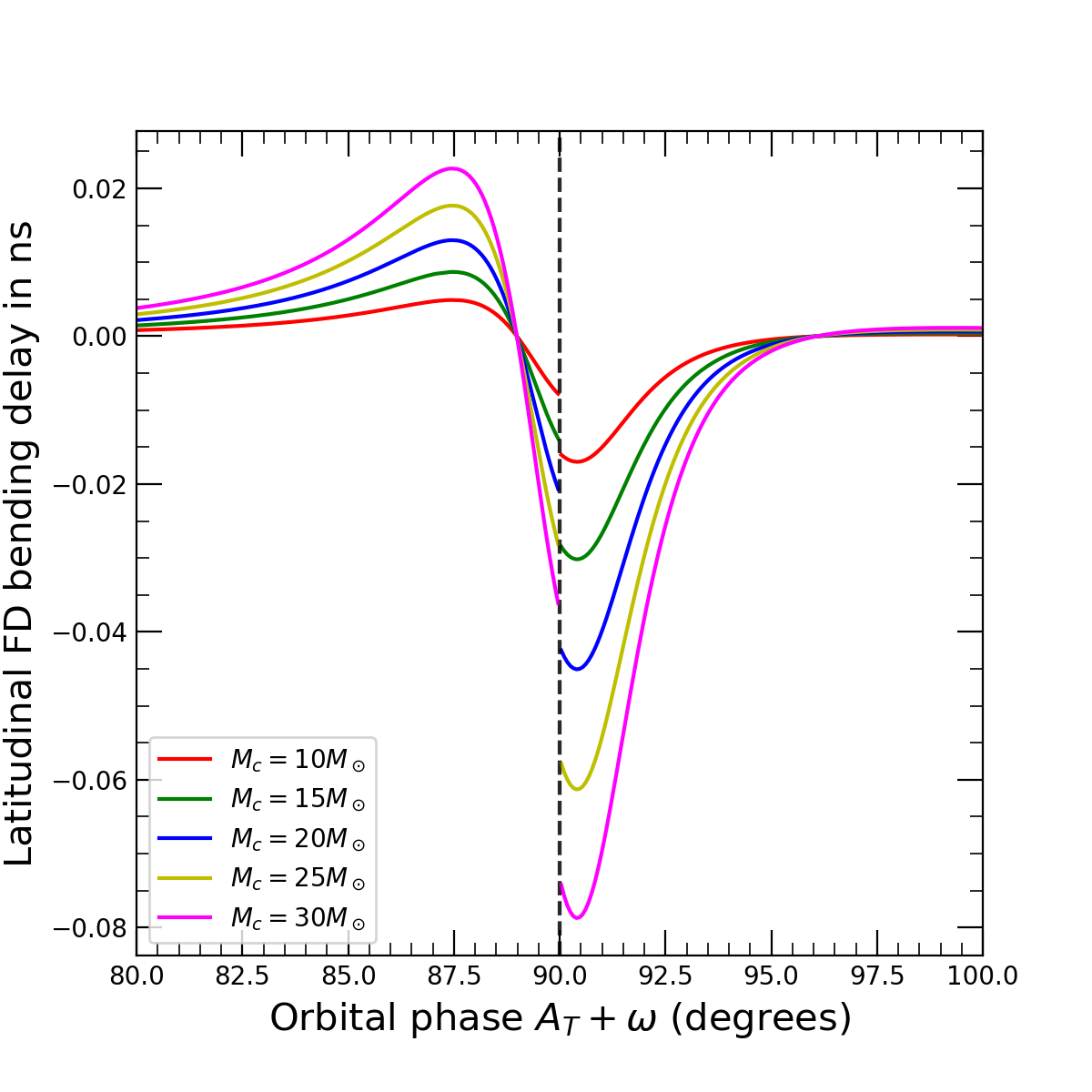}
    \subcaption{Latitudinal FD bending delays for different values of $M_{c}$.}
    \label{fig:diff_mass_lati}
  \end{minipage}
  \hfill
  \begin{minipage}{0.33\textwidth}
    \centering
    \includegraphics[width=\textwidth]{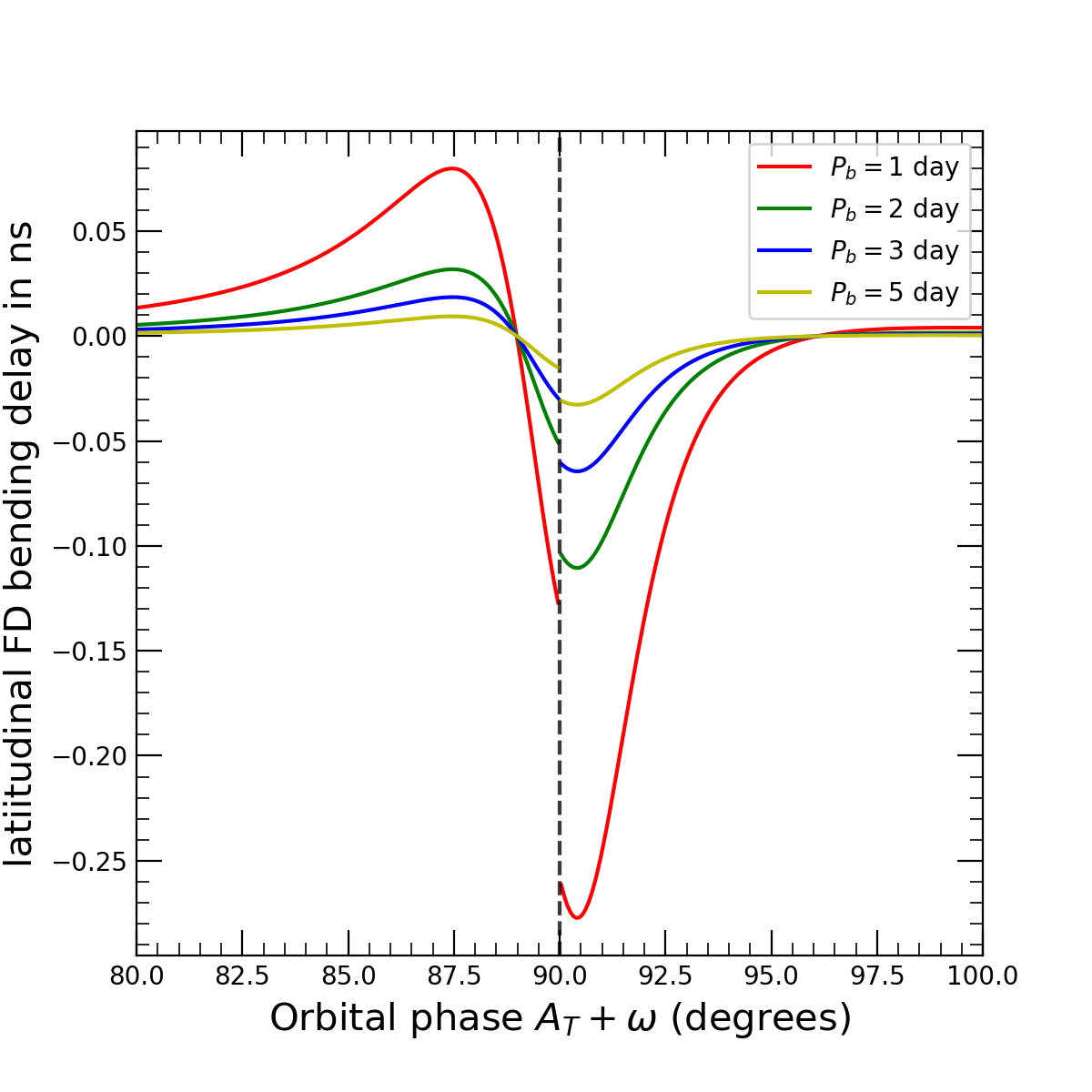}
    \subcaption{Latitudinal FD bending delays for different values of $Pb$.}
    \label{fig:diff_Pb_lati}
  \end{minipage}
  \hfill
  \begin{minipage}{0.33\textwidth}
    \centering
    \includegraphics[width=\textwidth]{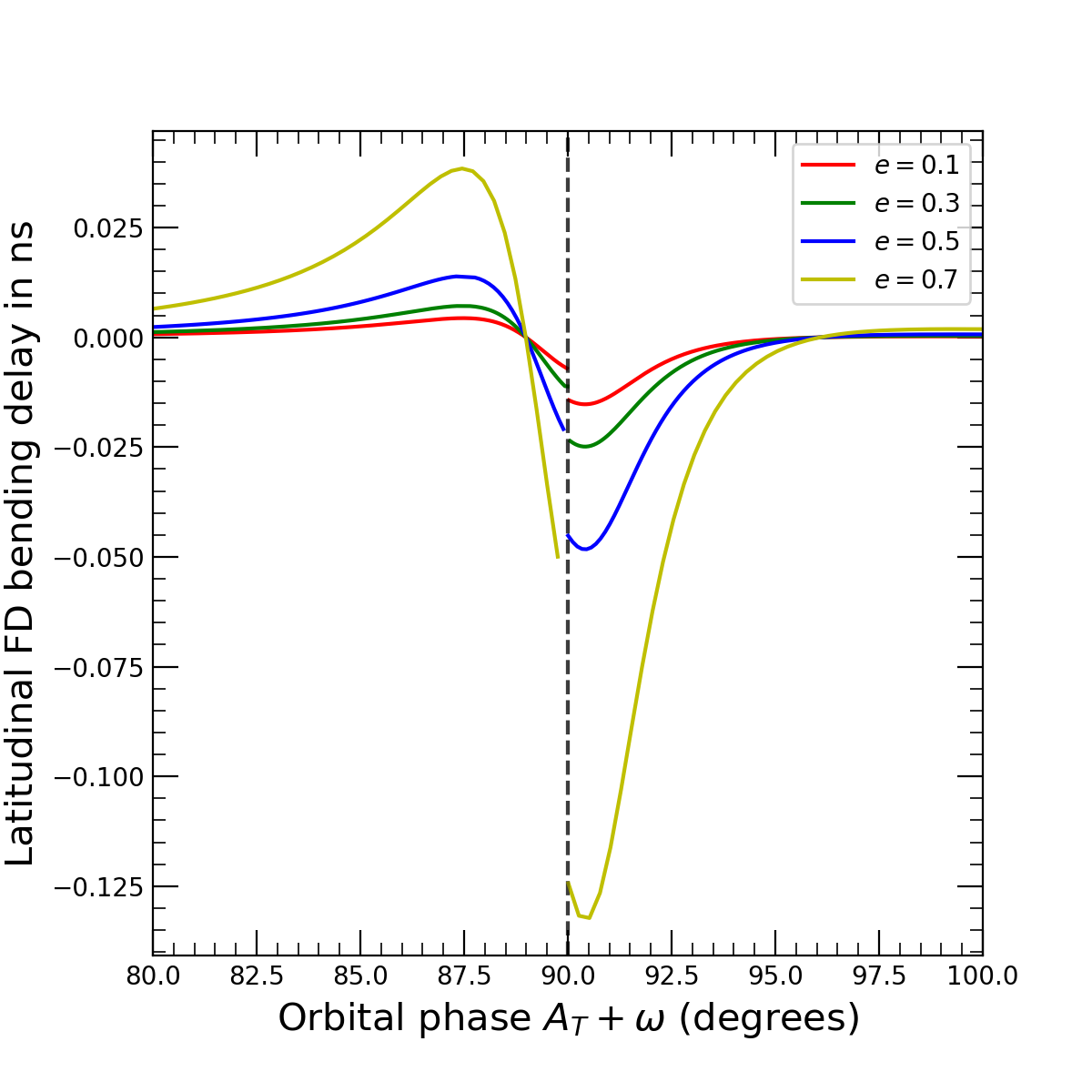}
    \subcaption{Latitudinal FD bending delays for different values of $e$.}
    \label{fig:diff_e_lati}
  \end{minipage}
      \begin{minipage}{0.33\textwidth}
    \centering
    \includegraphics[width=\textwidth]{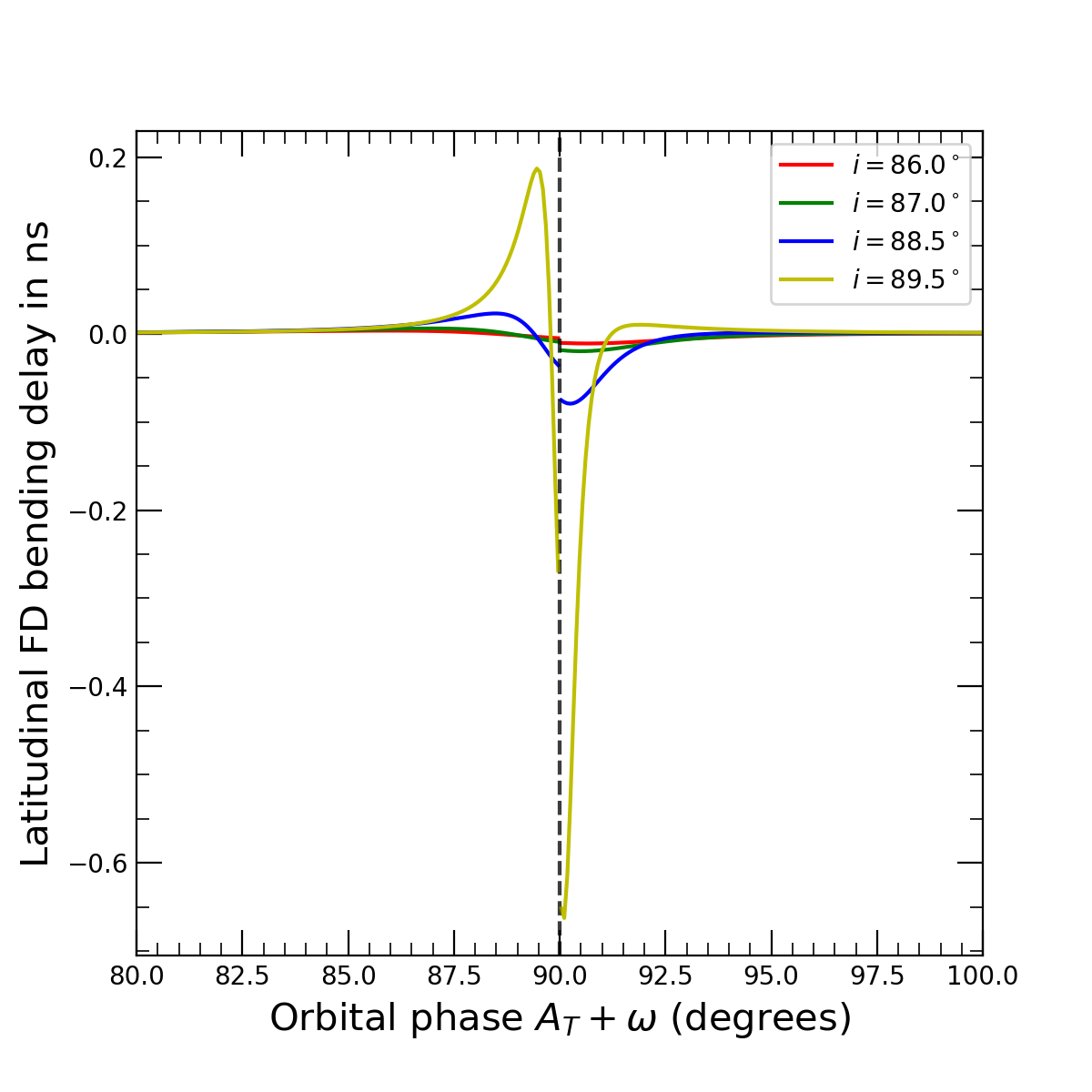}
    \subcaption{Latitudinal FD bending delays for different values of $i$.}
    \label{fig:diff_i_lati}
  \end{minipage}
  \hfill
  \begin{minipage}{0.33\textwidth}
    \centering
    \includegraphics[width=\textwidth]{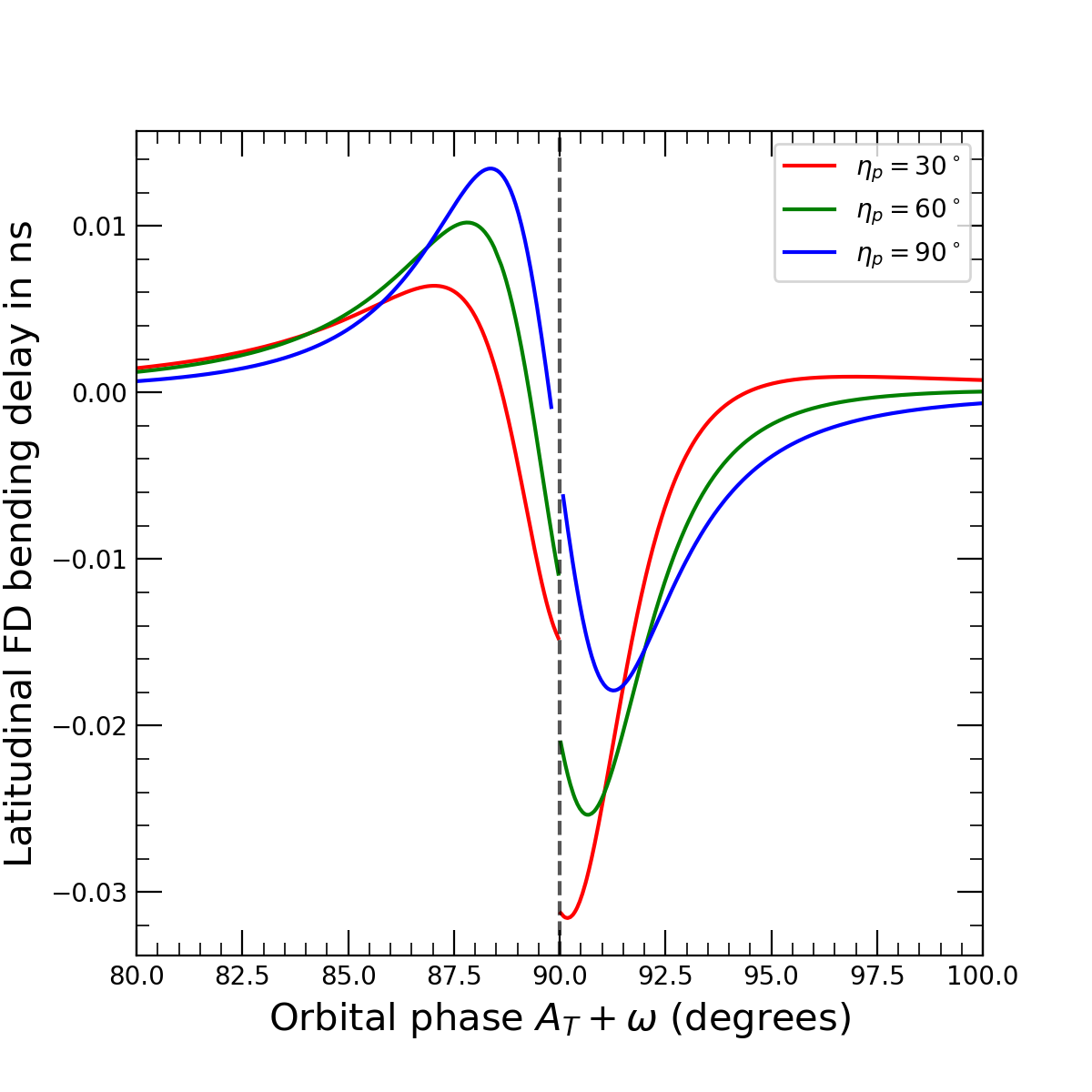}
    \subcaption{Latitudinal FD bending delays for different positive values of $\eta_p$.}
    \label{fig:diff_eta_pulsar_lati}
  \end{minipage}
  \hfill
  \begin{minipage}{0.33\textwidth}
    \centering
    \includegraphics[width=\textwidth]{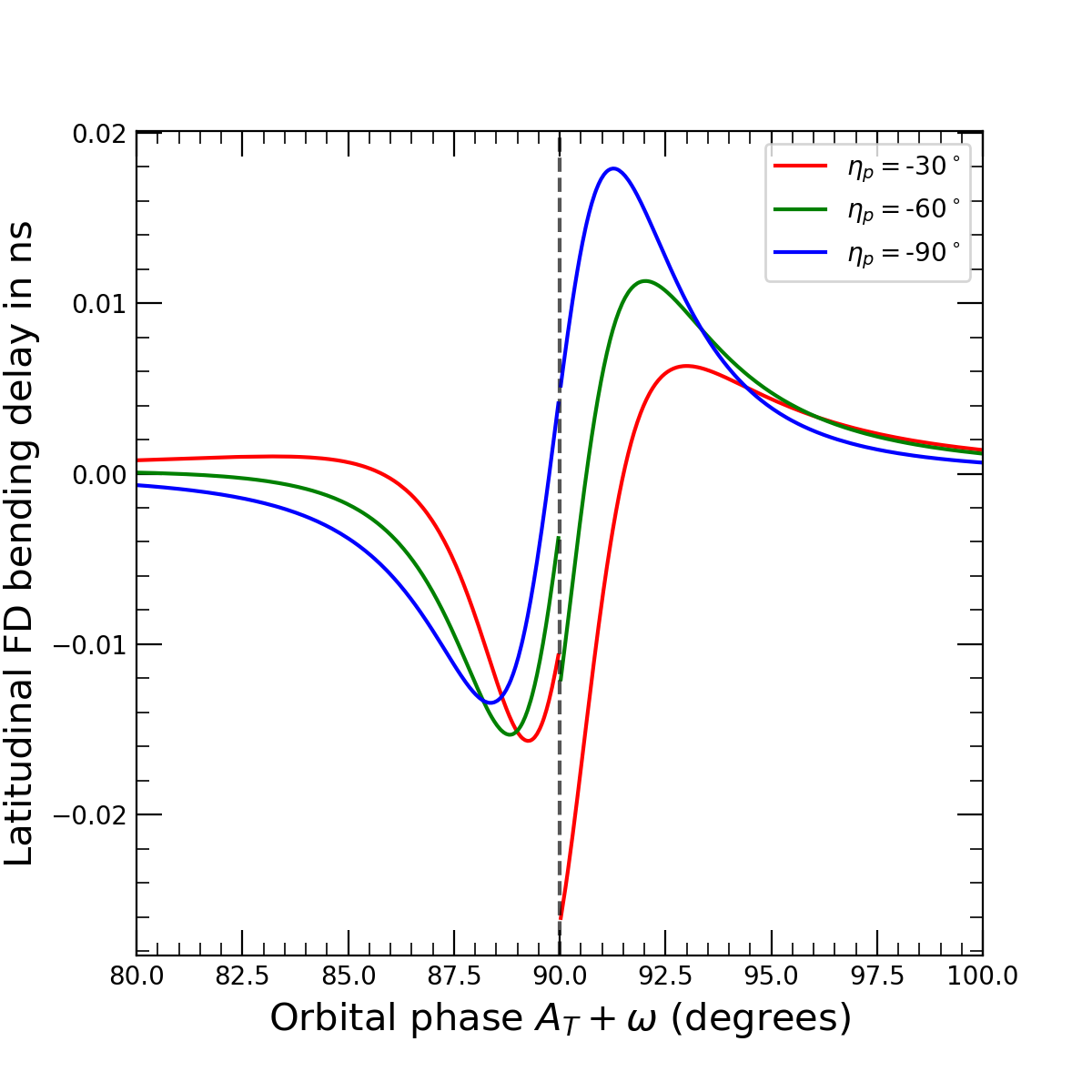}
    \subcaption{Latitudinal FD bending delays for different negative values of $\eta_p$.}
    \label{fig:diff_eta_pulsar_neg_lati}
  \end{minipage}
    \caption{The latitudinal FD bending delay curves for hypothetical pulsar-black hole binaries, changing one parameter in each panel and keeping all the other parameters the same as those mentioned in Table \ref{tab:PSRBH}. We have chosen $\tilde{a}=0.5$ except for panel (f) where $\tilde{a}$ is varied. Similarly, we set $\lambda_{\rm bh}=i=87.5^{\circ}$ except for panel (c) where $i$ is varied ($\lambda_{\rm bh}$ is still kept fixed at $87.5^{\circ}$). The vertical dashed lines represent the locations of the discontinuities. }
  \label{fig:diff_parameterOrbit_lati_delay}
\end{figure*}

\begin{figure*}
  \centering
  \begin{minipage}{0.33\textwidth}
    \centering
    \includegraphics[width=\textwidth]{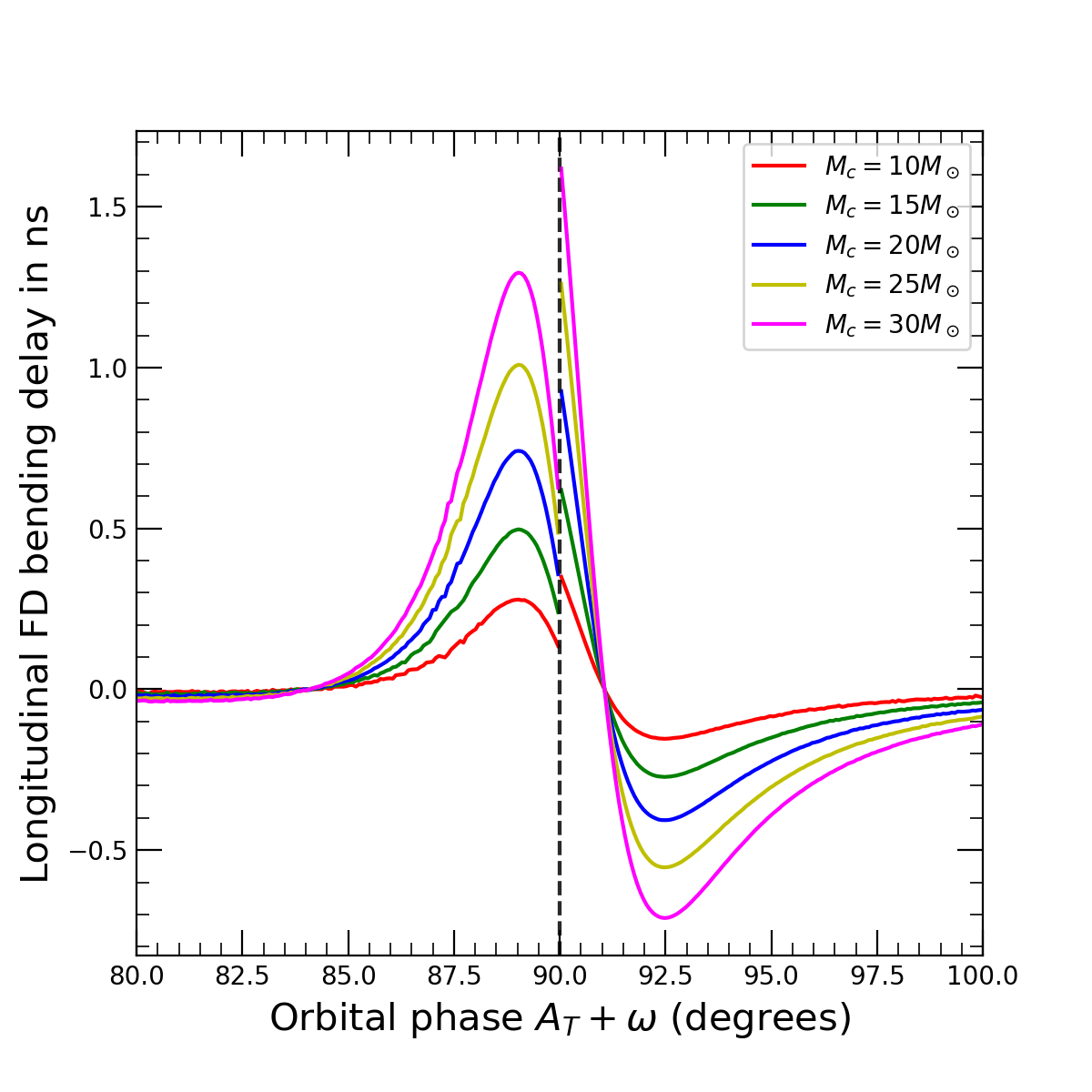}
    \subcaption{Longitudinal FD bending delays for different values of $M_{c}$.}
    \label{fig:diff_mass_long}
  \end{minipage}
   \hfill
  \begin{minipage}{0.33\textwidth}
    \centering
    \includegraphics[width=\textwidth]{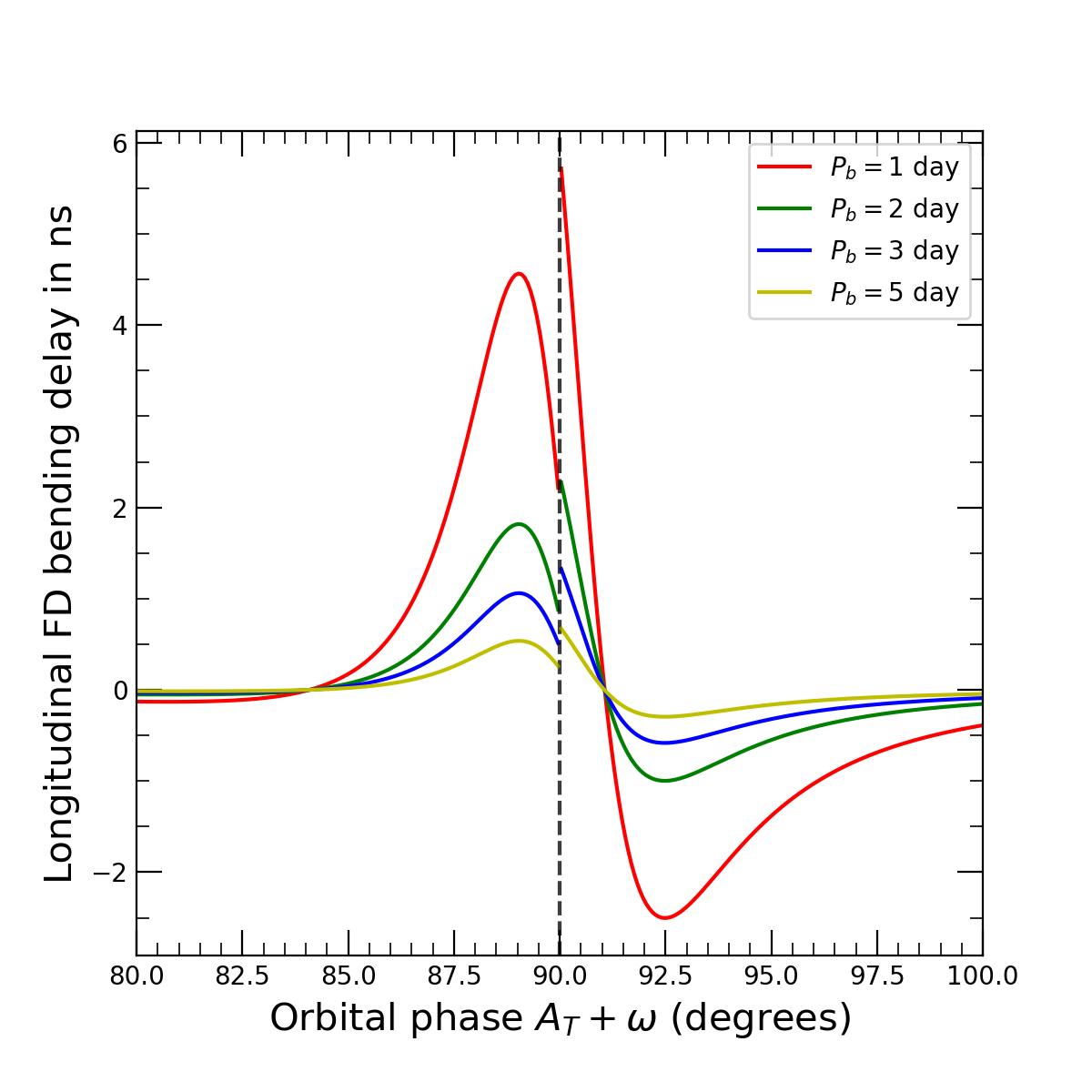}
    \subcaption{Longitudinal FD bending delays for different values of $P_b$.}
    \label{fig:diff_Pb_long}
  \end{minipage}
  \hfill
  \begin{minipage}{0.33\textwidth}
    \centering
    \includegraphics[width=\textwidth,height=6cm]{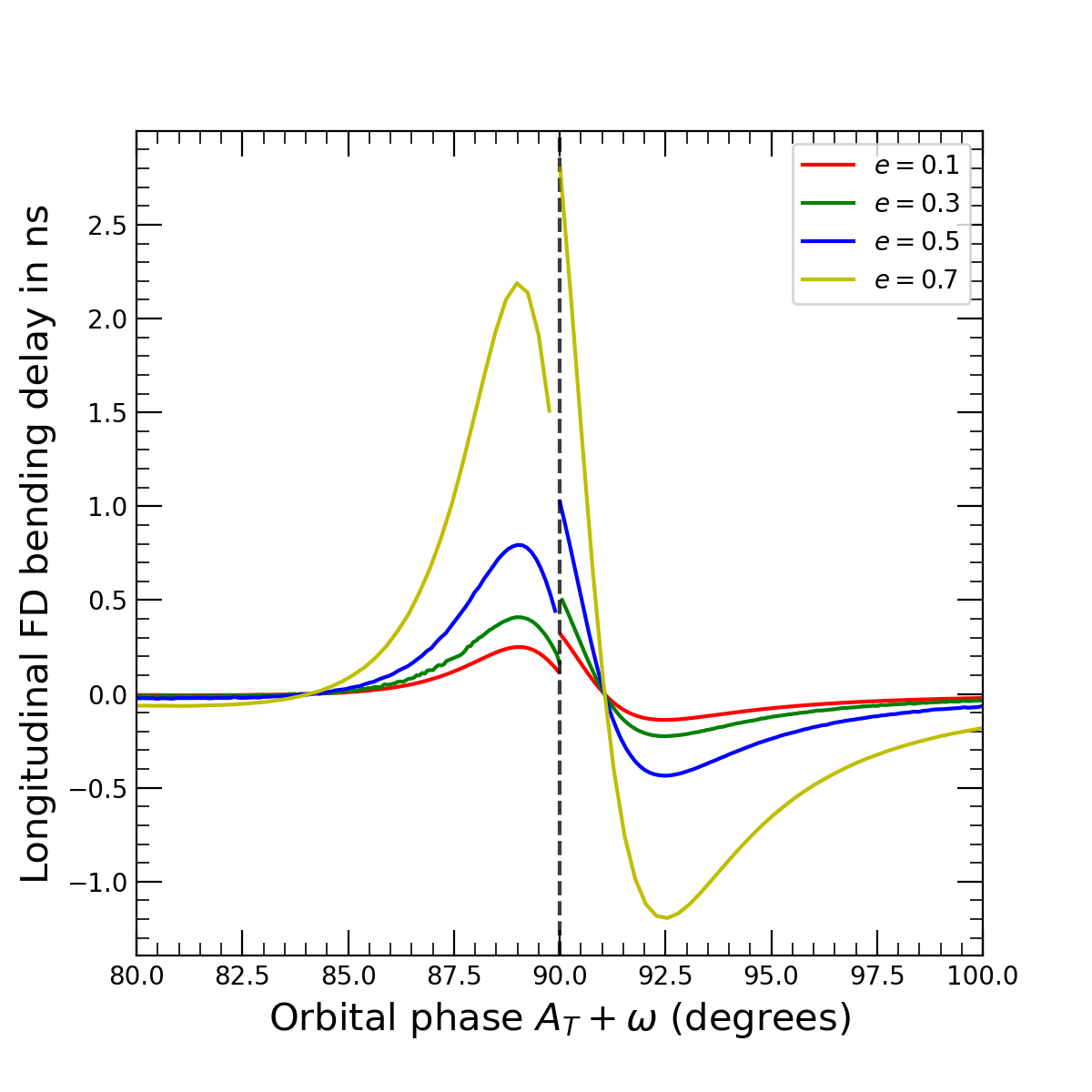}
    \subcaption{Longitudinal FD bending delays for different values of $e$.}
    \label{fig:diff_e_long}
  \end{minipage}
      \hfill
  \begin{minipage}{0.33\textwidth}
    \centering
    \includegraphics[width=\textwidth]{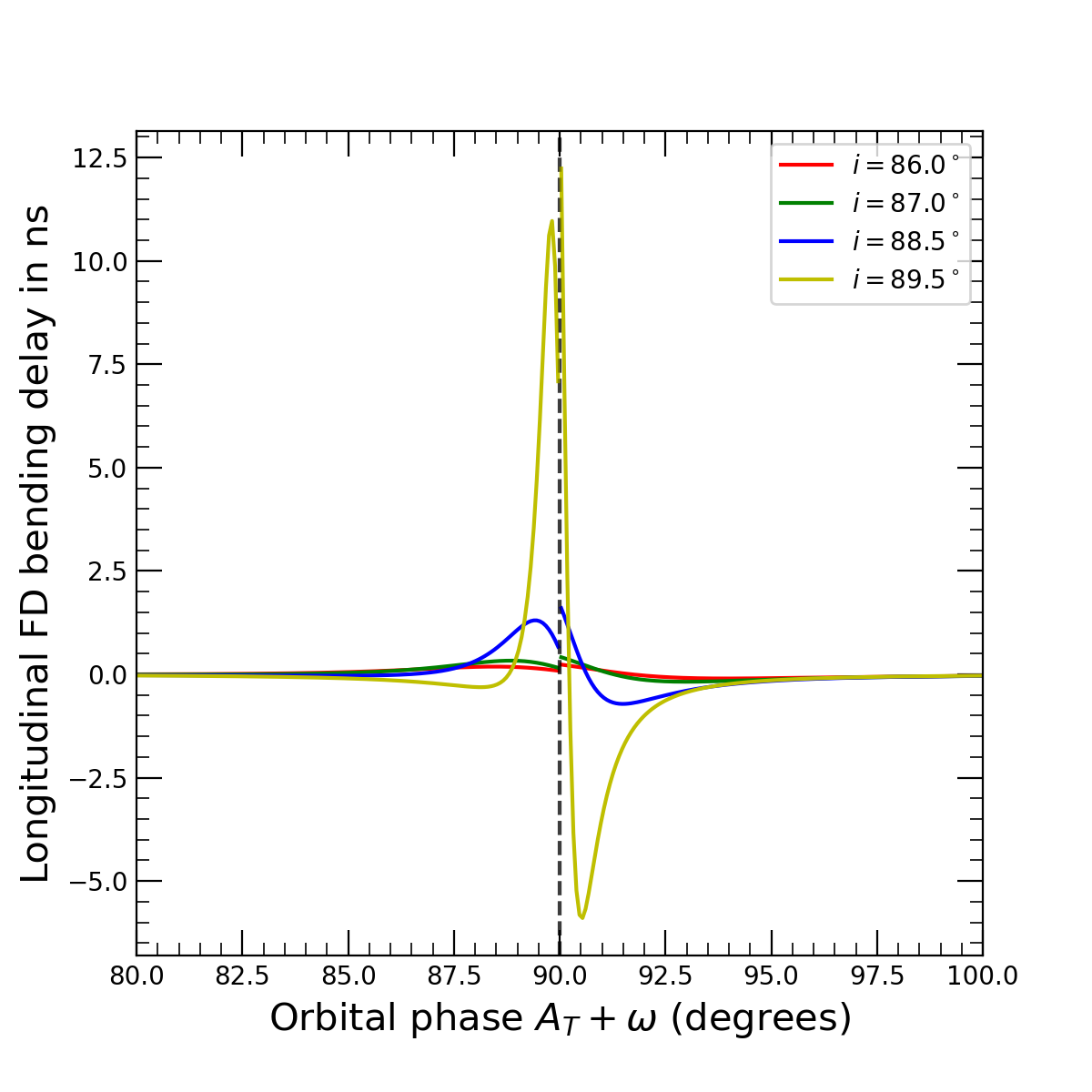}
    \subcaption{Longitudinal FD bending delays for different values of $i$.}
    \label{fig:diff_i_long}
  \end{minipage}
  \hfill
  \begin{minipage}{0.33\textwidth}
    \centering
    \includegraphics[width=\textwidth]{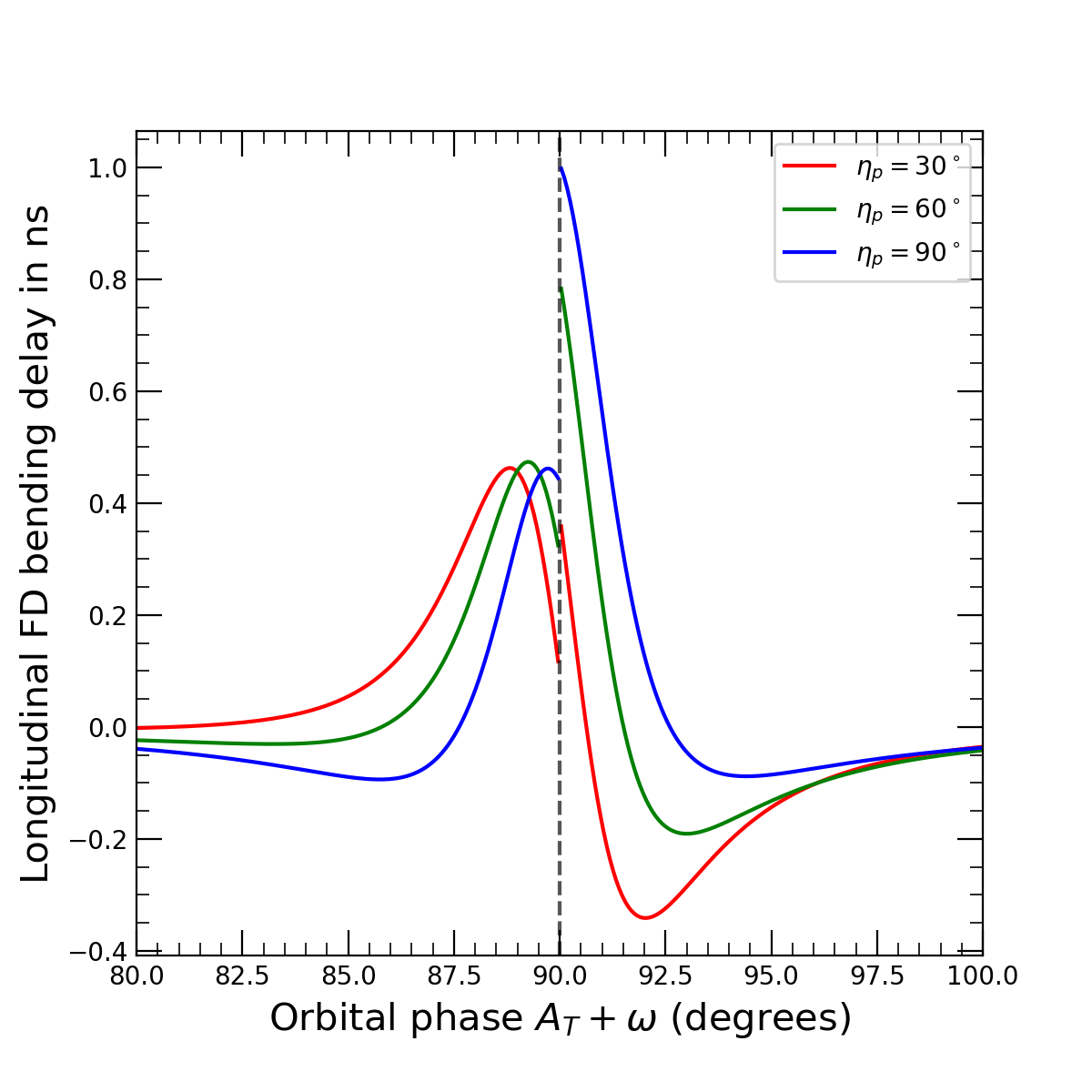}
    \subcaption{Longitudinal FD bending delays for different positive values of $\eta_p$.}
    \label{fig:diff_eta_pulsar_long}
  \end{minipage}
  \hfill
  \begin{minipage}{0.33\textwidth}
    \centering
    \includegraphics[width=\textwidth]{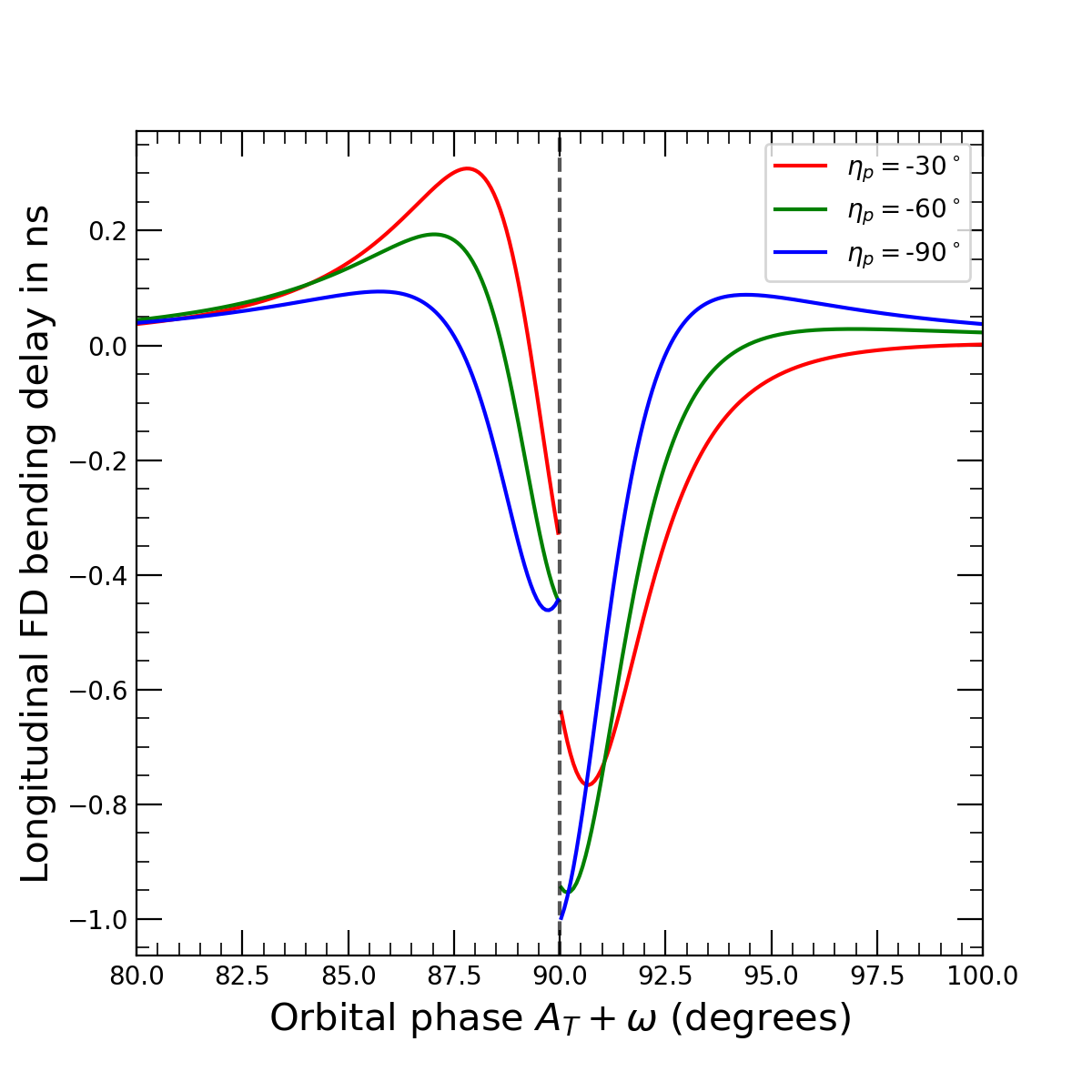}
    \subcaption{Longitudinal FD bending delays for different negative values of $\eta_p$.}
    \label{fig:diff_eta_pulsar_neg_long}
  \end{minipage}
    \caption{The latitudinal FD bending delay curves for hypothetical pulsar-black hole binaries, changing one parameter in each panel and keeping all the other parameters the same as those mentioned in Table \ref{tab:PSRBH}. We have chosen $\tilde{a}=0.5$ except for panel (f) where $\tilde{a}$ is varied. Similarly, we set $\lambda_{\rm bh}=i=87.5^{\circ}$ except for panel (c) where $i$ is varied ($\lambda_{\rm bh}$ is still kept fixed at $87.5^{\circ}$). The vertical dashed lines represent the locations of the discontinuities.}
  \label{fig:diff_parameterOrbit_long_delay}
\end{figure*}

First, in Figs. \ref{fig:diff_mass_lati} and \ref{fig:diff_mass_long}, we plot the latitudinal and the longitudinal FD bending delays against the orbital phase, respectively, for companion masses of 10, 15, 20, 25, and 30 solar masses. We see that as the mass of the companion increases, the values of the FD bending delays also increase. This happens because the spin angular momentum the black hole depends both on $\tilde{a}$ and mass of the black hole ($M_c$) as: $\mathcal{S}_{\rm bh} = \tilde{a} \, GM_c^2 / c$.

Next, in Figs. \ref{fig:diff_Pb_lati} and \ref{fig:diff_Pb_long}, we plot the latitudinal and the longitudinal FD bending delays against the orbital phase, respectively, for values of the orbital period as 1 day, 2 days, 3 days, and 5 days. As expected, we see that the values of the FD bending delays increases as the value of the orbital period decreases. This happens because for fixed values of masses of the pulsar and the companion, a smaller value of the orbital period means a smaller size of the orbit, hence, the light rays travel closer to the companions in the region where the spacetime curvature is larger.

Figs. \ref{fig:diff_e_lati} and \ref{fig:diff_e_long} show the latitudinal and the longitudinal FD bending delays for eccentricities 0.1, 0.3, 0.5, and 0.7. The maximum values of the FD bending delays are larger for higher eccentricities. This occurs because, for a higher eccentricity, the same size of the orbit (i.e., the same values of $P_b$, $M_p$, and $M_c$) results in a shorter distance between the pulsar and the companion near the periastron and nearby orbital phases, which increases the effect of the companion's spin on the geodesic path of the light rays. 

Next, in Fig. \ref{fig:diff_i_lati} and \ref{fig:diff_i_long}, we vary the inclination of the orbit in the range of $86.0^\circ$-$89.5^\circ$. As expected, the values of the FD bending delays increase as the inclination of the orbit increases. As the irregular region near the orbital phase of $90^{\circ}$ appear for $i > 89.7^{\circ}$, in this figure we use only the values of $i$ lower than that limit.

In all of the panels of Figs. \ref{fig:diff_parameterOrbit_lati_delay} and \ref{fig:diff_parameterOrbit_long_delay}, the vertical dashed lines at the orbital phase of $90^{\circ}$ represent the locations of the discontinuities in the delay curves that has been discussed already.

One might intuitively think that the FD bending delays would not depend on the orientation of the spin axis of the pulsar. However, as the longitudinal and the latitudinal directions of the pulsar are determined by the direction of its spin axis, if the orientation of the spin axis changes, the latitude and the longitude of all the light rays coming from the pulsar change, leading to variations in the FD bending delays. This effect can be seen in Figs. \ref{fig:diff_eta_pulsar_lati}, \ref{fig:diff_eta_pulsar_neg_lati}, \ref{fig:diff_eta_pulsar_long} and \ref{fig:diff_eta_pulsar_neg_long}. Fig. \ref{fig:diff_eta_pulsar_lati} shows the variation of the latitudinal FD delay with the variation of $\eta_p$, keeping it positive while Fig. \ref{fig:diff_eta_pulsar_neg_lati} shows the variation of the latitudinal FD delay with the variation of $\eta_p$, keeping it negative. Similarly, Fig. \ref{fig:diff_eta_pulsar_long} shows the variation of the longitudinal FD delay with the variation of $\eta_p$, keeping it positive, while Fig. \ref{fig:diff_eta_pulsar_neg_long} shows the variation of the longitudinal FD delay with the variation of $\eta_p$, keeping it negative. We see a horizontal flip in the shapes of the latitudinal FD delay curves and a vertical flip in the shapes of the longitudinal FD delay curves for a sign reversal of $\eta_p$. We did not attempt to investigate the variation of the FD bending delays with the variation of $\lambda_p$, as only a small range $134.51^{\circ} \geq \lambda_p \geq 125.49^\circ$ satisfy the visibility condition $| 180^{\circ} + \mathcal{W}_{1.4, 10}^{out} - \alpha | \leq | \lambda_p | \leq | 180^{\circ} - \mathcal{W}_{1.4, 10}^{out} - \alpha |$ for our choice of $\alpha = 50^{\circ}$, and our choice of $\lambda_p = 130^{\circ}$ lies almost middle of that range. 

\begin{figure*}
  \centering
  \begin{subfigure}[b]{0.49\textwidth} 
    \centering
    \includegraphics[width=\textwidth]{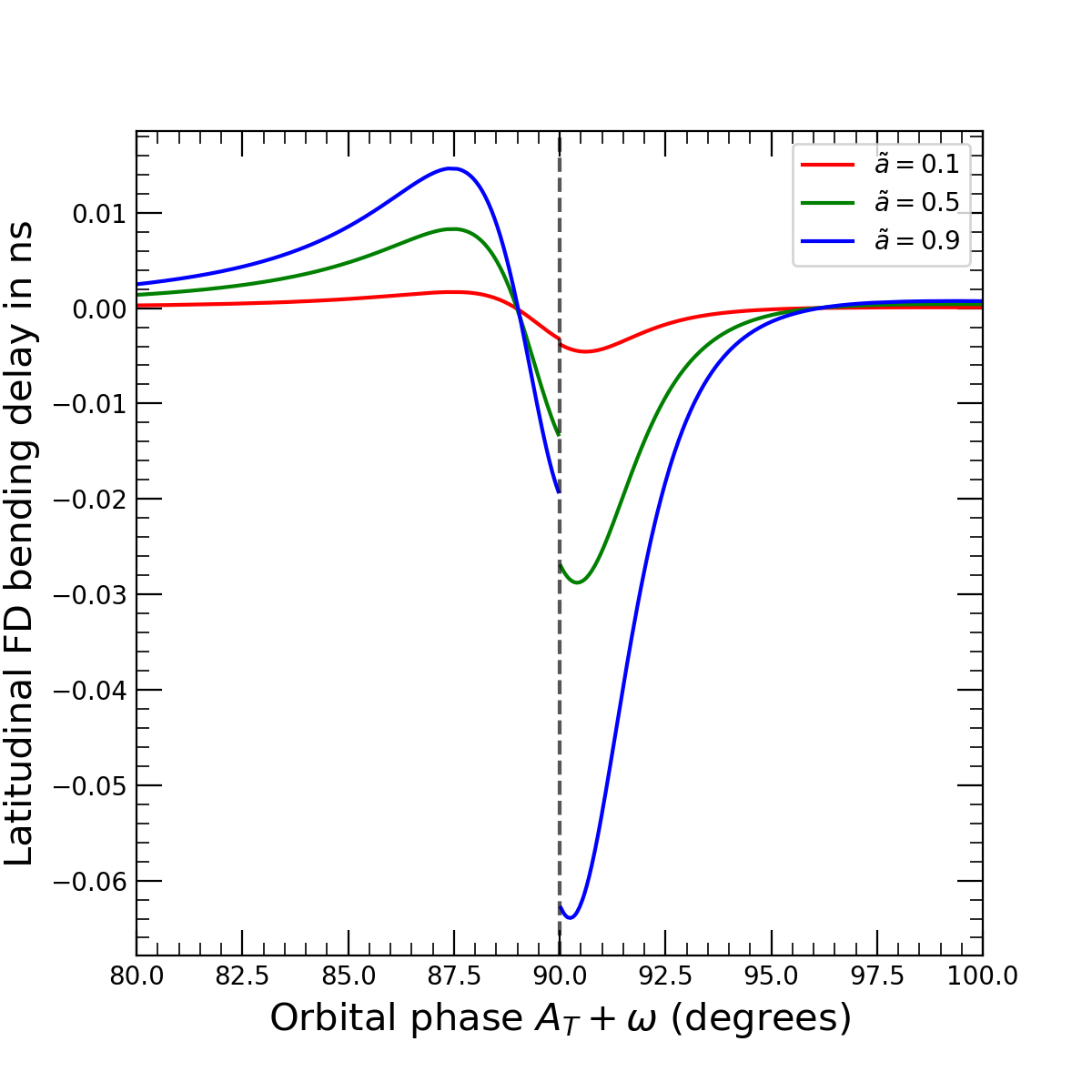}
        \subcaption{Latitudinal FD bending delays for different values of $\tilde{a}$.}
  \label{fig:Lat_bending_delay_spinvariation} 
  \end{subfigure}
  \hfill
  \begin{subfigure}[b]{0.49\textwidth}  
    \centering
    \includegraphics[width=\textwidth]{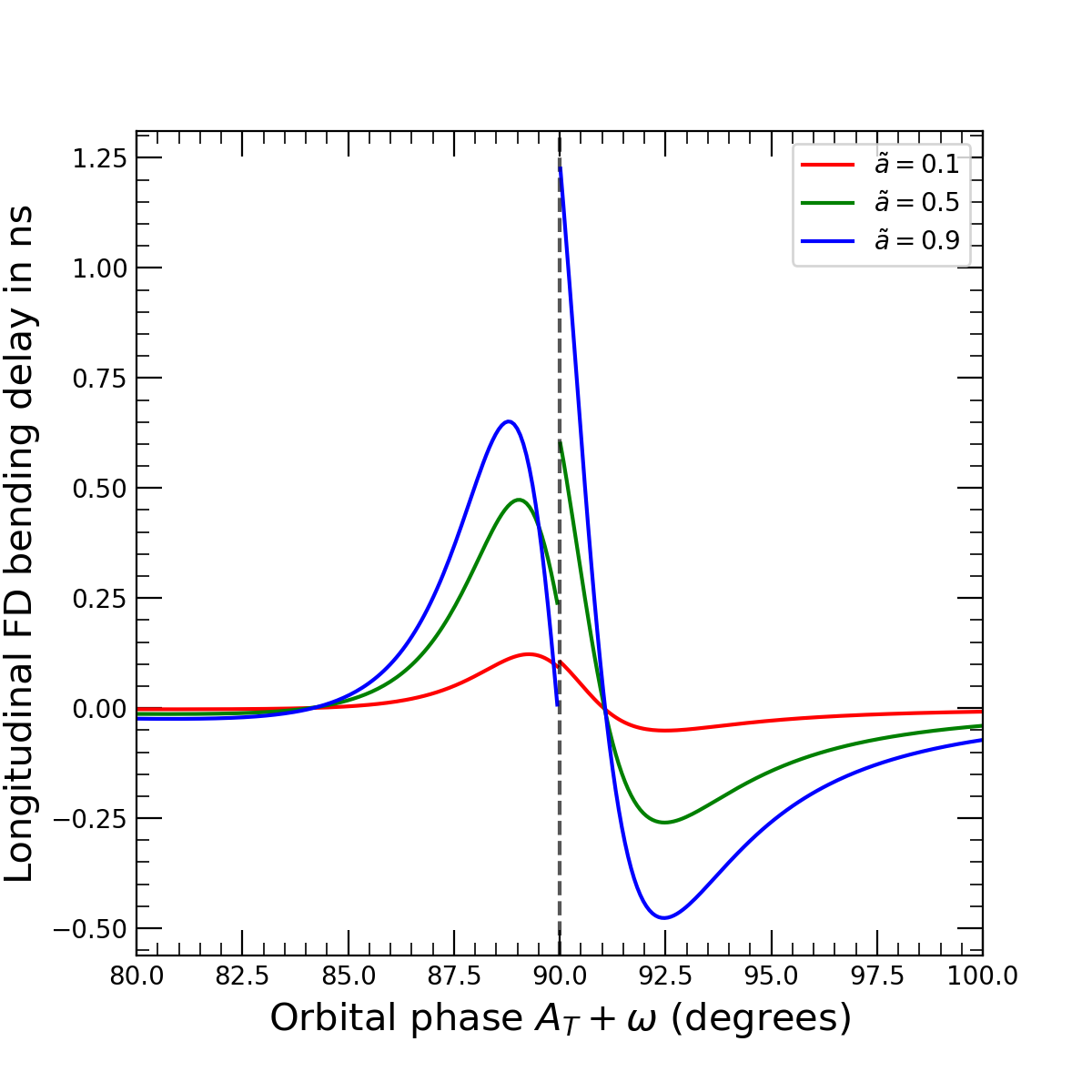}
     \subcaption{Longitudinal FD bending delays for different values of $\tilde{a}$.}
    \label{fig:Long_bending_delay_spinvariation}
  \end{subfigure}
  \caption{The latitudinal (panel a) and the longitudinal (panel b) FD bending delay curves for hypothetical pulsar-black hole binaries for different values of $\tilde{a}$. We use $\lambda_{\rm bh}=i=87.5^{\circ}$ and the values of all the other relevant parameters are taken the same as those mentioned in Table \ref{tab:PSRBH}. The vertical grey dashed lines represent the locations of the discontinuities. }
  \label{fig:bending_delay_spinvariation}
\end{figure*}

As the concept of $\tilde{a}$, $\eta_{\rm bh}$, and $\lambda_{\rm bh}$ does not arise when the pulsar is in a binary system with a Schwarzschild black hole as studied in \citetalias{dbb23}, the effect of these parameters on the latitudinal and the longitudinal bending delays is worth exploring.  We have already seen in Fig. \ref{fig:bending_delay_a_9} that the change in the latitudinal and longitudinal bending delays with $\tilde{a}$ is imperceptible. Hence in Fig. \ref{fig:bending_delay_spinvariation}, we demonstrate the effect of variation of $\tilde{a}$ on FD bending delays. We see that the values of both the latitudinal and the longitudinal FD delays decrease with the decrease of $\tilde{a}$ from 0.9 to 0.5 and to 0.1. 

\begin{figure*}
  \centering
  \begin{subfigure}[b]{0.49\textwidth}
    \centering
    \includegraphics[width=\textwidth]{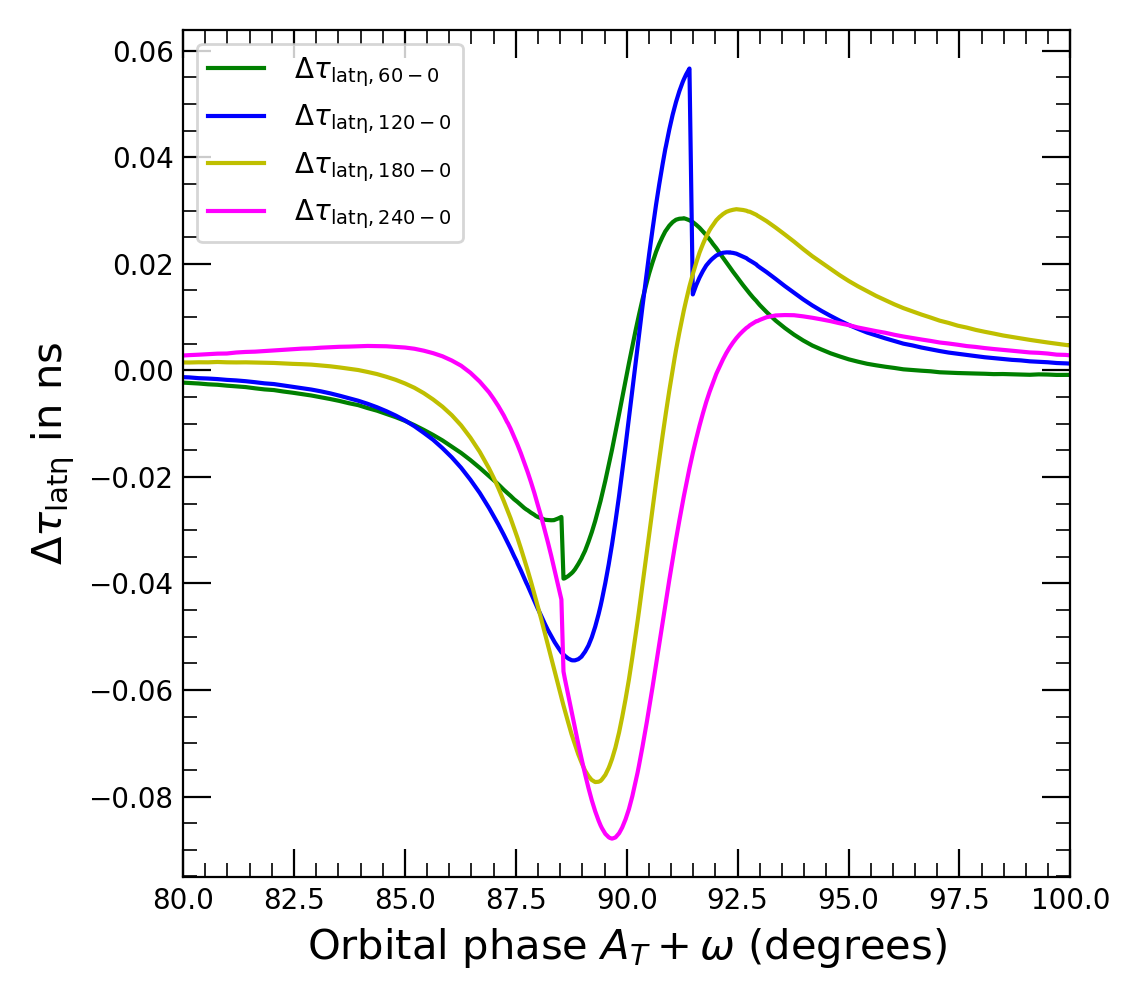}
    \caption{Difference in the latitudinal bending delays for various values of $\eta_{\rm bh}$.}
    \label{fig:lati_diffetavar}
  \end{subfigure}
  \hfill
  \begin{subfigure}[b]{0.49\textwidth} 
    \centering
    \includegraphics[width=\textwidth]{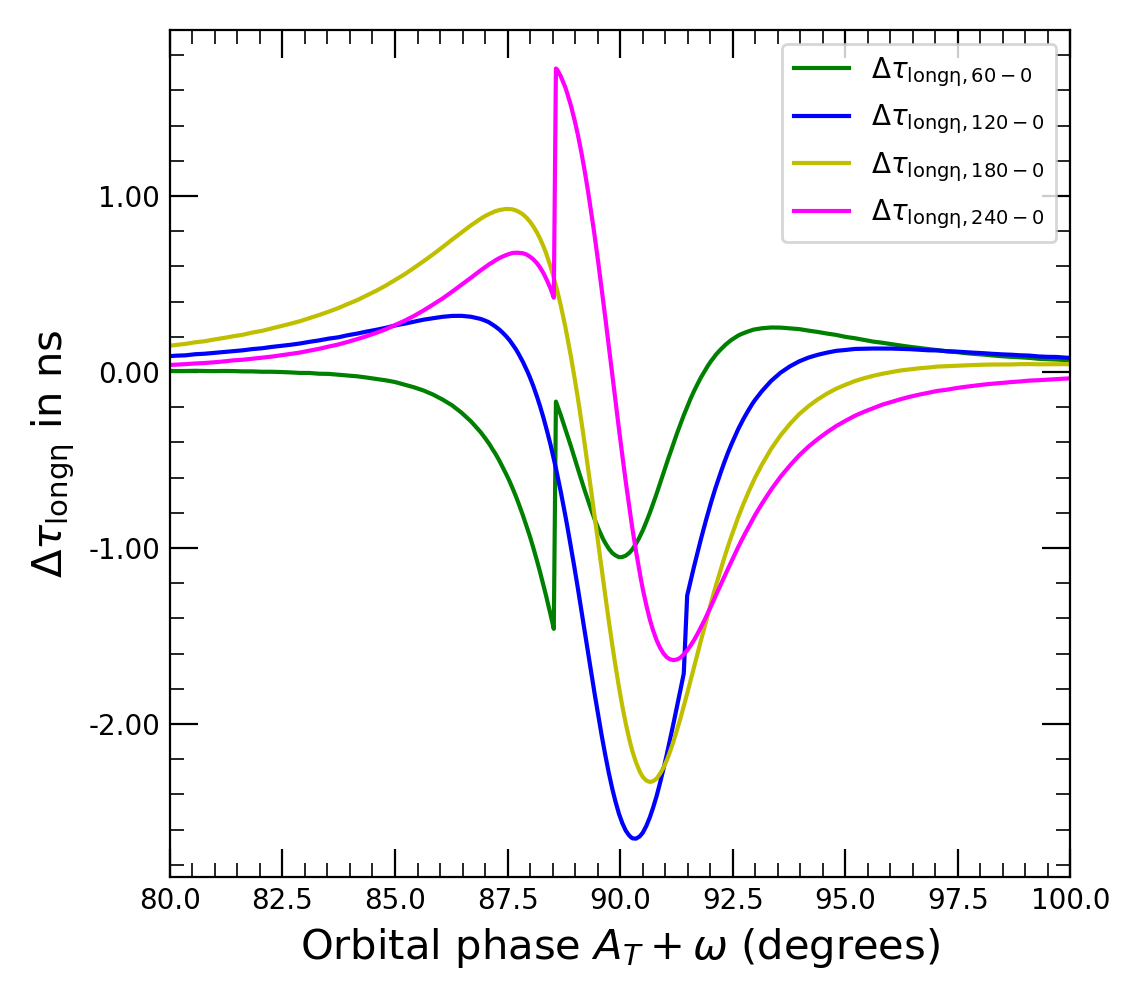}
    \caption{Difference in the longitudinal bending delays for various values of $\eta_{\rm bh}$.}
    \label{fig:long_diffetavar}
  \end{subfigure}
  \hfill
  \begin{subfigure}[b]{0.49\textwidth} 
    \centering
    \includegraphics[width=\textwidth]{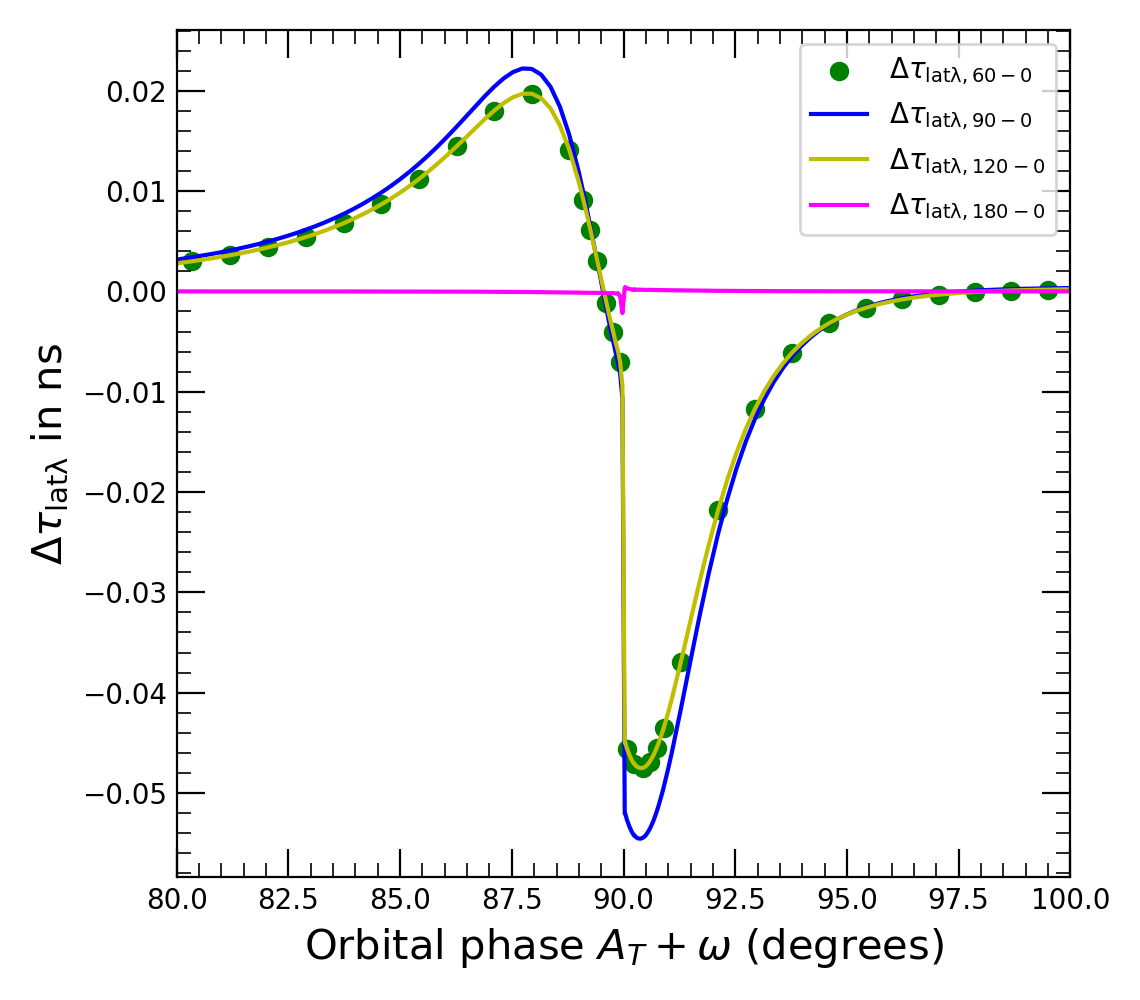}
    \caption{Difference in the latitudinal bending delays for various values of $\lambda_{\rm bh}$.}
    \label{fig:lati_difflambdavar}
  \end{subfigure}
  \hfill
  \begin{subfigure}[b]{0.49\textwidth} 
    \centering
    \includegraphics[width=\textwidth]{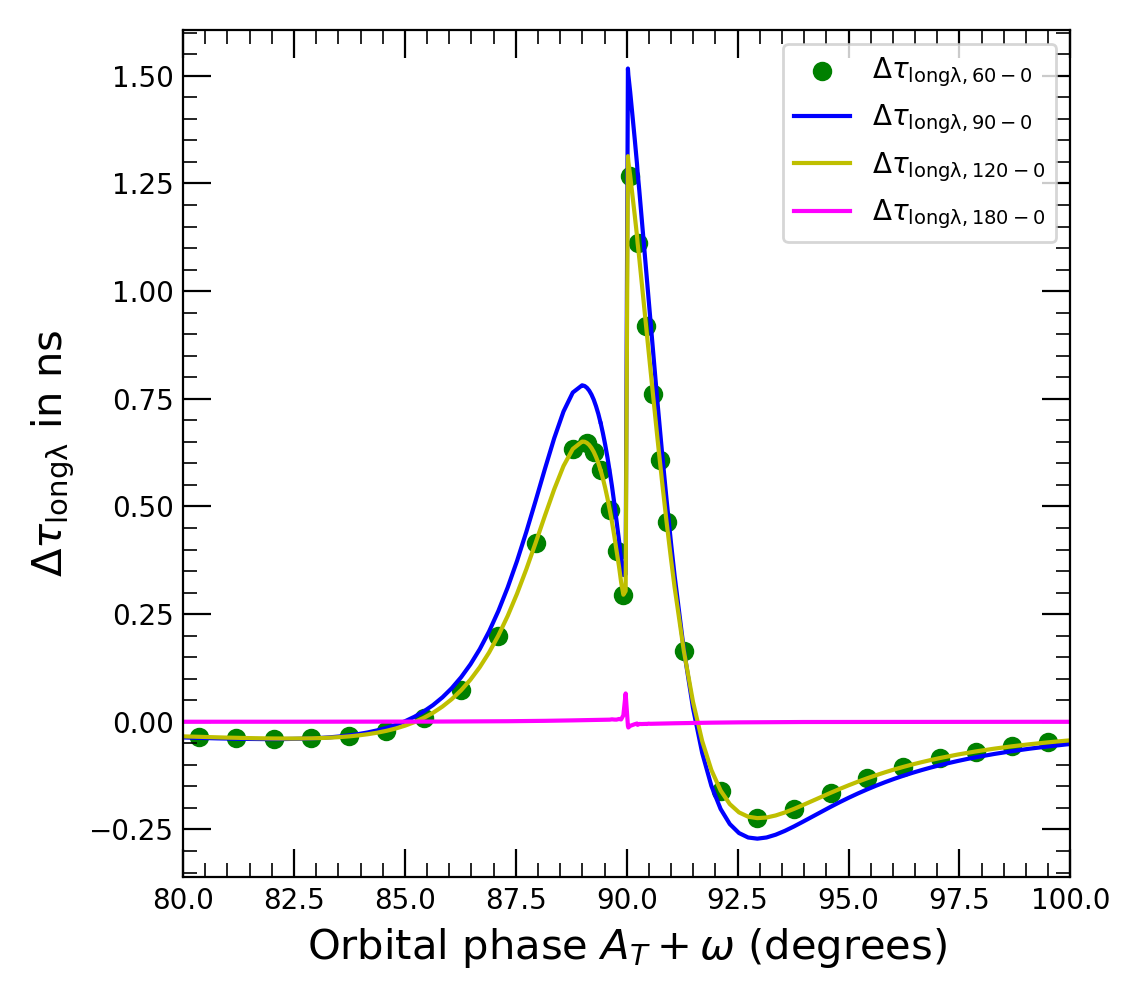}
    \caption{Difference in the longitudinal bending delays for various values of $\lambda_{\rm bh}$.}
    \label{fig:long_difflambdavar}
  \end{subfigure}
  \caption{The impact of change in the values of $\eta_{\rm bh}$ and $\lambda_{\rm bh}$ on the latitudinal and the longitudinal bending delays for $\tilde{a}=0.9$ for hypothetical pulsar-black hole binaries with $i=90^{\circ}$. In each panel, the orbital phase is plotted in degrees along the abscissa only in the range where the differences are comparatively large. The top-right panel (panel-a) has $\Delta \tau_{\text{lat} \eta, \eta_1-0}$ along the ordinate, the top-left panel (panel-b) has $\Delta \tau_{\text{long} \eta, \eta_1-0}$ along the ordinate, the bottom-right panel (panel-c) has $\Delta \tau_{\text{lat} \lambda, \lambda_1-0}$ along the ordinate, and the bottom-left panel (panel-d) has $\Delta \tau_{\text{long} \lambda, \lambda_1-0}$ along the ordinate. These parameters are defined in the text. For each panel, values of all the relevant parameters are taken as the ones mentioned in Table \ref{tab:PSRBH} except the ones being varied. In panels (a) and (b) where $\eta_{\rm bh}$ is varied, we choose $\lambda_{\rm bh}=i$. }
  \label{fig:impact_etabh_lambdabh_variation}
\end{figure*}

We find that a change in the value of $\eta_{\rm bh}$ changes the values of the latitudinal and the longitudinal bending delays in the nanosecond order and a change in the value of $\lambda_{\rm bh}$ changes the values of the latitudinal and the longitudinal bending delays in the sub-nanosecond order. The changes are maximum near the orbital phase of $90^{\circ}$. As these changes would not be visible if we plot the bending delays that are in the microsecond order, we rather plot the differences. For this purpose, we first define a set of parameters as: 
\begin{equation}
\label{longdelay_etadiff}
\Delta \tau_{\text{long} \eta, \eta_1-\eta_2} =   \tau_{\text{long} \eta, \eta_1} -  \tau_{\text{long} \eta, \eta_2} ,
\end{equation}  
\begin{equation}
\label{longdelay_lambdadiff}
\Delta \tau_{\text{long} \lambda, \lambda_1-\lambda_2} =   \tau_{\text{long} \lambda, \lambda_1} -  \tau_{\text{long} \lambda, \lambda_2} ,
\end{equation}
\begin{equation}
\label{latdelay_etadiff}
\Delta \tau_{\text{lat} \eta, \eta_1-\eta_2} =  \tau_{\text{lat} \eta, \eta_1} -  \tau_{\text{lat} \eta, \eta_2},
\end{equation} and 
\begin{equation}
\label{latdelay_lambdadiff}
\Delta \tau_{\text{lat} \lambda, \lambda_1-\lambda_2} =   \tau_{\text{lat} \lambda, \lambda_1} -  \tau_{\text{lat} \lambda, \lambda_2},
\end{equation} where $\tau_{\text{long} \eta, \eta_i}$ and $\tau_{\text{lat} \eta, \eta_i}$ mean the values of the longitudinal and the latitudinal bending delays, respectively, for $\eta_{\rm bh} = \eta_i$ degrees with all other parameters having constant pre-assigned values (here as in Table \ref{tab:PSRBH}). Similarly, $\tau_{\text{long} \lambda, \lambda_i}$ and $\tau_{\text{lat} \lambda, \lambda_i}$ mean the values of the longitudinal and the latitudinal bending delays, respectively, for $\lambda_{\rm bh} = \lambda_i$ degrees with all other parameters having constant pre-assigned values (here as in Table \ref{tab:PSRBH}). In Fig. \ref{fig:impact_etabh_lambdabh_variation}, we plot the orbital phase in the range of $80^{\circ}$-$100^{\circ}$ along the abscissa and $\Delta \tau_{\text{lat} \eta, \eta_i - 0}$, $\Delta \tau_{\text{long} \eta, \eta_i - 0} $, $\Delta \tau_{\text{lat} \lambda, \lambda_i - 0} $, and $\Delta \tau_{\text{long} \lambda, \lambda_1 - 0}$ along the ordinate in different panels by varying $\eta_{\rm bh} = \eta_i$ and $\lambda_{\rm bh} = \lambda_i$ over their full allowed ranges of $0^{\circ}$-$360^{\circ}$ and $0^{\circ}$-$180^{\circ}$, respectively. From Fig. \ref{fig:impact_etabh_lambdabh_variation}, it is clear that $\Delta \tau_{\text{lat} \eta, \eta_i-0}$, $\Delta \tau_{\text{long} \eta, \eta_i-0}$, $\Delta \tau_{\text{lat} \lambda, \lambda_i-0}$, and $\Delta \tau_{\text{long} \lambda, \lambda_i-0}$ remain in the nanosecond order. However, all these parameters depend strongly on the orbital phase in the range of $85^{\circ}$-$95^{\circ}$. In this range, they can even change the sign one or more times. The variation of $\eta_{\rm bh}$ has a stronger effect than the variation of $\lambda_{\rm bh}$.

\begin{figure*}
  \centering
  \begin{subfigure}[b]{0.49\textwidth}
    \centering
    \includegraphics[width=\textwidth]{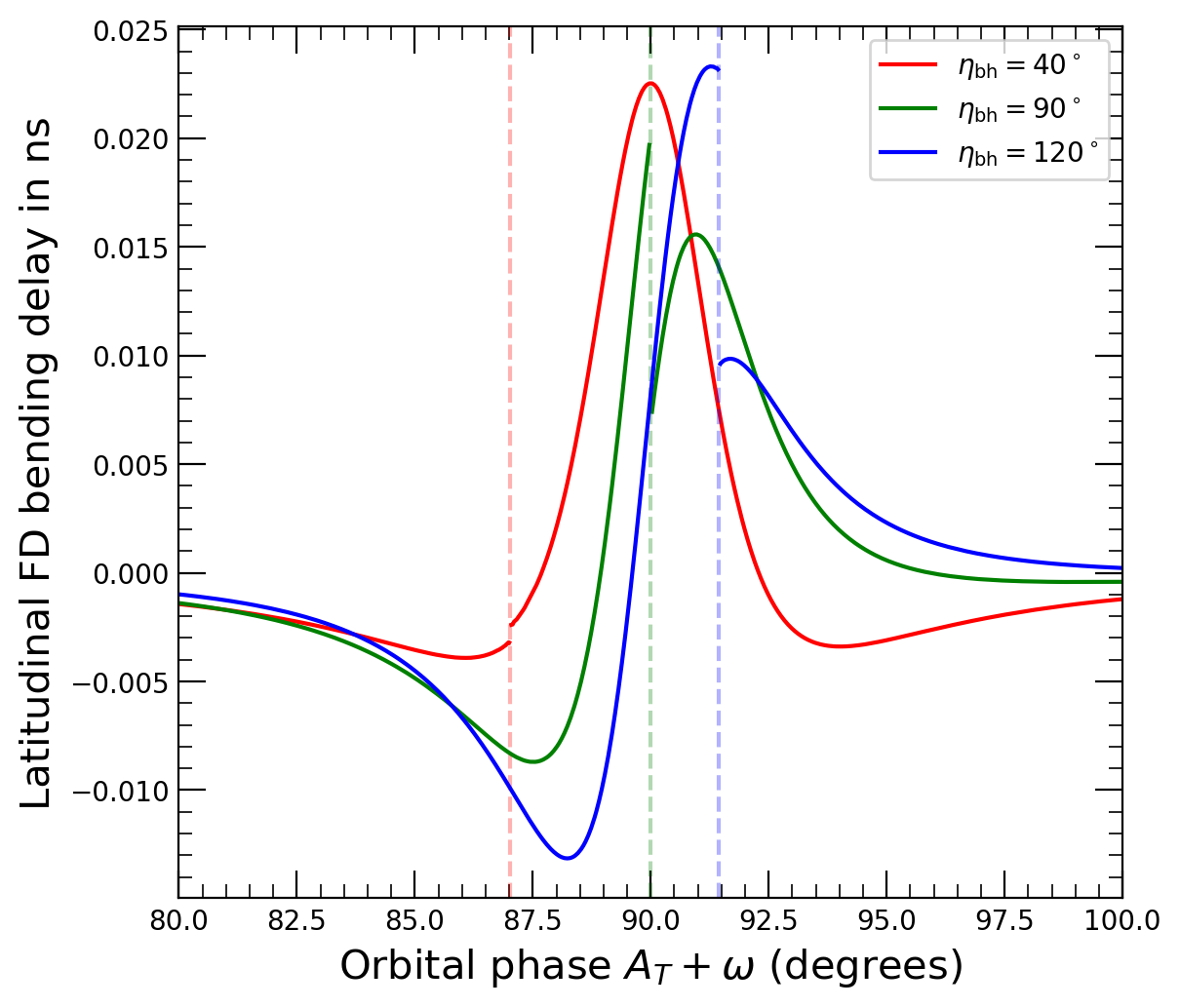}
    \caption{Latitudinal FD bending delays for different positive  values of $\eta_{\rm bh}$.}
    \label{fig:diff_eta_bh_lati}
  \end{subfigure}
  \hfill
  \begin{subfigure}[b]{0.49\textwidth} 
    \centering
    \includegraphics[width=\textwidth]{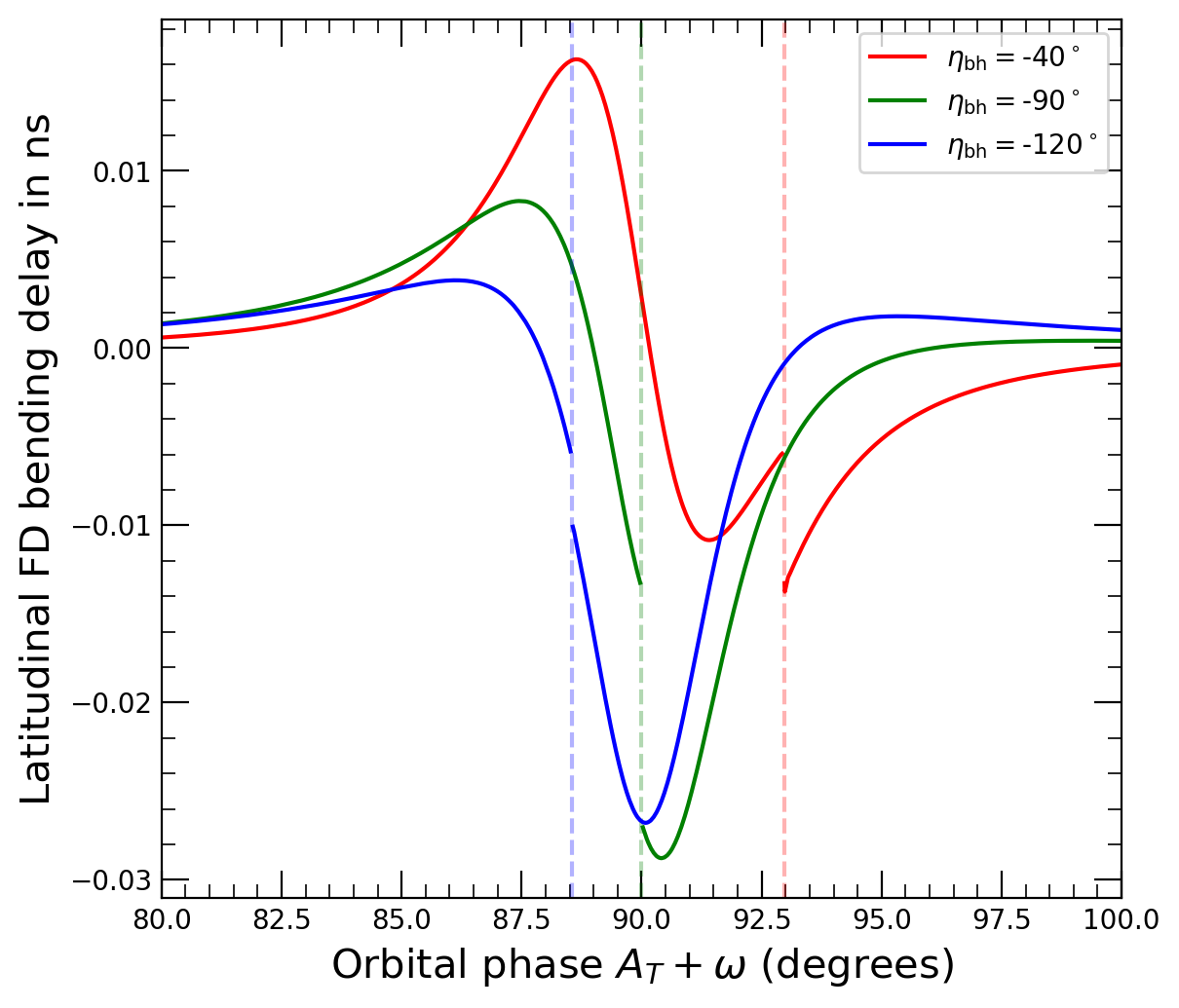}
    \caption{Latitudinal FD bending delays for different negative values of $\eta_{\rm bh}$.}
    \label{fig:diff_eta_bh_neg_lati}
  \end{subfigure}
  \hfill
  \begin{subfigure}[b]{0.49\textwidth} 
    \centering
    \includegraphics[width=\textwidth]{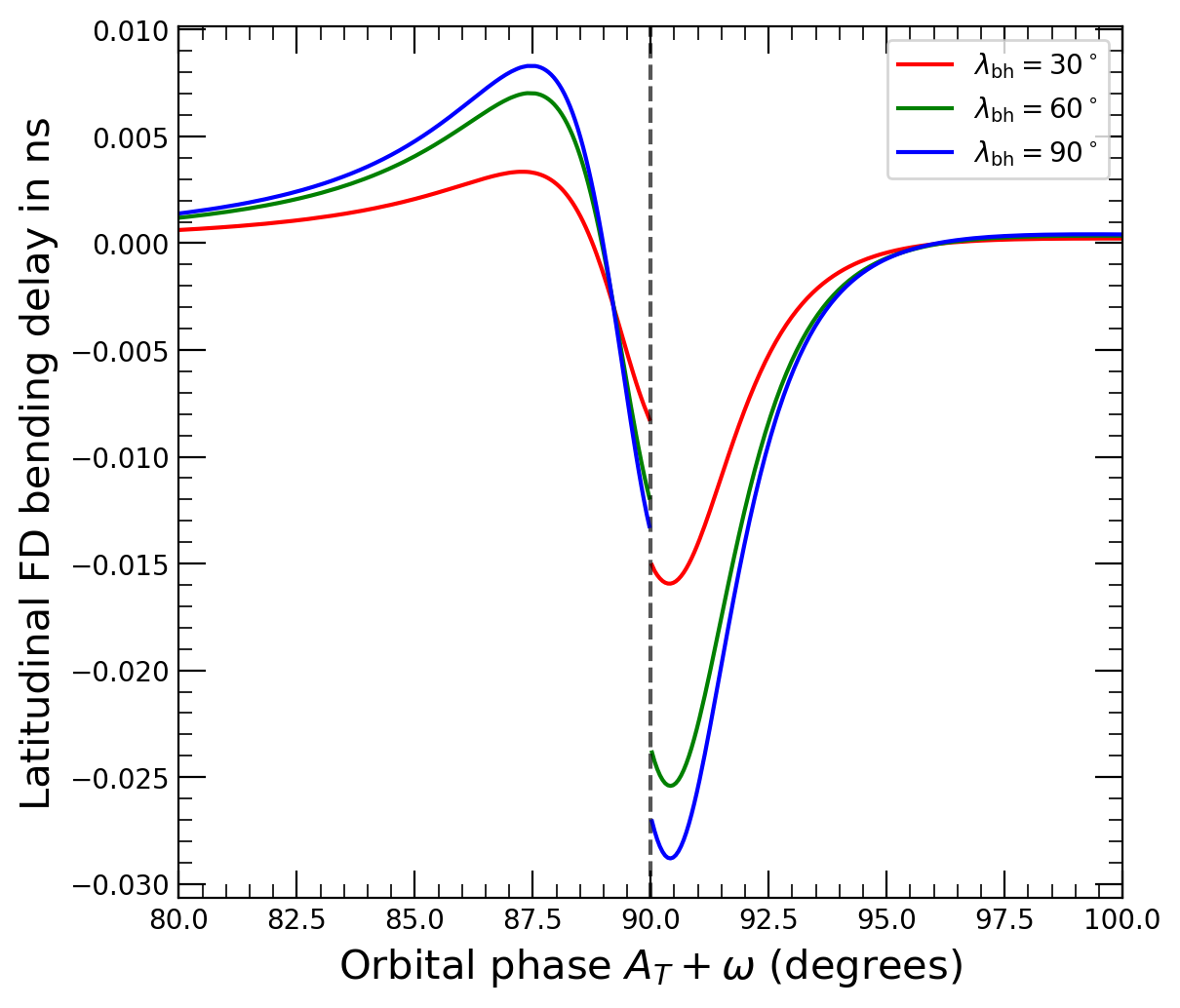}
    \caption{Latitudinal FD bending delays for different values of $\lambda_{\rm bh} \le 90^\circ$.}
    \label{fig:diff_lamda_bh_below90_lati}
  \end{subfigure}
  \hfill
  \begin{subfigure}[b]{0.49\textwidth} 
    \centering
    \includegraphics[width=\textwidth]{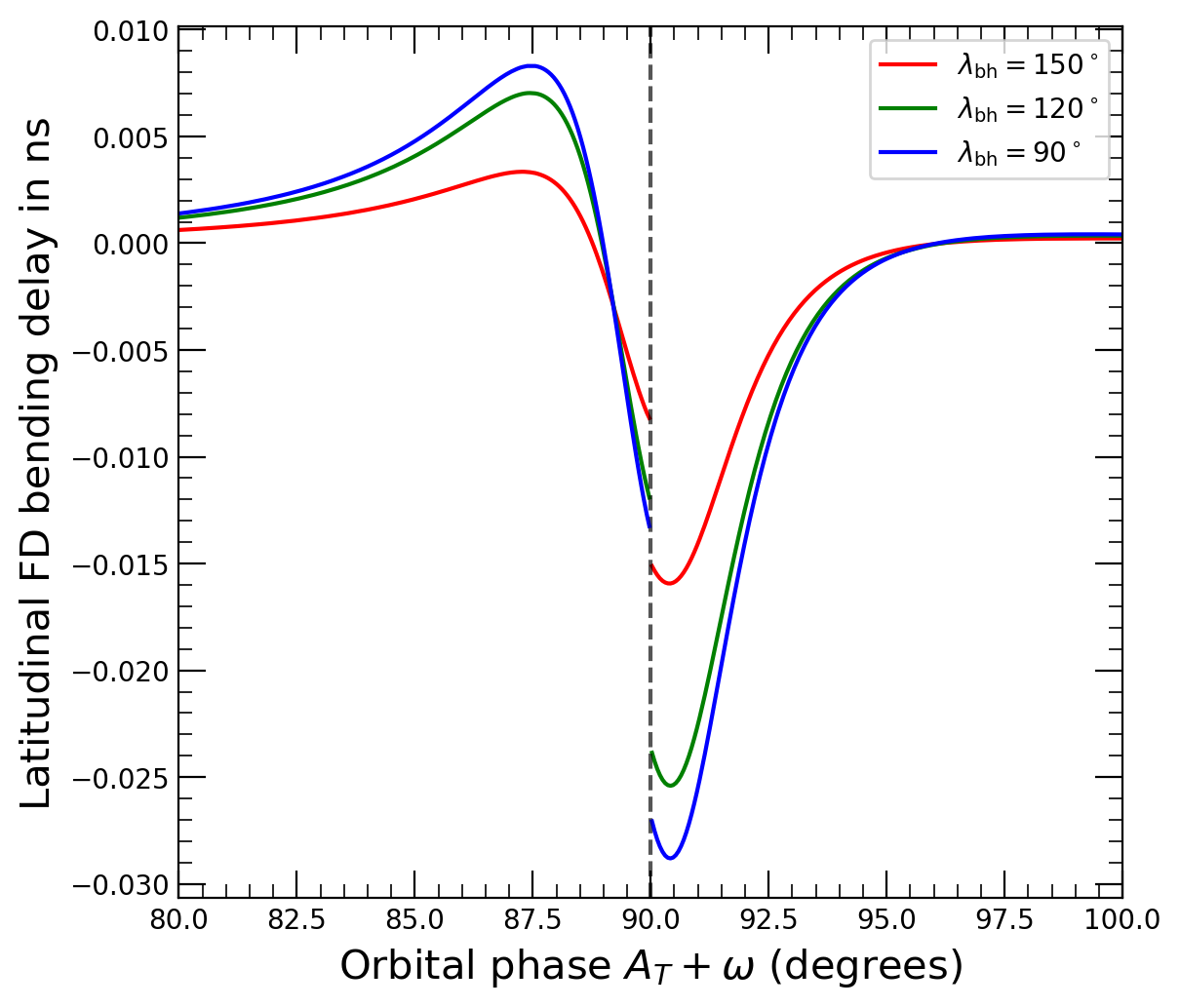}
    \caption{Latitudinal FD bending delays for different values of $\lambda_{\rm bh} \ge 90^\circ$.}
    \label{fig:diff_lamda_bh_above90_lati}
  \end{subfigure}
  \caption{The latitudinal FD bending delays for hypothetical pulsar-black hole binaries with $\tilde{a}=0.5$, for different orientations of the spin axis of the black hole. In each panel. Only one parameter is varied at a time. We take $i=87.5^{\circ}$ and other relevant parameters are kept the same as those mentioned in Table \ref{tab:PSRBH}. In panels (a) and (b) where $\eta_{\rm bh}$ is varied, we choose $\lambda_{\rm bh}=i$. }
  \label{fig:diff_parameter_lati_delay}
\end{figure*}

\begin{figure*}
  \centering
  \begin{subfigure}[b]{0.49\textwidth}
    \centering
    \includegraphics[width=\textwidth]{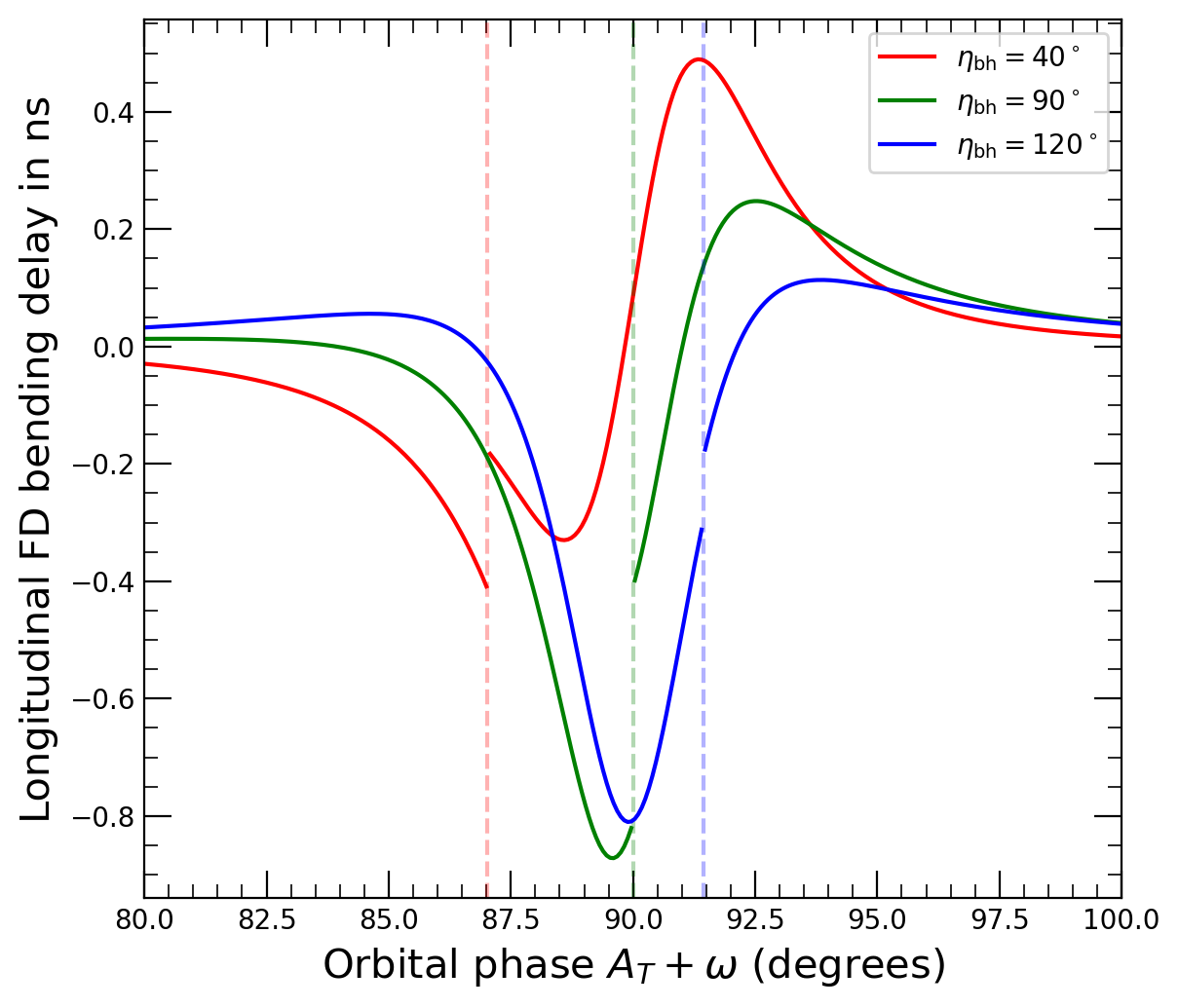}
    \caption{Longitudinal FD bending delays for different positive values of $\eta_{\rm bh}$.}
    \label{fig:diff_eta_bh_long}
  \end{subfigure}
  \hfill
  \begin{subfigure}[b]{0.49\textwidth} 
    \centering
    \includegraphics[width=\textwidth]{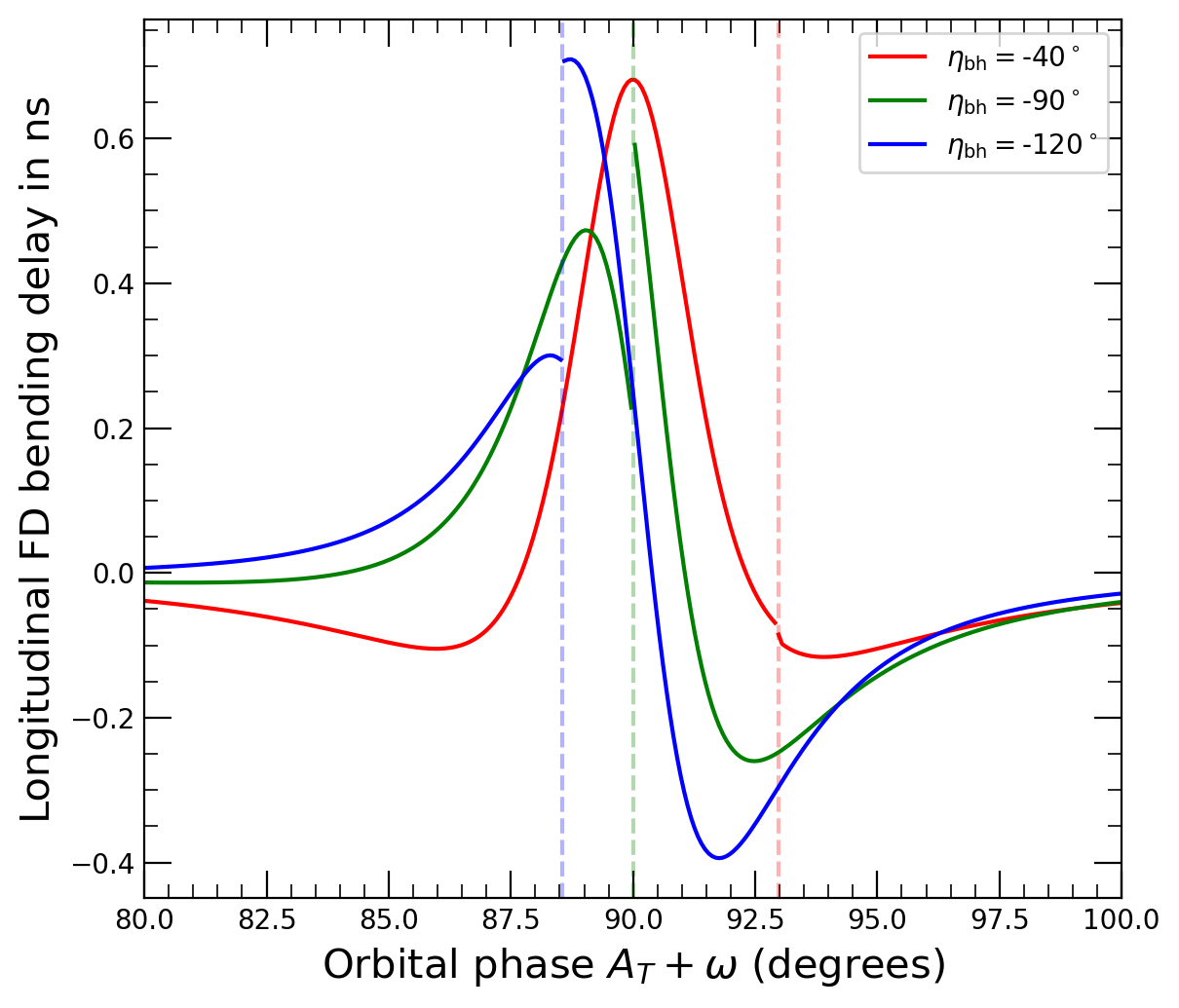}
    \caption{Longitudinal FD bending delays for different negative values of $\eta_{\rm bh}$.}
    \label{fig:diff_eta_bh_neg_long}
  \end{subfigure}
  \hfill
  \begin{subfigure}[b]{0.49\textwidth} 
    \centering
    \includegraphics[width=\textwidth]{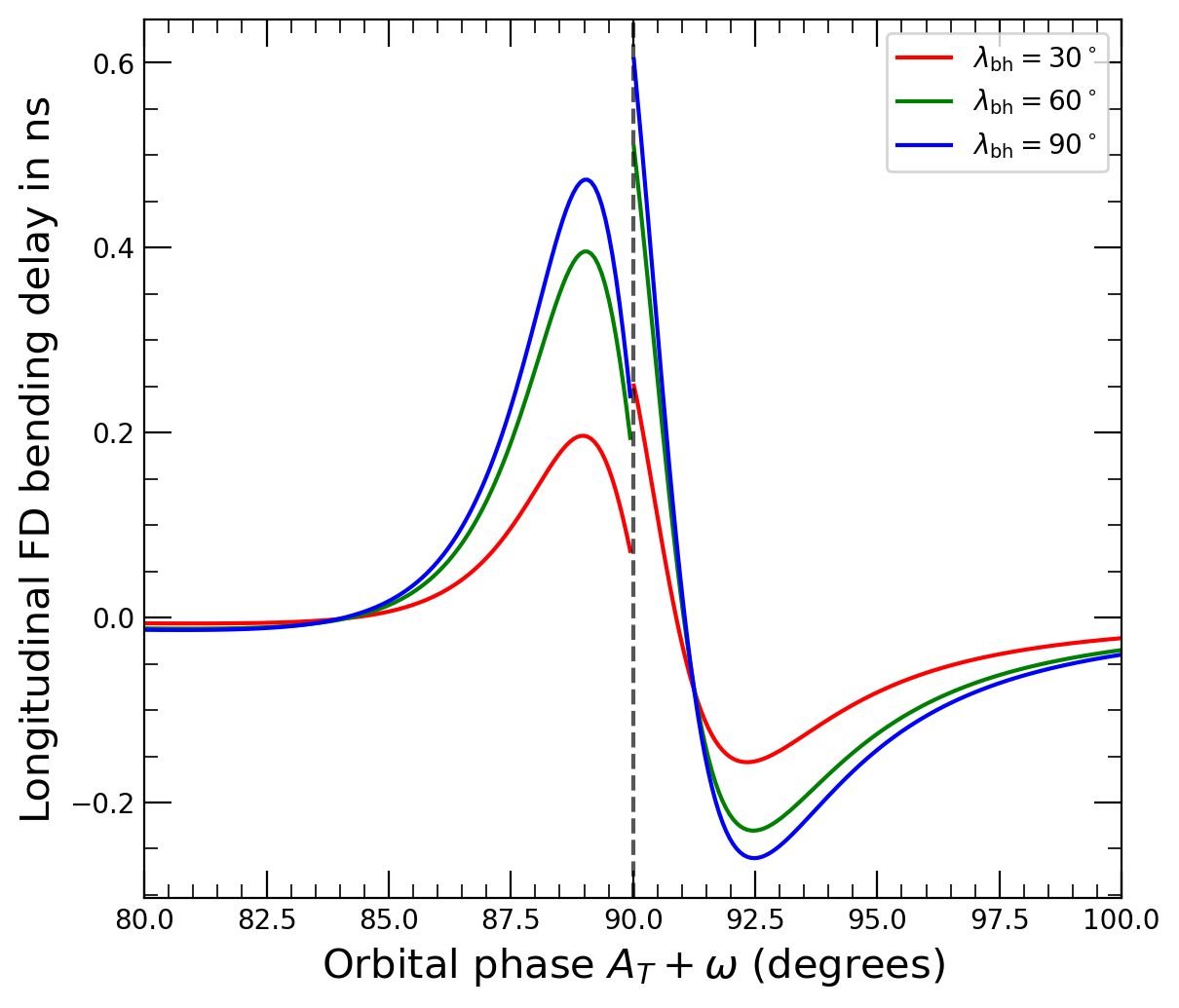}
    \caption{Longitudinal FD bending delays for different values of $\lambda_{\rm bh} \le 90^\circ$.}
    \label{fig:diff_lamda_bh__below90_long}
  \end{subfigure}
  \hfill
  \begin{subfigure}[b]{0.49\textwidth} 
    \centering
    \includegraphics[width=\textwidth]{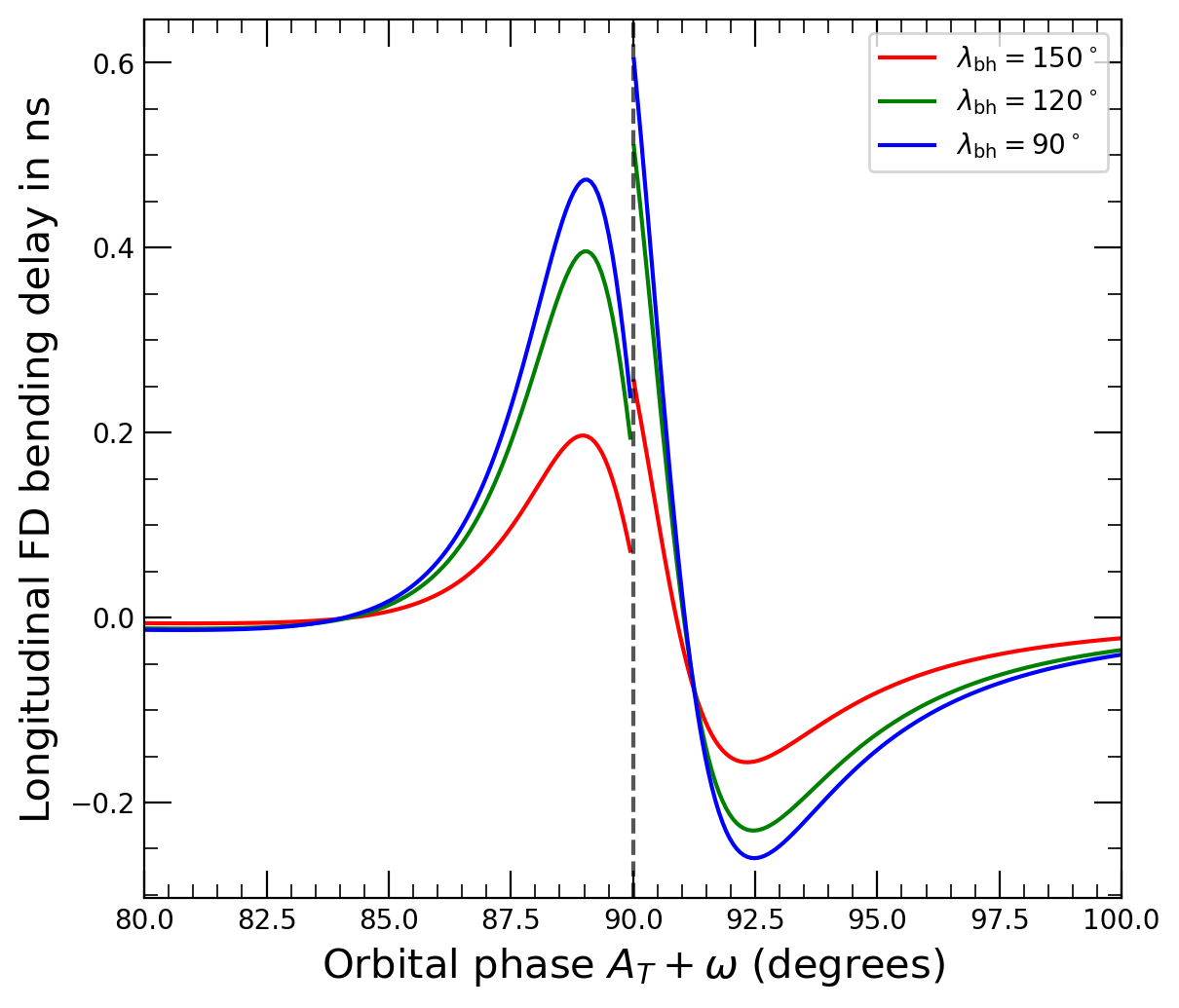}
    \caption{Longitudinal FD bending delays for different values of $\lambda_{\rm bh} \ge 90^\circ$.}
    \label{fig:diff_lamda_bh_above90_long}
  \end{subfigure}
  \caption{The longitudinal FD bending delays for hypothetical pulsar-black hole binaries with $\tilde{a}=0.5$, for different orientations of the spin axis of the black hole. In each panel. Only one parameter is varied at a time. We take $i=87.5^{circ}$ and other relevant parameters are kept the same as those mentioned in Table \ref{tab:PSRBH}. In panels (a) and (b) where $\eta_{\rm bh}$ is varied, we choose $\lambda_{\rm bh}=i$.}
  \label{fig:diff_parameter_long_delay}
\end{figure*}

Next, we explore the effect of the orientation of the spin axis of the black hole, i.e., the values of $\eta_{\rm bh}$ and $\lambda_{\rm bh}$ on the FD bending delays. In Figs. \ref{fig:diff_eta_bh_lati}, \ref{fig:diff_eta_bh_neg_lati}, \ref{fig:diff_eta_bh_long}, and \ref{fig:diff_eta_bh_neg_long}, we study the effect of different choice of the value of $\eta_{\rm bh}$ on the FD bending delays. In these figures, we plot the values of the FD bending delays along the ordinates and the orbital phase in the range of $80^{\circ}$-$100^{\circ}$ along the abscissa. Discontinuities occur at different orbital phases, as given by Eq. (\ref{eq:discontinuity}). For the value of $\eta_{\rm bh}$ being $40^{\circ}, \, 90^{\circ}, ~{\rm and} ~ 120^{\circ}$, the discontinuities occur at orbital phases of $(87.02^{\circ}, 267.02^{\circ})$, $(90.0^{\circ}, 270.0^{\circ})$, and $(91.44^{\circ}, 271.44^{\circ})$, respectively, while for the value of $\eta_{\rm bh}$ being $-40^{\circ}, \, -90^{\circ}, ~{\rm and} ~ -120^{\circ}$, the discontinuities occur at orbital phases of $(92.97^{\circ},272.97^{\circ})$, $(90.0^{\circ}, 270.0^{\circ})$, and $(88.55^{\circ}, 268.55^{\circ})$, respectively. In Figs. \ref{fig:diff_eta_bh_lati}, \ref{fig:diff_eta_bh_neg_lati}, \ref{fig:diff_eta_bh_long}, and \ref{fig:diff_eta_bh_neg_long}, only the first discontinuity from each pair can be seen as the second one is out of the range plotted, and would be imperceptibly small even if those ranges were included as the values of the FD bending delays themselves are small at those orbital phases.

In Figs. \ref{fig:diff_lamda_bh_below90_lati}, \ref{fig:diff_lamda_bh_above90_lati}, \ref{fig:diff_lamda_bh__below90_long}, and \ref{fig:diff_lamda_bh_above90_long}, we study the effect of different choice of the value of $\lambda_{\rm bh}$ on the FD bending delays. We see that the values of the FD bending delays are larger when the value of $\lambda_{\rm bh}$ comes closer to $90^\circ$. In fact, the FD bending delays are maximum when $\lambda_{\rm bh} = i$ ($i=87.5^\circ$ in our case). We also found that the value of the FD bending delays are symmetric around $\lambda_{\rm bh} = i$. As mentioned earlier, the location of the discontinuity does not depend on the value of $\lambda_{\rm bh}$. In all panels of Figs. \ref{fig:diff_parameter_lati_delay} and \ref{fig:diff_parameter_long_delay}, we use the values of the relevant parameters as those given in Table \ref{tab:PSRBH} with $\tilde{a}=0.5$ except the parameter that is varied.

In \citetalias{dbb23}, we showed that the photon distribution on the cross-section of the beam changes due to the bending resulting a change in the pulse profile. Although the overall shape of the profile remains unaltered, the heights of the peaks increase and sometimes small features appear. These effects were seen to be more prominent closer to the superior conjunction, i.e., for $i=90^{\circ}$ and $A_T + \omega = 90^{\circ}$. This happens because in this situation, the LoS contains the gravitating body (the companion black hole) and hence a large number of photons that would be away from the LoS in absence of the bending, get diverted towards the LoS. In the present paper, we investigate whether in this situation, the spin of the black hole play any significant role in the photon distribution across the beam. For this purpose, we choose $\lambda_{\rm bh}=i=90^\circ$, $\eta_{\rm bh}=-90^\circ$.

\begin{figure*}
  \centering
    \includegraphics[width=0.9\textwidth]{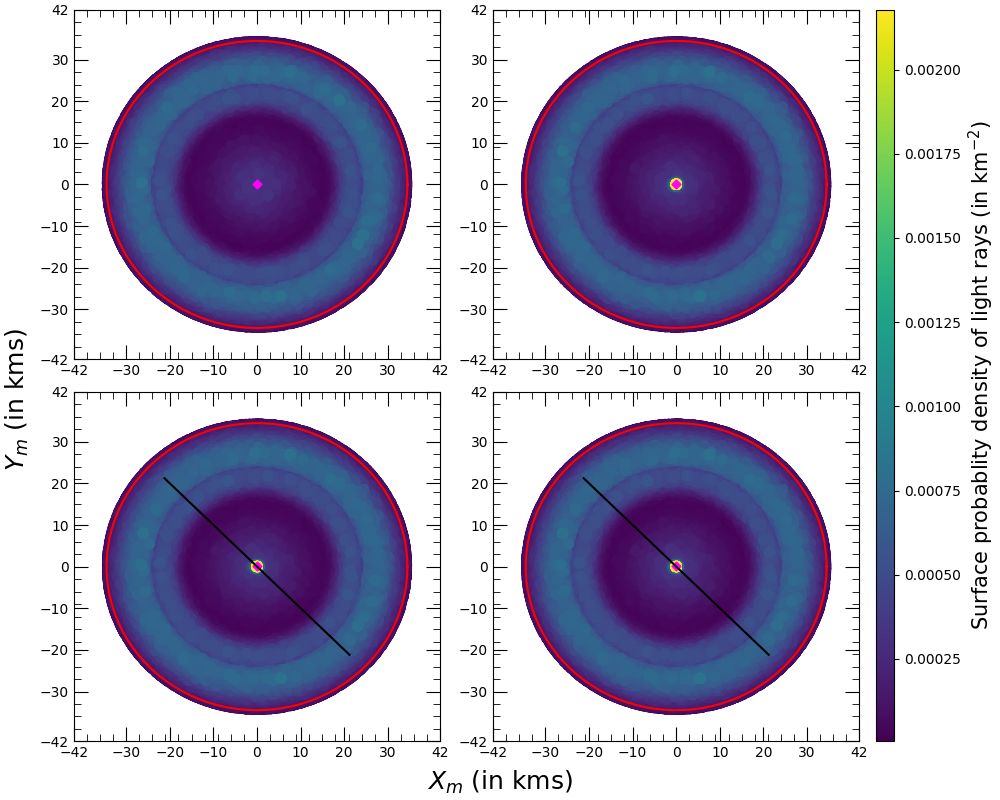}
\caption{The photon distribution on the cross-section of the beam of a pulsar with a black hole companion with $\lambda_{\rm bh}=i=90^{\circ}$. The top-left panel is the photon distribution of the original beam (without any bending), the top-right panel is the photon distribution of the distorted (due to bending) beam when $\tilde{a}=0$, the bottom-left panel shows the photon distribution of the distorted beam when $\tilde{a}=0.5$, and the bottom-right panel shows the photon distribution of the distorted (due to bending) beam when $\tilde{a}=0.9$. The values of all other relevant parameters are chosen the same as those mentioned in Table \ref{tab:PSRBH}. For each case, we show the intensity distribution when the LoS is at the middle of the beam. For the panels where $\tilde{a} \ne 0$, the black solid line represent the the projection of the spin axis on the cross-section of the beam. }
\label{distorted_beam}
\end{figure*}

\begin{figure*}
  \centering
    \includegraphics[width=0.9\textwidth]{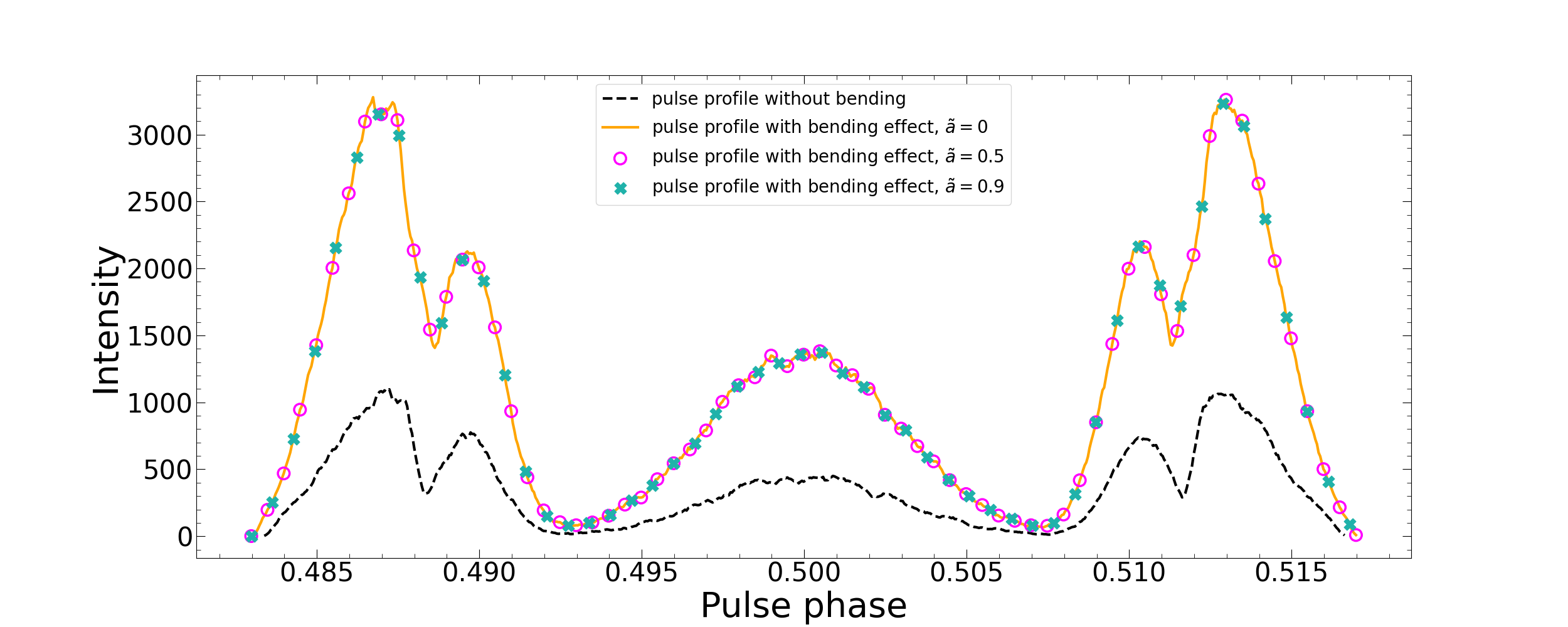}
     \caption{Comparison of the pulse profile of a pulsar with a black hole companion with $\lambda_{\rm bh}=i=90^{\circ}$ and $A_T+\omega = 90^{\circ}$. We show the profiles without bending (black dashed line) and with bending for various values of the spin parameter of the black hole, e.g., $\tilde{a}=0$ (solid orange line), $\tilde{a}=0.5$ (magenta unfilled circles), and  $\tilde{a}=0.9$ (sea-green crosses). The values of all other relevant parameters are taken the same as mentioned in Table \ref{tab:PSRBH}. }
    \label{fig:pulse_profileAll}
\end{figure*}

In Fig. \ref{distorted_beam}, we show the photon distributions across the beam (on the ${\rm X_m Y_m}$ plane). The top-left panel is the photon distribution of the original beam (without any bending), the top-right panel is the photon distribution of the distorted (due to bending) beam when $\tilde{a}=0$, the bottom-left panel shows the photon distribution of the distorted beam (due to bending) when $\tilde{a}=0.5$, and the bottom-right panel shows the photon distribution of the distorted (due to bending) beam when $\tilde{a}=0.9$. In these panels, in addition to the colour coded photon distributions, we see a solid black line when $\tilde{a} \ne 0$. This line represents the projection of the spin axis on the cross-section of the beam. For each case, we show the intensity distribution when the LoS is at the middle of the beam \footnote{In \citetalias{dbb23}, we showed that when the LoS sweep across the beam, at each position of the LoS, the photon distribution on the ${\rm X_m Y_m}$ plane is slightly different (see Figs. 13 and 14 of that paper).}. We see that the bending changes the intensity distribution in comparison to the original beam, however the spin of the black hole does not show any visible effect. 

To establish the above observation even strongly, in Fig. \ref{fig:pulse_profileAll}, we plot pulse profiles obtained from the same pulsar-black hole binary without and with the effect of the bending, and for various values of $\tilde{a}$ when the bending is considered. We use the same formalism as explained in \citetalias{dbb23} to get a dimensionless proxy for the intensity. This intensity is plotted along the ordinate and the pulse phase is plotted along the abscissa. The pulse phase is defined to vary from 0 to 1, covering a full rotation of the magnetic axis. For visualisation purposes, we define the phase in such a way that the peak of the core component of the profile without the bending is at the phase of 0.5. In other words, at a pulse phase of 0.5, the LoS is at the centre of the unbent beam. When plotting the profiles, we show only the range over which the LoS is within the beam, instead of the full pulse phase range. From Fig. \ref{fig:pulse_profileAll}, it is evident that the pulse becomes stronger, i.e., the intensity increases (due to the enhanced number of light rays along the LoS) due to bending, but the value of $\tilde{a}$ does not play any role in this enhancement.

While obtaining Figs. \ref{distorted_beam} and \ref{fig:pulse_profileAll}, we used the procedure as described in Sec. 3.3 of \citetalias{dbb23}

\section{A test case of super-massive BH}
\label{sec:testcaseSupermassive}

At this point, a logical curiosity might arise, what happens when a pulsar orbits around a super massive black hole? How does the light bending phenomenon effect the signal of the pulsar? As we do not have any such system discovered yet, we again choose a hypothetical system with parameters the same as the Table \ref{tab:PSRBH}, except the mass of the black hole $M_c$ is replaced by $10^6 ~{\rm M_{\odot}}$.

We first check the variation of different parameters over one orbit. With the choice of parameters as in Table \ref{tab:PSRBH} except $M_c=10^6 ~{\rm M_{\odot}}$ and with $i=\lambda_{\rm bh}=87.5^{\circ}$, we use the formalism presented in Sec. \ref{subsec:orbitandspinevolution} to find that the rate of change of different parameters are very small and can be ignored over a few orbit. Specifically, over one orbit, $\omega$ changes by $2.01^{\circ}, 1.95^{\circ},~{\rm and}~ 1.91^{\circ}$ for $\tilde{a}=0, 0.5,~{\rm and}~0.9$, respectively, $\eta_p$ changes by  $-0.54^{\circ}, -0.53^{\circ},~{\rm and}~ -0.52^{\circ}$ for $\tilde{a}=0, 0.5,~{\rm and}~0.9$, respectively, $\lambda_p$ changes by $-0.71^{\circ}, -0.70^{\circ},~{\rm and}~ -0.69^{\circ}$ for $\tilde{a}=0, 0.5,~{\rm and}~0.9$, respectively. Similarly, $P_b$ changes by $-2.37 \times 10^{-5}$ s, $e$ changes by $-1.23 \times 10^{-11}$ and $i$ changes by $(7.95 \times 10^{-12} )^{\circ}$ regardless of the spin of the black hole.

Note that, for our choice of parameters, the average separation between the pulsar and the black hole in the range of $401.5 - 833.8 ~ G M_c c^{-2}$ (for stellar mass black hole, it was $7.02 \times 10^5 - 1.45 \times 10^6 ~ G M_c c^{-2}$). That is why the evolution and precession of the orbit and the spin of the pulsar are not large over one orbit and the formalism presented in Sec. \ref{subsec:orbitandspinevolution} is valid. However, this orbital precession demands special care in the timing analysis as usually, pulsar timing algorithms fits for a particular value of $\omega$ when ToAs for only a few orbits are available and then adds $\dot{\omega}$ in the fitting algorithm when ToAs of a large number of orbits are available.

Moreover, we choose $P_s=$ 1.67 s, which was motivated by the detailed study of binary evolution by \citet{css21} for pulsar - stellar mass black hole binaries. Inspired by this, we use the same value of $P_s$ even for a pulsar - super massive black hole. For our choice of orbital parameters and masses, if we use $P_s=1$ ms, the only significant change would be in $i$ which would change by $(1.33 \times 10^{-8} )^{\circ}$ regardless of the spin of the black hole. This is still very small.

More rigorous calculation for tighter (hence more relativistic) binaries are available in the literature. In particular, \citet{Dsing2014}, \citet{KJLi2019}, and \citet{lwls22} explored properties of pulsars around super-massive and massive black holes. They modelled the spin-spin and spin-curvature using with the help of with the help of Mathisson-Papapetrou-Dixon (MPD) equations and Kerr metric around the black hole. They used a fast pulsar with $P_s=$ 1 ms and the average separation between the pulsar and the black hole in the range of $10 - 100 ~ G M_c c^{-2}$, the lower the separation, the higher is the general relativistic effects. Note that \citet{Dsing2014} showed that the motion of the pulsar out of the initial orbital plane is negligible when the average separation between the pulsar and the super-massive black hole is much larger than $40 ~ G M_c c^{-2}$. This agrees with our results of small change in $i$ (although obtained by a different formalism).

Our numerical results of spin precession following Eqs. (\ref{eq:spinprecfreq1PN}), (\ref{eq:spinprec1PN}), (\ref{eq:spinprecfreqLT}), and (\ref{eq:spinprecLT}) agree with recent results of \citet{majar2009} and \citet{majar2012}. Note that these equations do not give the scale of the nutation. However,  \cite{KJLi2019} (see their figure 5) showed that the scale of nutation is very small, in the order of $10^{-4}$ radian even when the separation is as small as $20 ~ G M_c c^{-2}$. Hence, we do not worry about the nutation. We first investigate how the strong bending affects the pulsar beam resulting in a change in the shape and strength of the pulse profile. 

\begin{figure*}
\begin{subfigure}[b]{0.86\textwidth}
	\includegraphics[width=\textwidth]{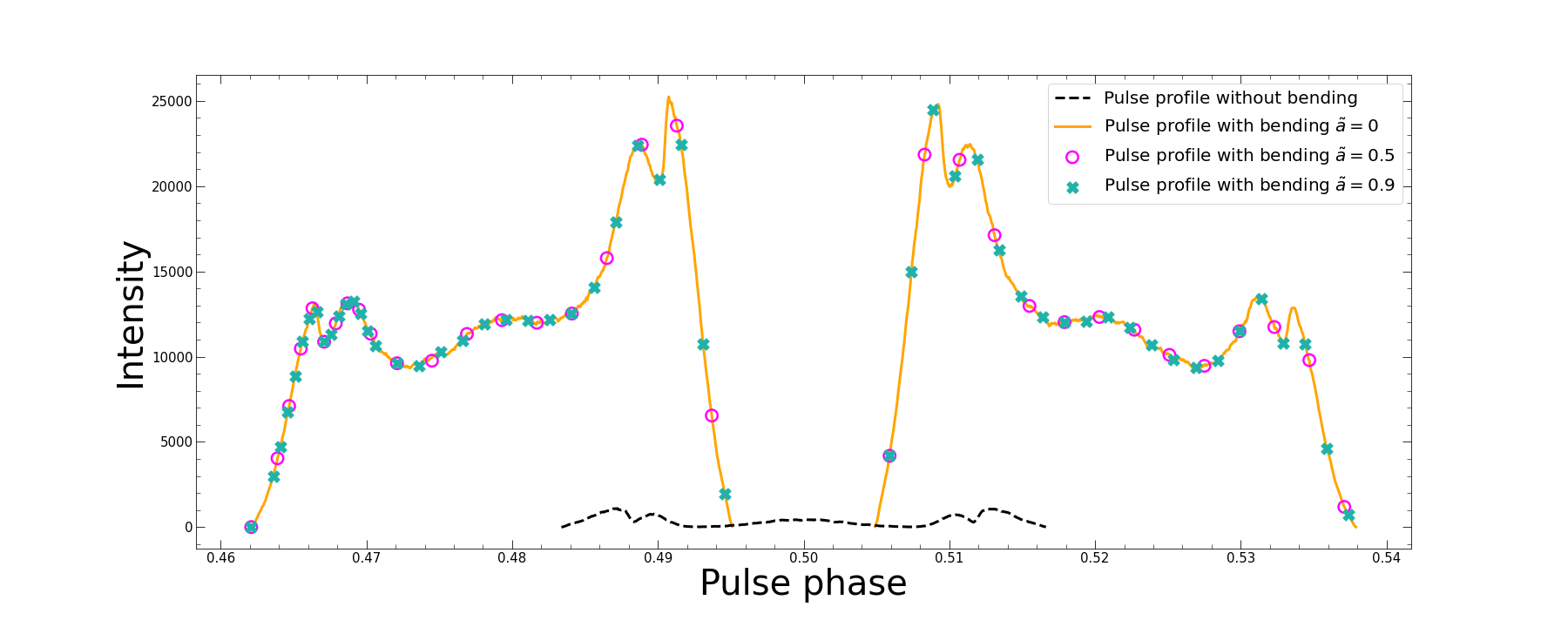}
	\caption{Pulse profiles for  $i=90^\circ$, $A_T+\omega = 90^\circ$. }
	\label{fig:supBH_pulse_phase_90_i_90}
\end{subfigure}

\begin{subfigure}[b]{0.86\textwidth}
	\includegraphics[width=\textwidth]{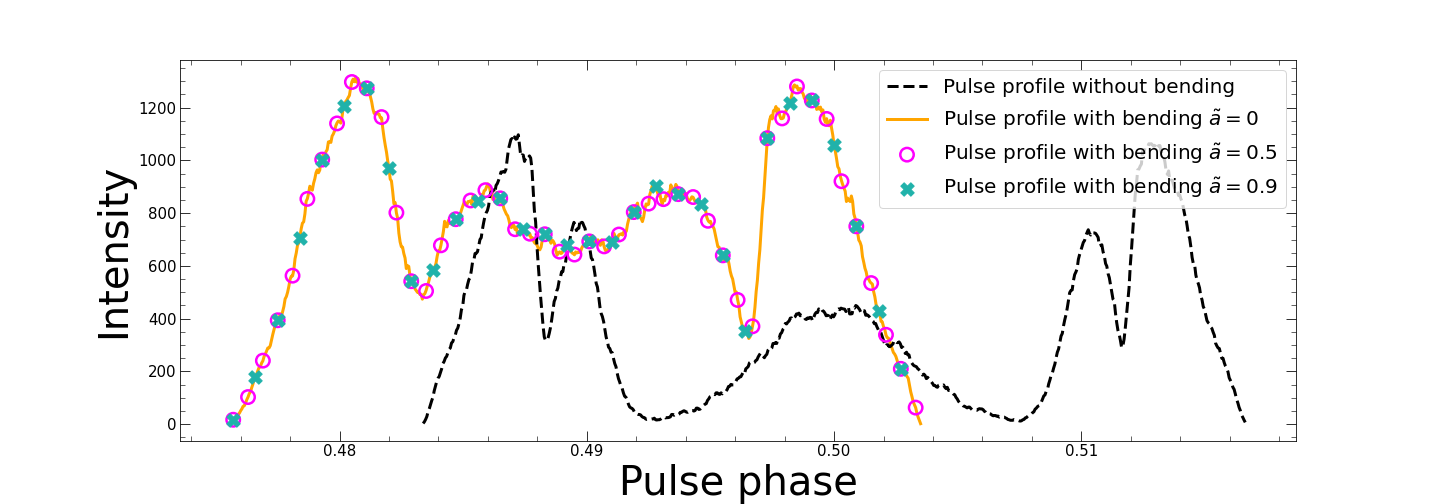}
	\caption{Pulse profiles for $i=90^\circ$, $A_T+\omega = 85^\circ$. }
	\label{fig:supBH_pulse_phase_85_i_90}
\end{subfigure}
\caption{Comparison of the pulse profile of a pulsar with a black hole companion with $\lambda_{\rm bh}=i=90^{\circ}$. We show the profiles without bending (black dashed line) and with bending for various values of the spin parameter of the black hole, e.g., $\tilde{a}=0$ (solid orange line), $\tilde{a}=0.5$ (magenta unfilled circles), and  $\tilde{a}=0.9$ (sea-green crosses). The values of all other relevant parameters are taken the same as mentioned in Table \ref{tab:PSRBH} except $M_c=10^6 {\rm M_\odot}$. }
\label{fig:supBH_pulse_orbital_inclination_90}
\end{figure*}

\begin{figure*}
  \centering
    \includegraphics[width=0.9\textwidth]{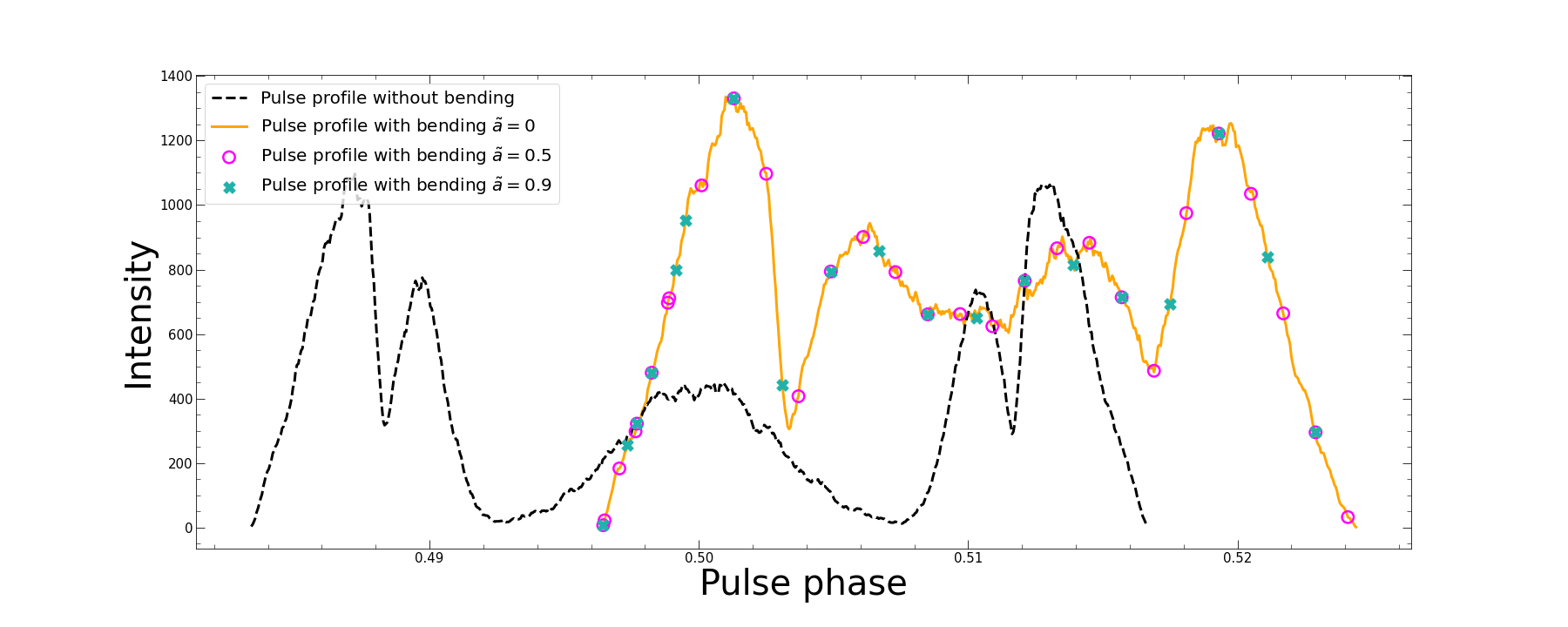}
     \caption{Comparison of the pulse profile of a pulsar with a black hole companion with $\lambda_{\rm bh}=i=85^{\circ}$, $A_T+\omega=90^{\circ}$. We show the profiles without bending (black dashed line) and with bending for various values of the spin parameter of the black hole, e.g., $\tilde{a}=0$ (solid orange line), $\tilde{a}=0.5$ (magenta unfilled circles), and  $\tilde{a}=0.9$ (sea-green crosses). The values of all other relevant parameters are taken the same as mentioned in Table \ref{tab:PSRBH} except $M_c=10^6 {\rm M_\odot}$. }
    \label{fig:supBH_pulse_phase_90_i_85}
\end{figure*}

In Fig. \ref{fig:supBH_pulse_orbital_inclination_90}, we show the pulse profile for two different orbital phases, $A_T+\omega=90^\circ$ (upper panel) and $A_T+\omega=85^\circ$ (lower panel), both for $i=90^\circ$. In each of the cases, we show the profiles if there was no bending, and with bending for $\tilde{a}=0$, $\tilde{a}=0.5$, and $\tilde{a}=0.9$. For $i=90^{\circ}$ and $A_T+\omega=90^\circ$, we see a drastically different shape of the pulse profile in comparison to the unbent profile. There are  two pulse components symmetrically separated from the center of the unbent pulse profile. This is quite different than what we see when the companion is a stellar mass black hole in Fig. \ref{fig:pulse_profileAll} where for the same values of $i$ and $A_T+\omega$, the shape of the pulse profile is similar to the unbent profile, only the strength of each components increases. For the super-massive black hole companion, even when we choose $i=90^{\circ}$ and $A_T+\omega=85^\circ$, the profile shape is significantly different than the unbent profile, the pulse profile undergoes moderate distortion. The core component of the pulse vanishes, while the peaks corresponding to the conal components become wider. Additionally, there is an overall leftward shift between the bent and the unbent profiles. This intrigued us to check what happens if we keep the value of $A_T+\omega$ fixed at $90^{\circ}$ but change the value of $i$ to $85^{\circ}$, as shown in \ref{fig:supBH_pulse_phase_90_i_85}. We see that the shape of the profile is somewhat similar to the case of $i=90^{\circ}$, $AT+\omega=85^{\circ}$ (Fig. \ref{fig:supBH_pulse_phase_85_i_90}), only the shift of the bent profile is now leftward in comparison to the unbent profile. Interestingly, in none of the cases, any difference between the profiles with bending for different values of $\tilde{a}$ is observed. We explain these features at and near the superior conjunctions with the help of simplified geometry in Sec. \ref{sec:appendix}.

\begin{figure*}
\begin{subfigure}[b]{0.86\textwidth}
	\includegraphics[width=\textwidth]{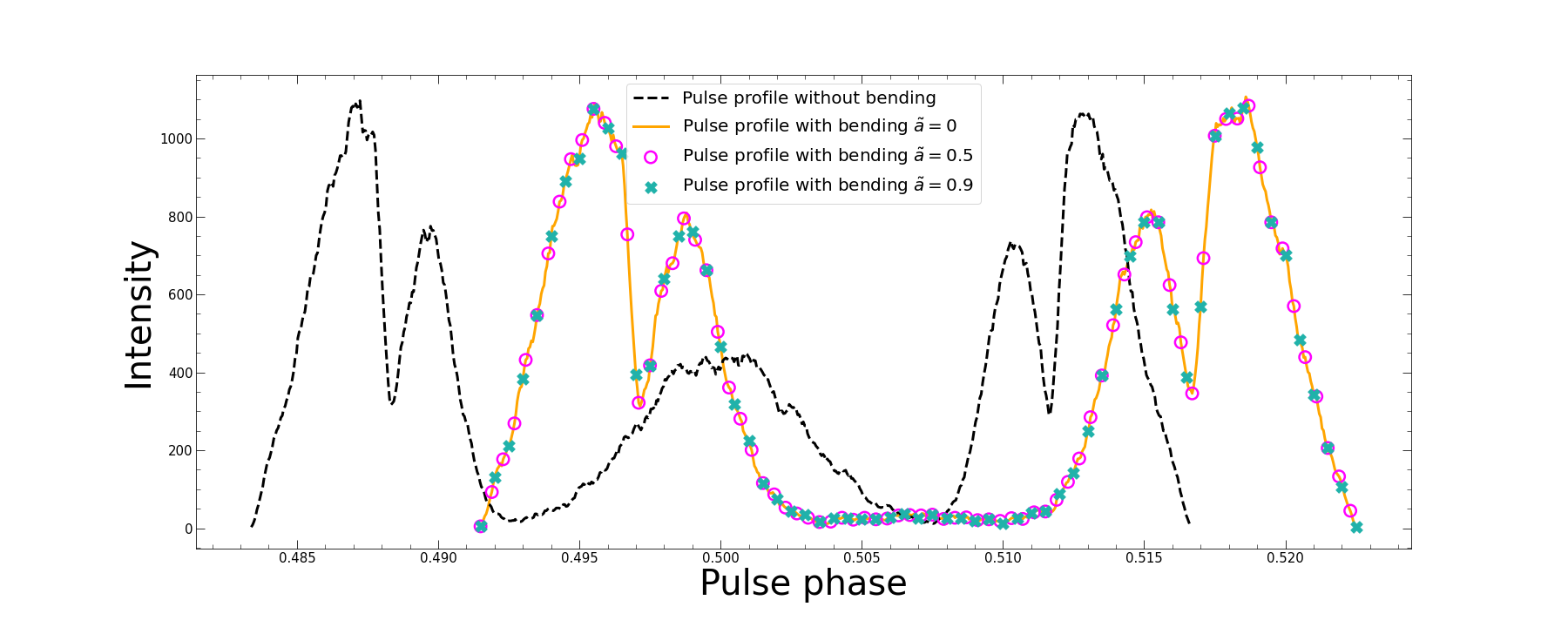}
	\caption{Pulse profiles for $i = 80^{\circ}$, $A_T+\omega=90^{\circ}$. }
	\label{fig:supBH_pulse_phase_90_i_80}
\end{subfigure}
\begin{subfigure}[b]{0.86\textwidth}
	\includegraphics[width=\textwidth]{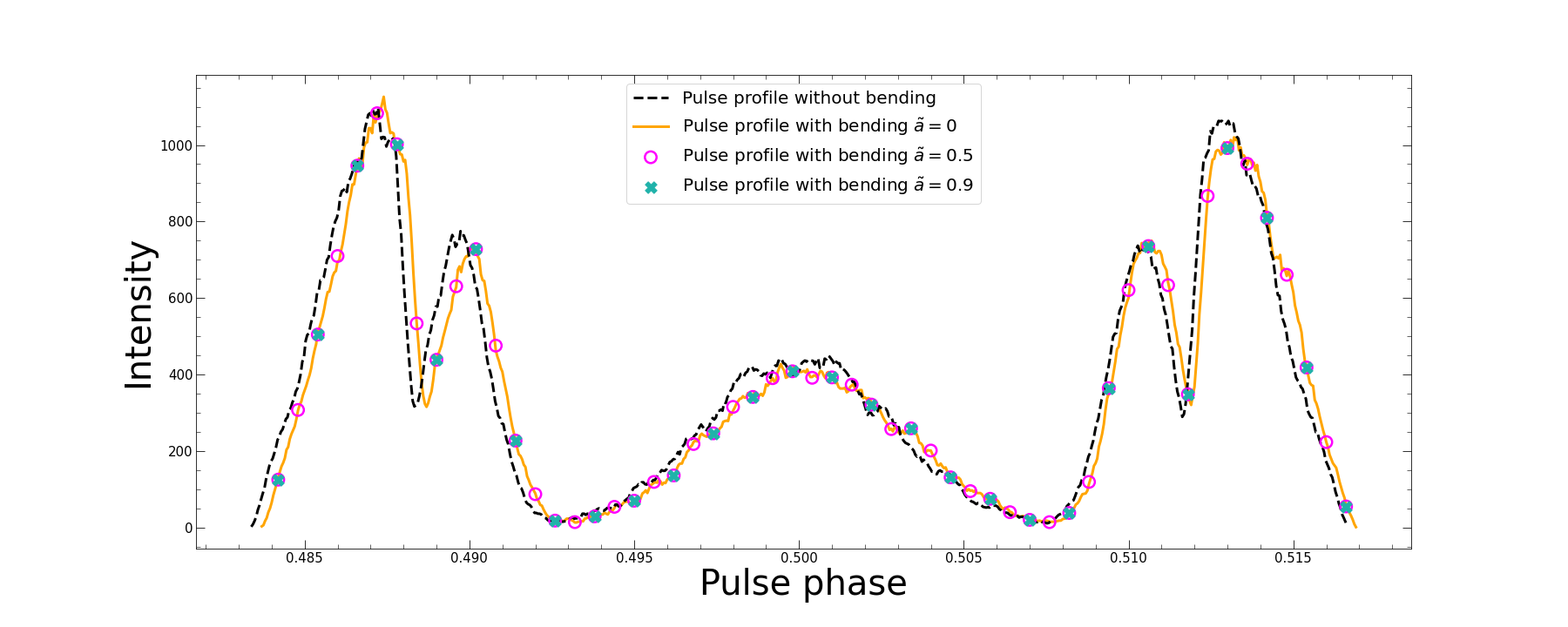}
	\caption{Pulse profiles for $i = 80{^\circ}$, $A_T+\omega=200^{\circ}$. }
	\label{fig:supBH_pulse_phase_200_i_80}
\end{subfigure}
\caption{Comparison of the pulse profile of a pulsar with a black hole companion with $\lambda_{\rm bh}=i=80^{\circ}$. We show the profiles without bending (black dashed line) and with bending for various values of the spin parameter of the black hole, e.g., $\tilde{a}=0$ (solid orange line), $\tilde{a}=0.5$ (magenta unfilled circles), and  $\tilde{a}=0.9$ (sea-green crosses). The values of all other relevant parameters are taken the same as mentioned in Table \ref{tab:PSRBH} except $M_c=10^6 {\rm M_\odot}$.}
\label{fig:supBH_pulse_inc_80}
\end{figure*}

This motivate us to investigate whether the shape and strength of the pulse profile come closer to the unbent profile if the configuration deviates significantly from the superior conjunction. In Fig. \ref{fig:supBH_pulse_phase_90_i_80}, we set $i=80^{\circ}$ and see that for $A_T+\omega=90^{\circ}$, although the enhancement of the conal components are very small, the core component is very weak and the pulse shape is still significantly different and shifted rightwards from the unbent profile. However, for $A_T+\omega=200^{\circ}$ the profile is nearly similar to the unbent one (Fig. \ref{fig:supBH_pulse_phase_200_i_80}). Interestingly, the bending is so weak at this orbital phase that the bent profile is nearly similar to the unbent one even if we set $i=90^{\circ}$. The reason is the fact that as the distance of the pulsar from the observer is smaller than the distance of the black hole from the observer, the light rays that reach the observer never pass very close to the black hole. We have also checked what happens for $i=60^{\circ}$ and see that for $A_T+\omega=90^{\circ}$ (not shown in the paper), the enhancement of the conal components are very small and the core component is visible. There is still a shift between the bent and the unbent profiles, but much less than the case of $i=80^{\circ}$ and see that for $A_T+\omega=90^{\circ}$ (Fig. \ref{fig:supBH_pulse_phase_90_i_80}).

\begin{figure*}
  \centering
  \begin{subfigure}[b]{0.49\textwidth} 
    \centering
    \includegraphics[width=\textwidth]{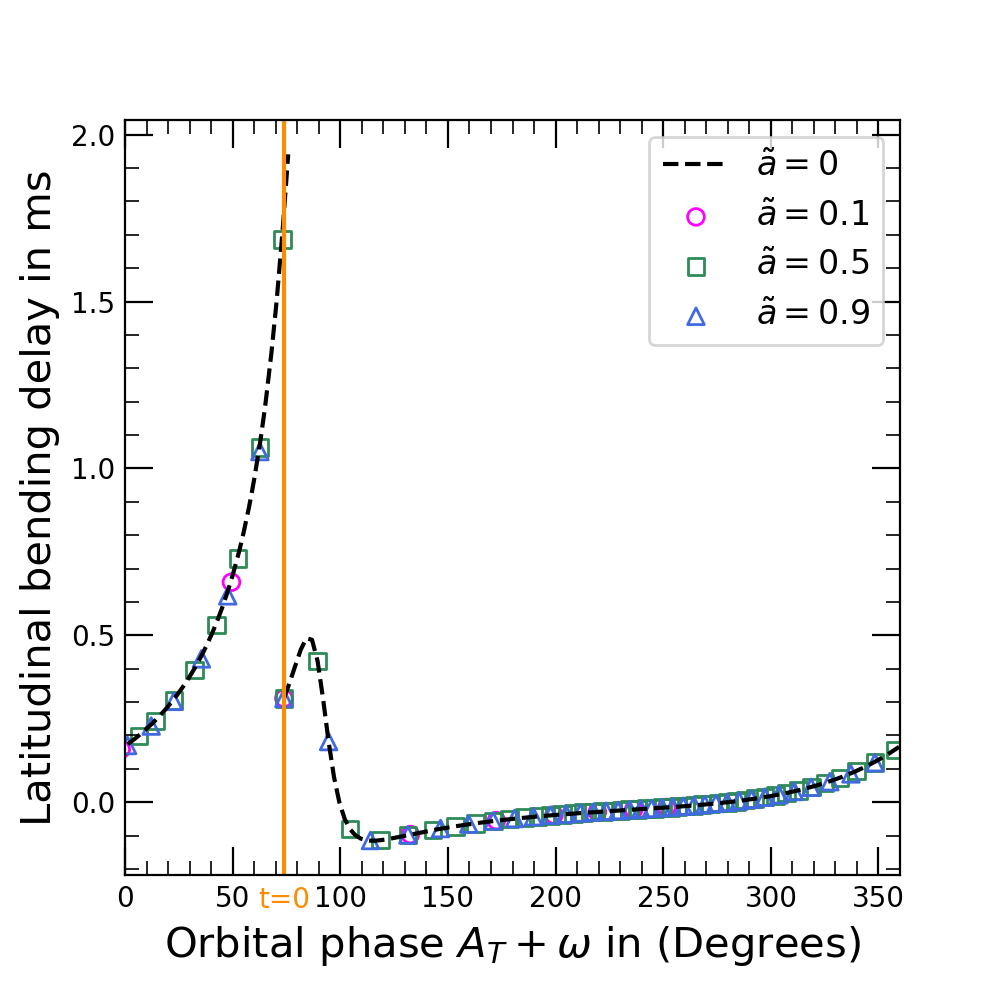}
       \phantomcaption{}
  \label{fig:supBH_lati_delay_i_80} 
  \end{subfigure}
  \hfill
  \begin{subfigure}[b]{0.49\textwidth}  
    \centering
    \includegraphics[width=\textwidth]{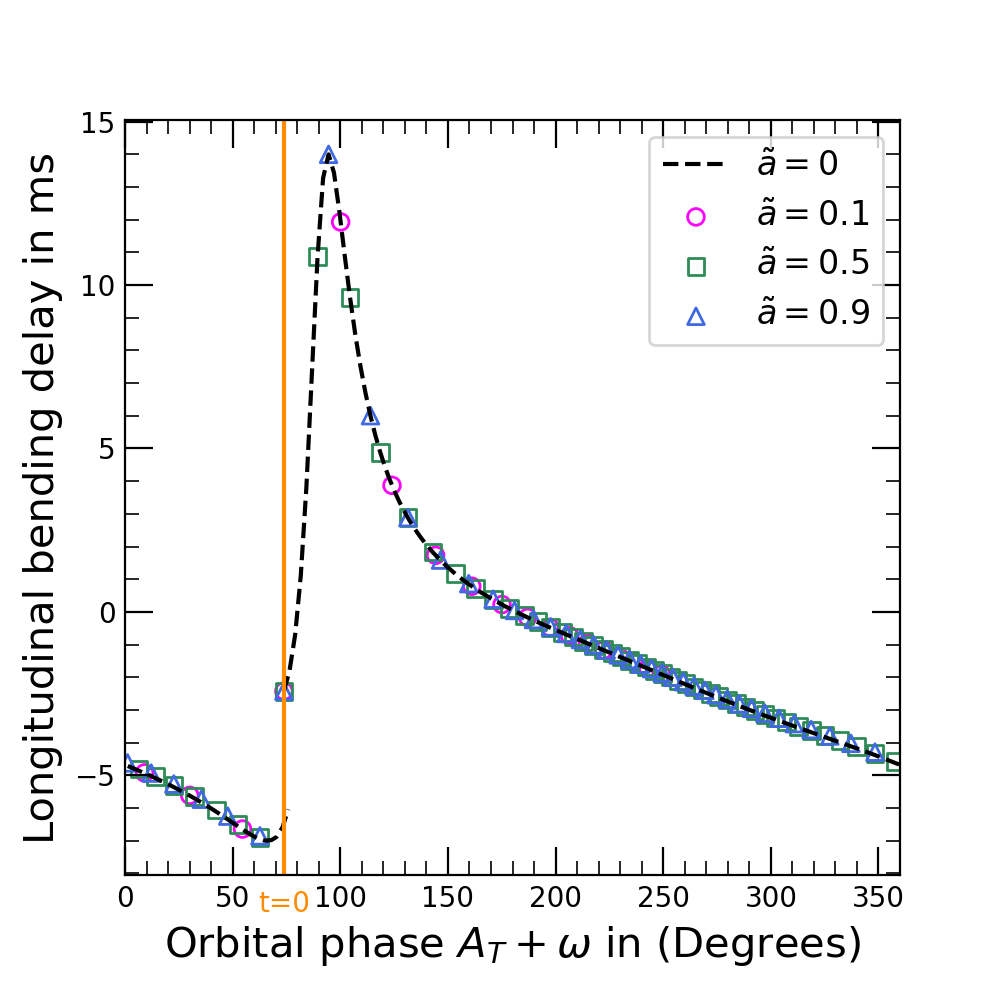}
    \phantomcaption{}
    \label{fig:supBH_long_delay_i_80}
  \end{subfigure}
  \caption{The bending delays over one full orbit for a hypothetical pulsar-super massive black hole binary with $\lambda_{\rm bh}=i=80^{\circ}$. The left panel shows the latitudinal bending delays and the right panel is for the longitudinal bending delay. In both of the cases, we have used different values of the spin parameter of the black hole, i.e., $\tilde{a}=0$ (solid black line), $\tilde{a}=0.1$ (magenta circles), $\tilde{a}=0.5$ (green squares), and $\tilde{a}=0.9$ (blue triangles) The values of all other relevant parameters are the same as those listed in Table \ref{tab:PSRBH} except $M_c=10^6 {\rm M_\odot}$. The left panel shows the latitudinal bending delay, and the right panel shows the longitudinal bending delay. In both of the panels, the vertical orange line is for $t=0$, i.e., where $A_T=0^{\circ}$, and its left side actually represents the orbital phases post $360^{\circ}$.}
  \label{fig:supBH_bending_delay}
\end{figure*}

\begin{figure*}
  \centering
  \begin{subfigure}[b]{0.49\textwidth} 
    \centering
    \includegraphics[width=\textwidth]{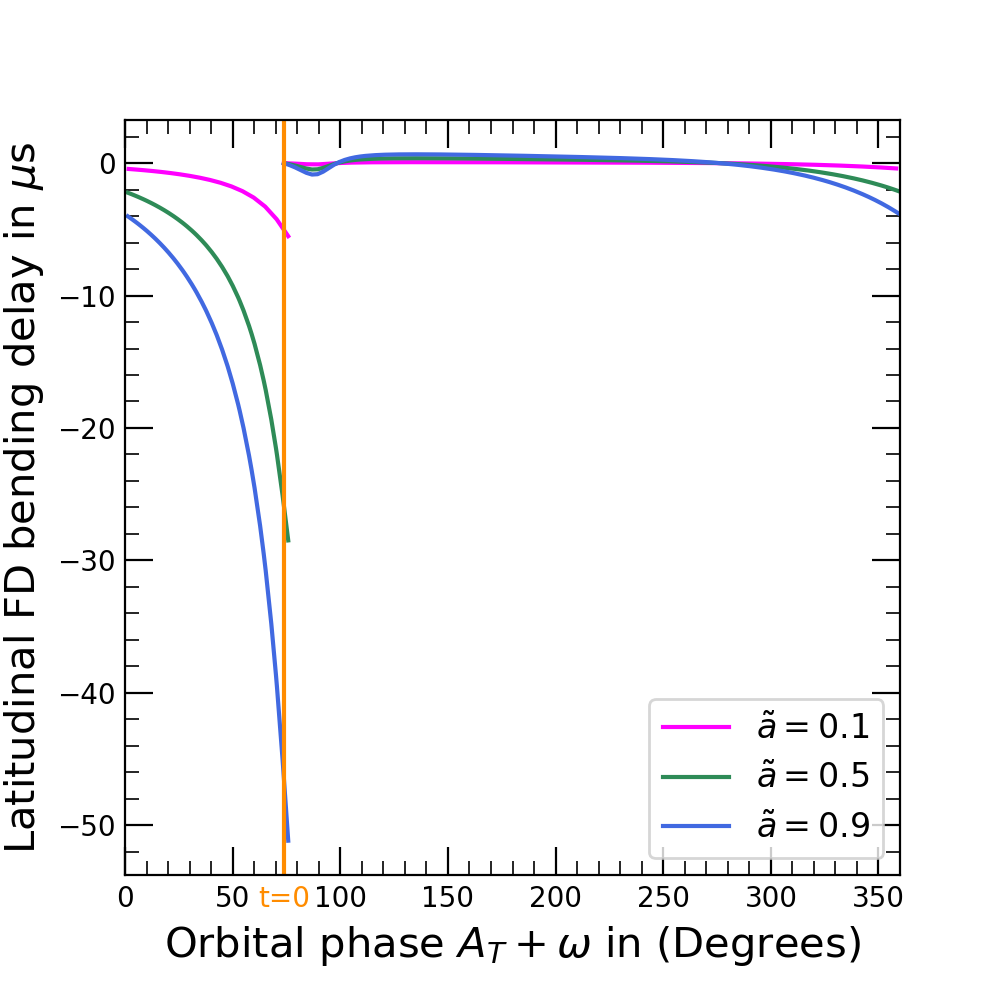}
       \phantomcaption{}
  \label{fig:supBH_FDlati_delay_i_80} 
  \end{subfigure}
  \hfill
  \begin{subfigure}[b]{0.49\textwidth}  
    \centering
    \includegraphics[width=\textwidth]{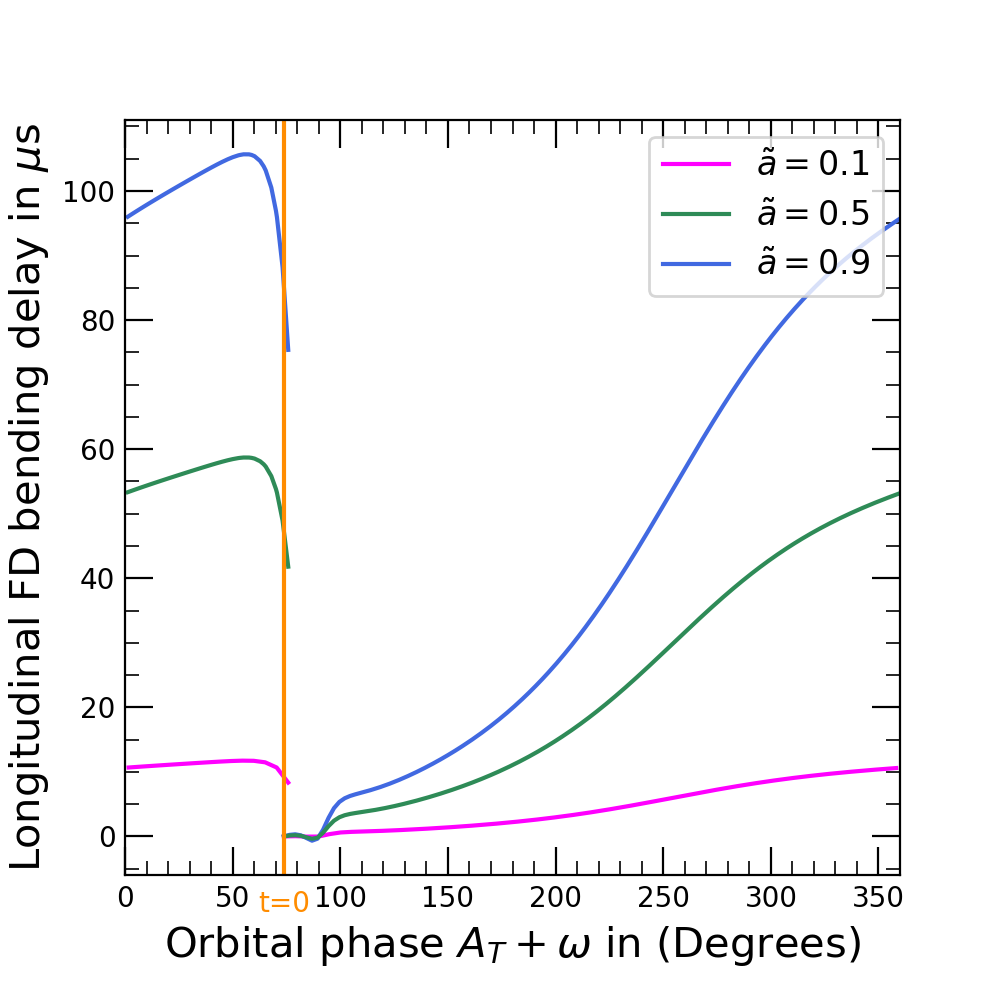}
    \phantomcaption{}
    \label{fig:supBH_lFDong_delay_i_80}
  \end{subfigure}
  \caption{The FD bending delays over one full orbit for a hypothetical pulsar-super massive black hole binary with $\lambda_{\rm bh}=i=80^{\circ}$. The left panel shows the latitudinal bending delays and the right panel is for the longitudinal bending delay. In both of the cases, we have used different values of the spin parameter of the black hole, i.e., $\tilde{a}=0$ (solid black line), $\tilde{a}=0.1$ (magenta circles), $\tilde{a}=0.5$ (green squares), and $\tilde{a}=0.9$ (blue triangles) The values of all other relevant parameters are the same as those listed in Table \ref{tab:PSRBH} except $M_c=10^6 {\rm M_\odot}$. The left panel shows the latitudinal bending delay, and the right panel shows the longitudinal bending delay. In both of the panels, the vertical orange line is for $t=0$, i.e., where $A_T=0^{\circ}$, and its left side actually represents the orbital phases post $360^{\circ}$.}
  \label{fig:supBH_FDbending_delay}
\end{figure*}

Now, we investigate what happens to bending delay curves. We choose $i=80^{\circ}$ where bending is strong near $A_T+\omega=90^{\circ}$ but weak at other orbital phases like $A_T+\omega=200^{\circ}$. In Fig. \ref{fig:supBH_bending_delay}, we plot the orbital phase ($A_T + \omega$) in degrees along the ordinate and the bending delays in millisecond along the abscissa. The left panel is for the latitudinal bending delay and the right panel is the longitudinal bending delay. In case of the stellar mass black hole companion, even for $i=87.5^{\circ}$ (Fig. \ref{fig:bending_delay_a_9}), the bending delays were in the order of microsecond. However, in case of a super-massive black hole companion, we see that the bending delays are much larger, in the order of millisecond. However, just like the case of stellar mass black hole, we can not distinguish bending delays for different values of $\tilde{a}$. Hence, we plot the value of the FD bending delays in Fig. \ref{fig:supBH_FDbending_delay}, where the orbital phase in degrees is plotted along the ordinate and the values of the FD bending delays in microsecond are shown along the abscissa. The left panel is for the latitudinal FD bending delay and the right panel is the longitudinal FD bending delay. We see that the values of FD bending delays for super-massive black hole is in the order of microsecond while those were in the order of nanosecond in the case of the stellar mass black hole (Fig. \ref{fig:FD_delay}).

Note that, in all calculations and figures of bending delays and FD bending delays, we set the time $t=0$ when $A_T=0^{\circ}$ or $A_T+\omega=73.804^{\circ}$. We represent $t=0$ by a vertical orange line in Figs. \ref{fig:bending_delay_a_9},  \ref{fig:supBH_bending_delay}, \ref{fig:supBH_FDbending_delay}. In each figure, the left side of this orange line represent orbital phases post $360^{\circ}$. We see  for the case of the super-massive black hole, at $t=P_b$, the value of $A_T+\omega$ is slightly larger than $73.804^{\circ}$ due precession of the orbit - visible from the spillage of the delay curves from left to right of the vertical orange lines. Moreover, at $t=P_b$, values of the bending delays and the FD bending delays are different than what what it was at $t=0$. This happens because spin precession affects the geometry determining the amount of bending. Interestingly, the discontinuities seen at $A_T+\omega=90^{\circ}$ (when $\eta_{\rm bh}=90^{\circ}$) in FD bending delay curves for the stellar mass black hole is not visible for the case of the super-massive black hole. We attribute this to comparatively larger changes in the configuration due to the comparatively larger spin and orbital precession for the second case.

\section{Conclusion}
\label{sec:conclu}

In this work, we studied the effect of the light-bending phenomenon on the signal of a pulsar in a binary system with a rotating black hole companion, focusing mainly on the stellar mass black holes. For this purpose, we solved the equations of the null geodesic in the Kerr spacetime using the formalism devised by \citet{KDL20} and \citet{GSL20}. As an initial guess, we adopted the orbital parameters and masses of the pulsar-black hole binary from \citet{css21}, although that work favored non-rotating black holes. Subsequently, we explored how variations in different parameters affect the bending delays. We found that the impacts of various parameters on the bending delays visually align with those observed for the non-rotating black hole as studied in \citetalias{dbb23}, because the magnitude of the spin as well as orientation of the spin axis of the black hole introduces changes in the nanosecond order and other parameters do so in the microsecond order. Hence, the effects are demonstrated by plotting the FD bending delays instead of the bending delays. We noticed nanosecond scale discontinuities in the FD bending delay curves when $\widehat{r}_s$, $\widehat{S}_{\rm bh}$, and $\widehat{N}_s$ align in the same plane. Moreover, as in the Schwarzschild case \citepalias{dbb23}, the bending delays become irregular (in the microsecond order) near the superior conjunction, where both the orbital phase and the orbital inclination approach ninety degrees. Additionally, the distortion of the beam and the resulting changes in the pulse shape are minimally influenced by the spin-related parameters of the black hole (both magnitude and orientation). \citet{km02} reported similar imperceptible differences in the propagation delays for rotating black hole companions compared to non-rotating ones. 

In our earlier work \citepalias{dbb23}, we demonstrated that for a non-rotating stellar-mass black hole companion of a pulsar, the bending delays calculated using a full general relativistic framework agree with the approximate analytical expressions given by \citet{RL06} and \citet{dk95}, except when both the orbital inclination and the orbital phase are close to ninety degrees.

Hence, we conclude that if a pulsar-black hole (stellar-mass) binary is discovered in the future, the analytical expressions of \citet{RL06} or \citet{dk95} would suffice to model the light-bending delays, regardless of the spin of the black hole, provided the timing accuracy does not reach the nanosecond scale. However, care should be taken during the superior conjunction, where changes in the shape of the profiles could lead to discrepancies with the stable template profile used in timing analysis.

We have also explored some cases of bending phenomenon for pulsars in a binary with a super-massive black hole, and found some interesting features in the pulse profile due to strong bending that are absent in the case of a stellar mass black hole companion.  We find significant enhancement and change in the shape of the profiles at and near the superior conjunction in comparison to the rotating stellar mass black hole. Moreover, bending delays are also about three orders of magnitude higher than those in case of the stellar mass black holes, when all other parameters are the same. These facts would make timing of such binaries difficult.

Note that, we have chosen the orbital period as 5.5 days, i.e., our system is much more relativistic than the pulsar-super massive binaries studied in general, where being guided by the orbits of the S0 and S2 stars, the orbital period is chosen in the order of years \citep{liu2012,zs2017}. However, as we discussed earlier, we can still use post-Newtonian formalism to model the evolution of the orbit and the precession of the spin of the pulsar instead of MPD formalism that becomes essential for even tighter binaries with faster neutron stars. We do not aim to study such extremely computationally challenging problem presently, as our aim is to provide some theoretical results to observers studying pulsar - black hole binaries in the future. Timing analysis of relativistic binaries is difficult for many reasons. One of them is the uncertainty in the measurements of pulse `Time of Arrivals' (ToAs) that are the building block of pulsar timing analysis and detecting any gravitational phenomena in the time domain. \citet{ekl2017} showed that if a light ray travels close to a black hole, the ToA uncertainty increases with the increase of the mass of the black hole. Moreover, the observer needs a stable template profile to extract ToAs \citep[chapter 8]{lorimer04} and if the shape of the profile vary significantly with the orbital phase (as it happens in case of the pulsar - supermassive black hole binary), ToA extraction becomes erroneous.

\section*{Data Availability}

No observational data have been used in this work. Numerical results and codes underlying this article will be shared on reasonable request to the corresponding author.

\bibliographystyle{mnras}

\bibliography{debnathKerr}

\section{Appendix}
\label{sec:appendix}

\begin{figure*}
  \centering
    \includegraphics[width=\textwidth]{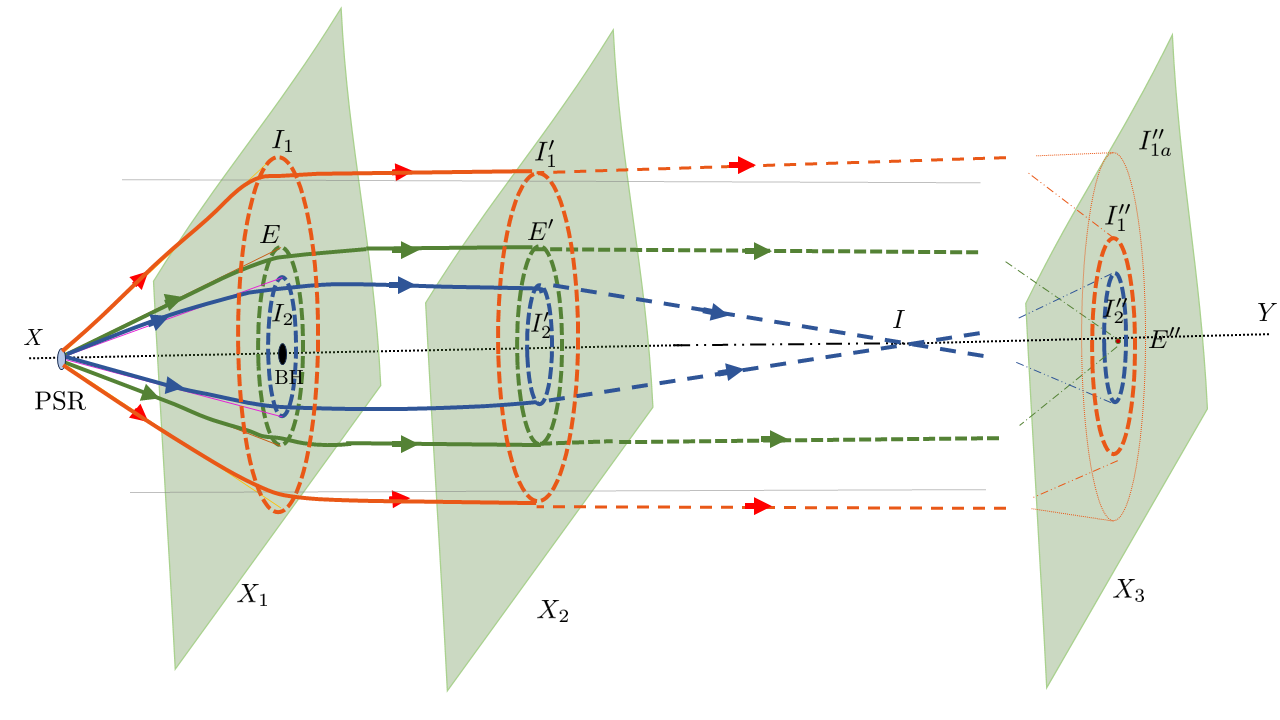}
    \caption{A schematic diagram showing bending and propagation of light rays at the superior conjunction configuration.}
    \label{fig:supBH_schematic_null_geodesic}
\end{figure*}

Here we try to understand the distortion of the pulse profile near the superior conjunction with a simplified geometry as show in Fig. \ref{fig:supBH_schematic_null_geodesic}. 

In this figure, we define a straight line $XY$ that connects the pulsar (denoted by `PSR') and the black hole (denoted by `BH') and extends throughout the space. Three planes, $X_1$, $X_2$, and $X_3$, all perpendicular to the line $XY$, are shown in a light green colour. The plane $X_1$ contains the black hole, $X_2$ is an intermediate plane located along the path of the light propagation slightly away from the black hole and towards the observer in such a way that the bending of the rays are completed before entering the $X_2$-plane, i.e., the directions of the light rays do not change after they enter the $X_2$-plane. $X_3$ is a plane on which the observer is located, it can be considered at infinity for the purpose of geometry. Note that, this geometry would not be valid for all configuration, e.g., for almost face on orbits (low values of $i$), the pulsar and the black hole would be in the same plane as seen from the observer. Even for $i=90^{\circ}$, for a range of the values of the orbital phase, the pulsar would be between the black hole and the observer. However, such cases are simply explained with the ray tracing and the passage of the LoS through the beam as explained in detailed in \citetalias{dbb23}.

Let us consider a set of light rays emitted from the pulsar in such a way that in the absence of the bending, they would intersect the plane $X_1$ on a circle $E$. In Fig. \ref{fig:supBH_schematic_null_geodesic}, only two such light rays are shown for the sake of simplicity. After the experiencing the gravitational bending, these rays become parallel to the line $XY$ and intersect the plane $X_2$ in a circle denoted by $E^{\prime}$. This circle $E^{\prime}$ is smaller than $E$, as it results from the light rays affected by the bending. However, the size of $E^{\prime}$ does not change with the position of the plane $X_2$. As we can assume that parallel light rays converge at infinity, let us assume that the rays forming the circle $E^{\prime}$ meet at the point $E^{\prime \prime}$ on the $X_3$-plane. This is a valid assumption as the distance of the observer is much larger than the size of the orbit of the binary. The point $E^{\prime \prime}$ is located on the line $XY$. In short, all the light rays initially directed towards the circle $E$ converge to the point $E^{\prime \prime}$ after bending. The circle $E$ corresponds to the well-known Einstein ring in gravitational lensing theory. The radius of $E$ is known as the Einstein radius $R_E$ and is given by \citep{RL06}:
\begin{equation}
\label{eq:RE}
R_E=\left[\left(\frac{4G M_c}{c^2}\right)  a_R |\sin i| (1-e^2)/(1+e\sin \omega)\right]^{1/2} ,
\end{equation} where $a_R$ is the semi major axis of the relative orbit.

Next, we consider light rays that are initially directed such a way that without bending, they would intersect the plane $X_1$ on a circle $I_1$ which is larger than $E$. Again, only two such rays are shown in the figure. After bending, these rays intersect the plane $X_2$ on the circle $I_1^{\prime}$. As the light rays that fall on $I_1$ are farther from the gravitating body (the black hole) than the light rays that fall on $E$, the rays emerging from $I_1^{\prime}$ would still be diverging (but less diverging than the original rays that fell on $I_1$). As a result, the size of $I_1^{\prime}$ increases with the increasing distance between the planes $X_1$ and $X_2$, and $I_1^{\prime}$ is always larger than $E^{\prime}$. On the $X_3$-plane, these light rays form a circle $I_{1a}^{\prime \prime}$ around the point $E^{\prime \prime}$. It is clear that $I_{1a}^{\prime \prime}$ is larger than $I_1^{\prime}$ and $I_1^{\prime}$ is larger than $I_1$. However, when we assume the $X_3$-plane to be located at the infinity, the light rays diverging from $I_1^{\prime}$ can be assumed to be converging at a smaller radius $I_1^{\prime \prime}$ (in the same logic parallel rays from $E^{\prime}$ have been assumed to be converging at the point $E^{\prime \prime}$).

Now, we consider another set of light rays that initially directed such a way that without bending, they would intersect the plane $X_1$ on a circle $I_2$ which is smaller than $E$. Again, only two such rays are shown in the figure. After bending, these rays intersect the plane $X_2$ on the circle $I_2^{\prime}$. These light rays are converging as they are closer to the black hole than the rays that fell on $E$ and become parallel. As a result, the size of $I_2^{\prime}$ decreases with increasing distance between the planes $X_1$ and $X_2$. $I_2^{\prime}$ is always smaller than $E^{\prime}$. Hence, the light rays would converge at a point $I$ on the line $XY$ in such a way that $I$ lies between the $X_2$-plane and the $X_3$-plane. Afterwards, these rays diverge from $I$ and cut the $X_3$-plane on the circle $I_2^{\prime \prime}$ around the point $E^{\prime \prime}$. As the rays that were above the $XY$ line before $I$ are below the $XY$ line in the $X_3$-plane (and vice versa), $I_2^{\prime \prime}$ might be termed as an inverted image of $I_2$. If the size of $I_2$ lies very close to $E$, then $I_2^{\prime \prime}$ would be a small circle around the point $E^{\prime \prime}$. In this situation $I_2^{\prime \prime}$ is smaller than $I_2$. However, if $I_2$ is much smaller than $E$, then the point $I$ would be closer to the $X_2$-plane and $I_2^{\prime \prime}$ would be larger than $I_2$ as the distance between the $X_2$-plane and the $X_3$-plane is much larger than the distance between the $X_1$-plane and the $X_2$-plane. If $I_2$ is too small corresponding to an impact parameter smaller than $3 \sqrt{3} G M_{c} /c^2$, the light rays would be trapped by the black hole. 

We can summarise the above discussions as follows:
\begin{enumerate}

\item Rays initially directed towards $E$ become parallel after bending and appear to converge to the point $E^{\prime \prime}$ on the plane containing the observer.

\item Rays initially more diverging than those forming $E$, form a circle $I_1^{\prime \prime}$ around $E^{\prime \prime}$  on the plane containing the observer.

\item Rays initially less diverging than those forming $E$ form an inverted circle $I_2^{\prime \prime}$ around $E^{\prime \prime}$  on the plane containing the observer.

\end{enumerate}
The observer's location on the $X_3$-plane depends on the inclination of the orbit as well as the orbital phase of the pulsar. For instance, the observer lies at $E^{\prime \prime}$ when the pulsar is exactly at the superior conjunction, i.e., when the pulsar, the black hole, and the observer are aligned along the line $XY$. At other orbital phases and orbital inclinations, the position of the observer shifts from $E^{\prime \prime}$ in the $X_3$-plane.

\begin{figure*}
  \centering
  \begin{subfigure}[b]{0.49\textwidth}
    \centering
    \includegraphics[width=\textwidth]{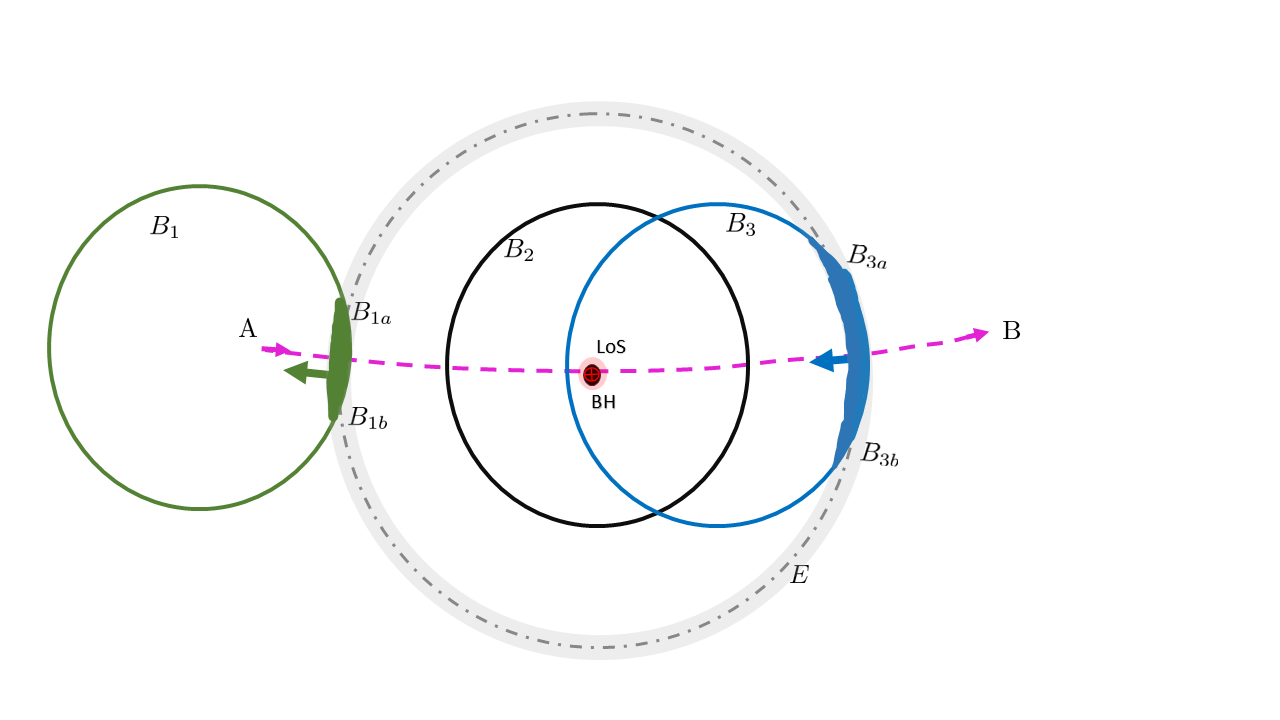}
    \caption{Schematic diagram for $i=90^\circ$, $A_T+\omega=90^\circ$}
    \label{fig:supBH_schematic_i_90_phase_90}
  \end{subfigure}
  \hfill
  \begin{subfigure}[b]{0.49\textwidth} 
    \centering
    \includegraphics[width=\textwidth]{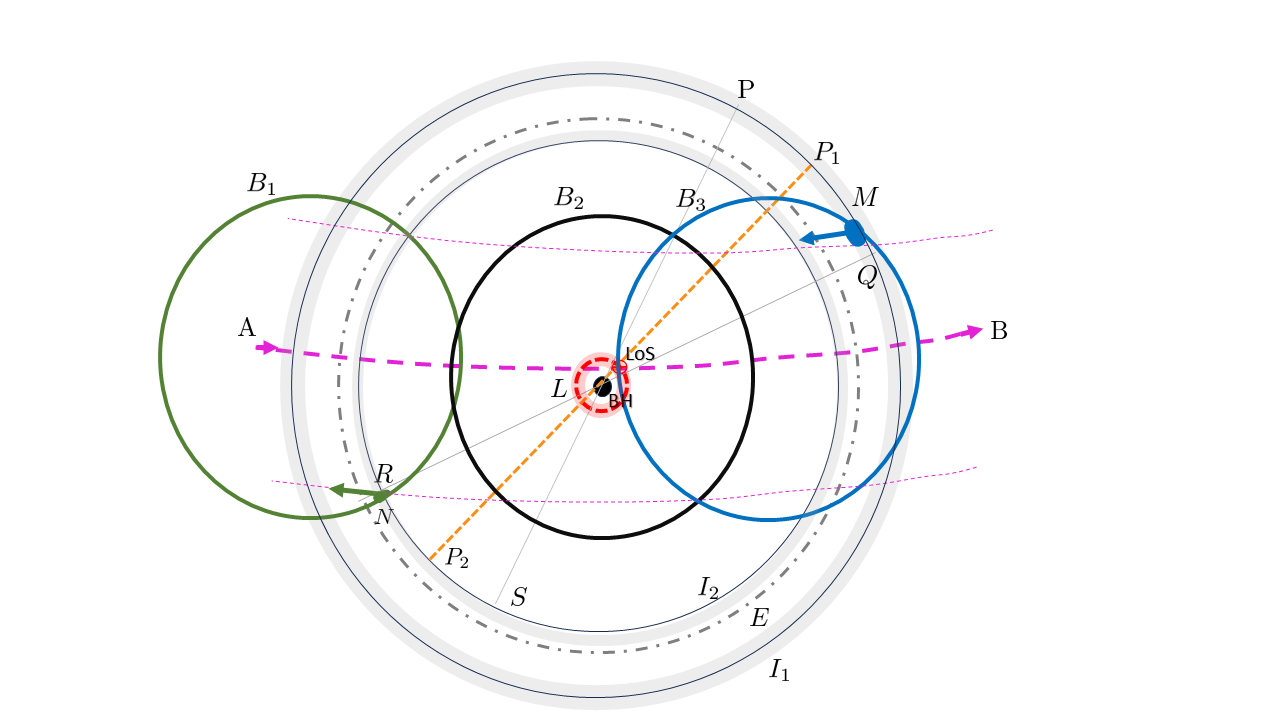}
    \caption{Schematic diagram for $i=89^\circ$, $A_T+\omega=90^\circ$.}
    \label{fig:supBH_schematic_i_89_phase_90}
  \end{subfigure}
  \hfill
  \begin{subfigure}[b]{0.49\textwidth} 
    \centering
    \vskip 0.1cm
    \includegraphics[width=\textwidth]{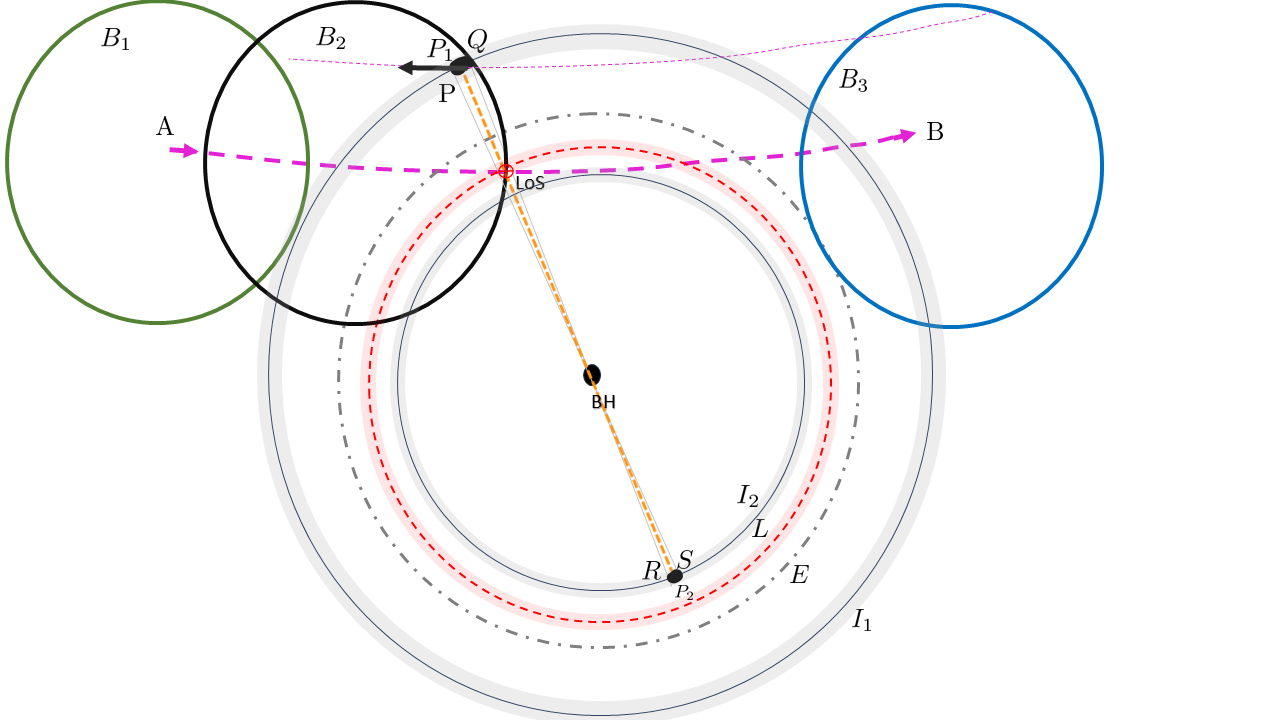}
    \caption{Schematic diagram for $i=90^\circ$, $A_T+\omega=85^\circ$.}
    \label{fig:supBH_schematic_i_90_phase_85}
  \end{subfigure}
  \hfill
  \begin{subfigure}[b]{0.49\textwidth} 
    \centering
    \vskip 0.1cm
    \includegraphics[width=\textwidth]{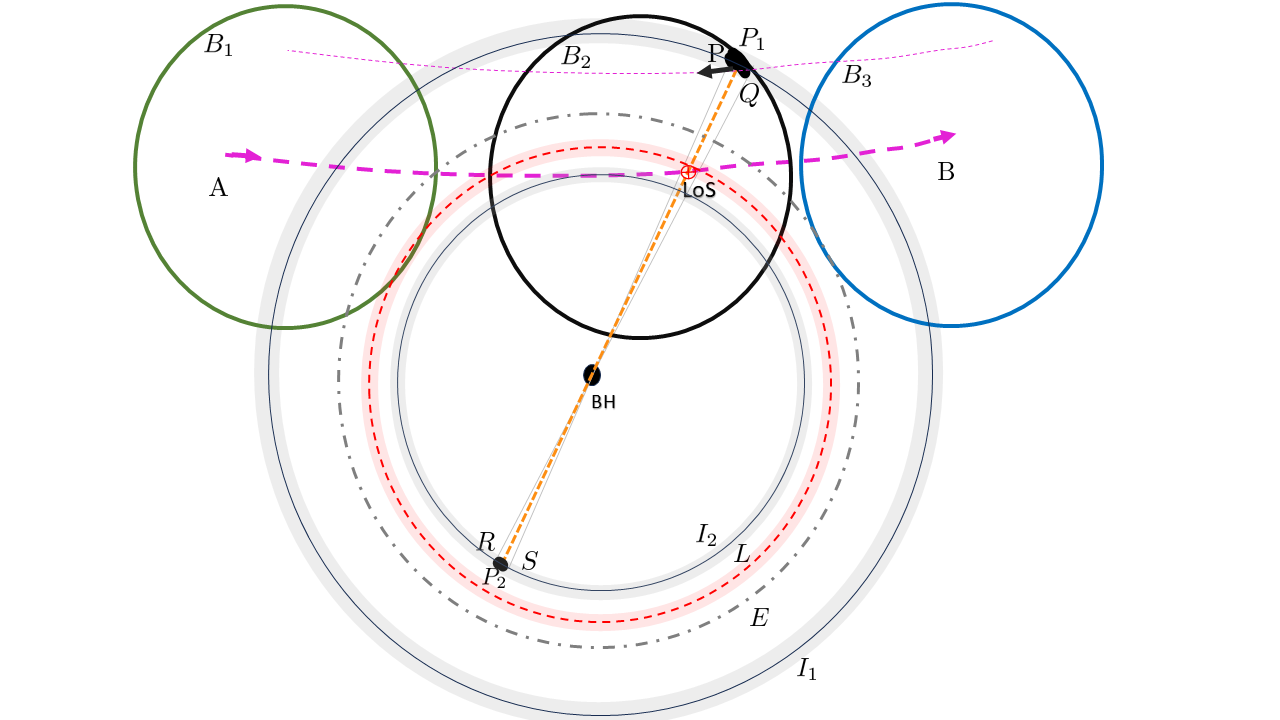}
    \caption{Schematic diagram for $i=85^\circ$, $A_T+\omega=90^\circ$.}
    \label{fig:supBH_schematic_i_85_phase_90}
  \end{subfigure}
  \caption{Understanding strong bending for a pulsar - super massive black hole binary system when the pulsar is near the superior conjunction for different combinations of $i$ and $A_T + \omega$ in the observer's frame, where the centre of the beam moves along the magenta line `$AB$' that passes through the Line-of-Sight (`LoS'). The arrows on $AB$ denotes the direction of motion of the beam. Three positions of beam $B_1$, $B_2$, and $B_3$ have been shown in each panel. The red ring marked $L$ is the ring on which the light rays from the rings $I_1$ and $I_2$ can come due to strong bending as explained in the text. As the red ring is chosen to pass through LoS, the observer can receive light rays from the arcs $PQ$ on $I_1$ and $RS$ on $I_2$, or more specifically from the segments of those arcs that fall on the beam. The direction of the motion of the segments have been shown with small arrows of the same colour used for that segment. The area of the beam traversed by such a segment results the pulse profile. }
  \label{fig:supBH_schematic_diagrams}
\end{figure*}

We are now ready to explain the pulse profiles near the superior conjunction based on how the cross-section of the beam of the pulsar moves in the $X_1$-plane (plane of the paper with the $X_3$-plane lying above it - towards the reader) as shown in  Fig. \ref{fig:supBH_schematic_diagrams}. The line of sight, which is from the pulsar to the observer, intersects this plane at the location marked by a small unfilled red circle labeled as `LoS'. In our work, when we construct a pulse profile, instead of counting light rays that falls onto a single point LoS, we consider all the light rays that make an angle less than $0.09^\circ$ with the LoS on the ${\rm X_m Y_m}$-plane (see Fig. \ref{fig:spin_axis_frame} of the present paper as well as \citetalias{dbb23}). This translates to a finite size of the LoS on the $X_3$ plane too, which is projected with a finite size on the $X_1$-plane in Figs. \ref{fig:supBH_schematic_diagrams}. Hence, the circles $I_1$ and $I_2$ becomes annular rings of finite widths as shown in various panels of Fig. \ref{fig:supBH_schematic_diagrams}. 

Different panels in Fig. \ref{fig:supBH_schematic_diagrams} correspond to different combinations of $i$ and $A_T + \omega$, all not very far from the superior conjunction. For these combinations, the value of $R_E$ remains constant $5.94 \times 10^7$ km for the pulsar - super massive black hole binary we are studying. In the rest frame of the observer, the centre of the beam of the pulsar moves (due to the spin of the pulsar) along the magenta dashed line labeled `$AB$'. The arrows in this line represent the motion of the beam. We call this line as the Line-of-Beam-Rotation (LoBR). Three near-by positions of the cross-section of the beam on the $X_1$-plane are shown with green ($B_1$), black ($B_2$), and blue ($B_3$) circles. We have chosen our parameters $\alpha$ and $\zeta_L$ (see Fig. \ref{fig:spin_axis_frame}) in such a way that the LoS falls on the LoBR. However, depending on the values of $i$ and $A_T+\omega$, both the LoBR and LoS shift on the $X_1$-plane with respect to the black hole.

As we know, the beam is generated near the pulsar with the radius of the outer cone as $34.49$ km (\citetalias{dbb23}). This conal beam propagates outward intersects the $X_1$-plane with a larger radius. As an example, for the pulsar super-massive black hole binary, radius of the beam on the $X_1$-plane is $4.7 \times 10^7$ km when the pulsar is at and near the superior conjunction. This implies that the size of the cross-section of the beam is slightly smaller than $E$ which has a radius of $5.94 \times 10^7$ km for the case of the pulsar - super-massive black hole binary at and near the superior conjunction configuration. If the configuration deviates from the superior conjunction, the radius of the beam on the $X_1$-plane decreases slightly but the radius of $E$ remains the same (as the changes of $a_R$, $e$ and $\omega$ within one orbit is negligible).

First, we consider the case of the superior conjunction, i.e., $i = 90^\circ$ and $A_T + \omega = 90^\circ$, as shown in Fig. \ref{fig:supBH_schematic_i_90_phase_90}. At this configuration, the observer is at the point $E^{\prime\prime}$ on the $X_3$-plane, i.e., the LoS coincides with the black hole on the $X_1$-plane. The light
rays from the portion of the beam intersecting the circle $E$ bend towards the observer (who is at $E''$). $B1$ and $B3$ are the two illustrative positions of the beam on two different sides of the circle $E$. The beam intersects the circle $E$ along the arc $B1a-B1b$ for the position $B1$. All light rays originating from this segment are bent toward the LoS and are received by the observer. As the beam continues to move, this arc sweeps through the beam's width, and the observer receives a highly intense pulse profile due to contributions from the entire beam. When the cross-section of the beam is fully inside $E$, the observer does not receive any signal (as the light rays fall onto $I_2^{\prime \prime}$ on the $X_3$ plane which is away from the observer $E^{\prime \prime}$). As the beam continues to move, it starts to cut $E$ again. One such position of the beam is $B3$, where the beam intersects the circle $E$ along the arc $B3a-B3b$ from where the observer receives the signal. As the beam keeps on moving, this arc sweeps through the beam's width, again producing a strong signal. As a result of the above facts, at the superior conjunction, the observer detects two highly intense pulse components, symmetrically located around the center of the unbent pulse profile. This explains the structure observed in the pulse profile shown in Fig. \ref{fig:supBH_pulse_phase_90_i_90}. 

Next, we consider the situation where the pulsar is not at the superior conjunction, but also not very far from it, i.e., the geometry of Fig. \ref{fig:supBH_schematic_null_geodesic} is still valid. In such a situation, the  observer would be at a point different than $E^{\prime \prime}$ on the $X_3$-plane implying that the point LoS would be shifted from the location of the black hole on the $X_1$-plane. However, as we have already mentioned, we have chosen parameters ($\alpha$ and $\zeta_L$) in such a way that the LoS always falls on the LoBR. The rings $I_1$ (larger than $E$) and $I_2$ (smaller than $E$) on the $X_1$-plane can be chosen in such a way that the circles $I_1^{\prime \prime}$ and $I_2^{\prime \prime}$ on the $X_3$-plane coincide. As we aim to explain the pulse profile seen by the observer, we choose $I_1$ and $I_2$ rings in such a way that the observer lies on this common ring. As a result, the observer can receive light rays from both $I_1$ and $I_2$ provided these rings fall on the beam of the pulsar. The projection of this common ring on the $X_1$-plane is denoted by the ring $L$ ($=I_2^{\prime \prime} = I_1^{\prime \prime}$) which contains the LoS. The radius of $L$ increases as the pulsar moves away from the superior conjunction. However, all the configurations shown in Figs. \ref{fig:supBH_schematic_diagrams}, $L$ remains smaller than $E$. Additionally, $L$ is always smaller than $I_1$, but can be either larger or smaller than $I_2$. Each point on the ring $L$ can receive light from a point on $I_1$ and a point on $I_2$ provided those points are on the beam of the pulsar. When this particular point on $L$ is the LoS, the corresponding point on $I_1$ is called $P_1$ and the point on $I_2$ is called $P_2$. The locations of $P_1$ and $P_2$ on those rings depend on the location of the LoS on the $X_1$-plane. The finite size of the LoS results in finite sizes of $P_1$ and $P_2$, too.

Now, as we have already seen that the effect of the spin of the black hole on the bending of the light rays is negligible, the light rays remain confined in a plane containing the the pulsar (the source), the black hole (the gravitating body), and the initial direction vector of the light ray; as it happens in the case of a Schwarzschild black hole as the binary companion \citepalias{dbb23}. In Fig. \ref{fig:supBH_schematic_null_geodesic}, this plane for any chosen light ray would be the plane containing the $XY$-line and the initial direction of the ray, which would intersect the $X_1$-plane. In each panel of Fig. \ref{fig:supBH_schematic_diagrams}, the red dashed line represents this intersection. It is obvious that $P_1$, $P_2$, LoS, BH, all lie on this line. As we have seen in Fig. \ref{fig:supBH_schematic_null_geodesic}, the light rays do not change the side of the $XY$-line when the travel from $I_1$ to $I_1^{\prime \prime}$, $P_1$ and LoS is located on the same side of the BH in Figs. \ref{fig:supBH_schematic_diagrams}. On the other hand, as $I_2^{\prime \prime}$ is an inverted image, i.e., the light rays swap sides of the $XY$-line when they travel from $I_2$ to $I_2^{\prime \prime}$, $P_2$ and LoS are located on the opposite sides of the BH in Figs. \ref{fig:supBH_schematic_diagrams}.

In addition to replacing the circles $I_1$ and $I_2$ by annular rings, the finite size of the LoS would also make the observer receive signals from an arc $PQ$ on $I_1$ symmetric around $P_1$ and an arc $RS$ on $I_2$ symmetric around $P_2$. The length of the arc $PQ$ depends on the difference in sizes of $L$ and $I_1$. Similarly, the length of the arc $RS$ depends on the difference in sizes of $L$ and $I_2$. The differences in the sizes of $L$ and $I_1$ as well as in the sizes of $L$ and $I_2$ depend on the values of $i$ and $A_T + \omega$, and will be discussed later.

Note that, as the arc $RS$ is far from the LoS (it falls on the part of $I_2$ located on the other side of the BH), it rarely traverses the beam, contributing insignificantly to the profile. On the other hand, as the arc $PQ$ is closer to the LoS which falls on the LoBR, it traverses comparatively larger area of the beam. Hence, the observer at the near superior conjunction receives signal mostly from the arc $PQ$.

When we calculate the values of the bending delay, we consider only the point LoS, and hence use only $P_1$ not $P_2$ as the second one rarely falls on the beam. The difference in the longitudes of $P_1$ and the LoS in the $I$-frame of the pulsar (as defined in Sec. \ref{subsec:beamgeometry}) gives $\Delta \phi $ needed to calculate the longitudinal bending delay (Sec. \ref{subsec:deflightbending}). Similarly, the difference in the co-latitudes of $P_1$ and the LoS in the $I$-frame of the pulsar gives $\Delta \zeta $ needed to calculate the latitudinal bending delay (Sec. \ref{subsec:deflightbending}). This definition of bending delays fail exactly at the superior conjunction when the LoS receives light rays from multiple points at a time (all lying on the arc of the circle $E$ that traverses the beam as seen in \ref{fig:supBH_schematic_i_90_phase_90}.

\begin{enumerate}

\item The geometry for $i=89^{\circ}$, $A_T + \omega = 90^{\circ}$ has been shown in Fig.\ref{fig:supBH_schematic_i_89_phase_90}. This is closest to the superior conjunction configuration among our three choices of near superior conjunction configurations. In this configuration, light rays from both of the arcs $PQ$ and $RS$ can reach the observer (LoS on the $X_1$-plane). When the beam is at the position $B_1$, it intersects the arc $RS$ in a small segment labeled $RN$, from which light rays reach the observer. As the beam moves, the segment $RN$ moves on the beam - traversing a part on the bottom portion of the beam. When the beam is at the location $B2$, it neither cuts $I_1$ nor $I_2$. Hence, the observer does not receive any light. Afterward, when the beam is at the position $B_3$, it intersects the arc $PQ$ in a segment $MQ$ from which the observer receives the rays. As the beam moves, the segment $MQ$ moves on the beam - traversing a part on the top portion of the beam. As the length of the segment $RN$ is much smaller than the length of the arc $MQ$, we see a weak component at the beginning of the profile due to the beam area traversed by $RN$, then zero intensity (when the beam is at $B2$) and then a strong single component due to the beam area traversed by $MQ$. 

\item The geometry for $i=90^{\circ}$, $A_T + \omega = 85^{\circ}$ has been shown in Fig. \ref{fig:supBH_schematic_i_90_phase_85}. In this configuration, light rays from both of the arcs $PQ$ and $RS$ can reach the observer (LoS on the $X_1$-plane). When the pulsar beam is at the position $B_1$, neither the arc $PQ$ nor the arc $RS$ falls on the beam, hence the observer does not receive any signal. After some time, when the beam is at the position $B2$, the arc $PQ$ falls on it and the observer receives light from that arc. As the beam moves, the arc $PQ$ moves on the beam - traversing a part on the top portion of the beam, from which the observer continues to receive light. When the beam moves to the position $B_3$, it stops intersecting the arc $PQ$ and the observer does not receive any light anymore. The arc $RS$ never falls on the beam. Hence, we see a shifted pulse profile as shown in Fig. \ref{fig:supBH_pulse_phase_85_i_90}.  We see a four-component profile as the arc $PQ$ travels through the outer and the inner cones of the beam (but not the core).

\item The geometry for $i=85^{\circ}$, $A_T + \omega = 90^{\circ}$ has been shown in Fig. \ref{fig:supBH_schematic_i_85_phase_90}. It is very similar to the above case, only the location of the LoS with respect to the black hole is different. Here also, when the pulsar beam is at the position $B_1$, neither the arc $PQ$ nor the arc $RS$ falls on the beam, hence the observer does not receive any signal. After some time, when the beam is the position $B2$, the arc $PQ$ falls on it and the observer receives light from that arc. As the beam moves, the arc $PQ$ moves on the beam - traversing a part on the top portion of the beam, from which the observer continues to receive light. When the beam moves to the position $B_3$, it stops intersecting the arc $PQ$ and the observer does not receive any light anymore. The arc $RS$ never falls on the beam. Hence, we see a shifted pulse profile as shown in Fig. \ref{fig:supBH_pulse_phase_90_i_85}. We see a four-component profile as the arc $PQ$ travels through the outer and the inner cones of the beam (but not the core).

\end{enumerate}

Let us now try to understand how the situation changes as the configuration deviates more from the superior conjunction configuration, but only moderately so that the geometry of Fig. \ref{fig:supBH_schematic_null_geodesic} is still valid. As the deviation from the superior conjunction increases, the observer on the $X_3$-plane shifts away from $E^{\prime \prime}$, leading to a larger ring $L$ on the $X_1$-plane.

As we have need to consider $I_2^{\prime \prime} = L$ for the LoS to be able to receive light rays from $I_2$, the increase in the size of $L$ implies the increase of the size of $I_2^{\prime \prime}$. This is possible if the point $I$ comes closer to the $X_2$-plane, or in other words, if both $I_2^{\prime}$ and $I_2$ becomes smaller. As we already mentioned, being on the far side of $I_2$ from the LoS, the arc $RS$ rarely traverses the beam resulting very little contribution to the profile. This little contribution decreases even more when the size of $I_2$ itself decreases.

On the other hand, an increase in $L=I_1^{\prime \prime}$ is accompanied by an increase in the sizes of $I_1^{\prime}$ and $I_1$. However, as the phenomenon of bending is responsible for $I_1^{\prime}$ being smaller than $I_1$ and  $I_1^{\prime \prime}$ being smaller $I_1^{\prime}$, the differences in sizes of $I_1$, $I_1^{\prime}$, $I_1^{\prime \prime}$ decreases as $I_1$ becomes larger. In short, on the $X_1$-plane, the ring $I_1$ starts coming closer to the ring $L$. Hence, the length of the arc $PQ$ decreases as it comes closer to the LoS, leading to a decrease in the area this arc traverses on the beam (if any). Hence, the contribution of $I_1$ on the resulting pulse profile also decreases.

The above discussions confirm the fact that as the configuration keeps on deviating from the superior conjunction configuration, the `strong bending' keeps on becoming weaker and the enhancement of the profile strength decreases and the profile starts to show similarity with the unbent profile. Eventually, the strong bending as explained with the help of Fig. \ref{fig:supBH_schematic_null_geodesic} is no longer valid. Little enhancement in the pulse profile due to the weak bending might be still observable, which is the case always for the stellar mass black hole as discussed next.

Next we try to understand why such intriguing features are not seen when the binary companion of the pulsar is a stellar mass black hole instead of a supper-massive black hole. 

As we have mentioned earlier, in case of the stellar mass black hole, the radius of the circle $E$ is $3.61 \times 10^4$ km at and near the superior conjunction as the contribution from varying $i$ slightly is very small and in one orbit, $\omega$ is almost constant (see Eq. (\ref{eq:RE})). Similarly, the radius of the pulsar beam on the $X_1$-plane remains around $1.2 \times 10^6$ km at and near the superior conjunction. Hence, in a figure like Fig. \ref{fig:supBH_schematic_diagrams}, one would basically have $E$ and $I_2$ as tiny rings around the black hole. On the other hand, $L$ becomes a point (we make it of a finite size of computation purposes) in case of a superior conjunction, however, with a slight deviation from the superior conjunction $L$ becomes larger than $E$. As an example, for $i=89^{\circ}$, $A_T + \omega= 90^{\circ}$, the radius of $L$ is $2.65 \times 10^5$ km.

At the superior conjunction (when the LoS coincides with the black hole), the observer would always receive the light rays from a tiny (with respect to the beam) circle $E$ around the LoS as the beam is larger than $E$.

At near superior conjunction configurations, the observer would receive almost no light rays from $I_2$ as it is very tiny. On the other hand, $I_1$ is larger than $E$, but as we have already mentioned $I_1^{\prime \prime}$ (or $L$) is smaller than $I_1$. As gravitational bending is responsible for this difference in sizes of $I_1^{\prime \prime}$ and $I_1$, for stellar mass black hole, we can consider them to be of nearly equal size. As a result, the length of the arc $PQ$ would be very small (as it is very close to the LoS) and even when it traverses the beam, it would sweep only a very small part, contributing insignificantly to the profile.

Hence, strong bending never happens. Only if the LoS passes close enough to the black hole, it receives some extra light rays due to the weak bending. This explains the difference in the profile shapes of Fig. \ref{fig:pulse_profileAll} and Fig. \ref{fig:supBH_pulse_phase_90_i_90}, for both of the cases, the pulsar is at the superior conjunction, however, for the first case (stellar mass black hole), the profile is much similar to the unbent profile than it is for the second case (super massive black hole).

\bsp	
\label{lastpage}
\end{document}